\documentclass{article}

\usepackage{arxiv}

\usepackage[utf8]{inputenc} % allow utf-8 input
\usepackage[T1]{fontenc}    % use 8-bit T1 fonts
\usepackage{hyperref}       % hyperlinks
\usepackage{url}            % simple URL typesetting
\usepackage{booktabs}       % professional-quality tables
\usepackage{amsfonts}       % blackboard math symbols
\usepackage{nicefrac}       % compact symbols for 1/2, etc.
\usepackage{microtype}      % microtypography
\usepackage{lipsum}		% Can be removed after putting your text content
\usepackage{graphicx}
\usepackage{doi}
\usepackage{tabularx} % Import the tabularx package
\usepackage{amssymb}% http://ctan.org/pkg/amssymb
\usepackage{pifont}% http://ctan.org/pkg/pifont
\usepackage{url}
\usepackage{bbding}
\newcommand{\cmark}{\ding{51}}%
\newcommand{\xmark}{\ding{55}}%

\usepackage{color, soul}
\usepackage{microtype}
\soulregister\cite7
\soulregister\ref7
\soulregister\pageref7

\NewDocumentCommand{\rot}{O{60} O{1em} m}{\makebox[#2][l]{\rotatebox{#1}{#3}}}%

\usepackage[colorinlistoftodos]{todonotes}

\usepackage[printonlyused]{acronym}
\newacro{NER}[NER]{Named Entity Recognition}
\newacro{CTI}[CTI]{Cyber Threat Intelligence}
\newacro{NLP}[NLP]{Natural Language Intelligence}

\title{NLP-Based Techniques for Cyber Threat Intelligence}

\author{%
  Marco Arazzi \textsuperscript{1},
  Dincy R. Arikkat\textsuperscript{2},
  Serena Nicolazzo\textsuperscript{3},
  Antonino Nocera\textsuperscript{1}, \\
  Rafidha Rehiman K. A.\textsuperscript{2},
  Vinod P. \textsuperscript{2,4,*},
  Mauro Conti\textsuperscript{4}
}

\affiliation{%
  \bf{\textsuperscript{1}} Department of Electrical, Computer and Biomedical Engineering, University of Pavia, Italy\\
  \bf{\textsuperscript{2}} Department of Computer Applications, Cochin University of Science and Technology, India \\
  \bf{\textsuperscript{3}} Department of Computer Science, University of Milan, Italy \\
  \bf{\textsuperscript{4}} Department of Mathematics, University of Padua, Italy \\
  \bigskip
  * Corresponding author: \texttt{vinod.p@cusat.ac.in}, \texttt{vinod.puthuvath@unipd.it}
}

\date{}

\hypersetup{
pdftitle={NLP-Based Techniques for Cyber Threat Intelligence},
pdfsubject={Survey},
pdfauthor={Marco Arazzi, Dincy R. Arikkat, Serena Nicolazzo, Antonino Nocera, Rafidha Rehiman K. A., Mauro Conti, Vinod P.},
pdfkeywords={Cyber Threat Intelligence, Natural Language Processing, Web Crawling, Named Entity Recognition, Knowledge Graph},
}

\begin{document}
\maketitle

\begin{abstract}
	In the digital era, threat actors employ sophisticated techniques for which, often, digital traces in the form of textual data are available. Cyber Threat Intelligence~(CTI) is related to all the solutions inherent to data collection, processing, and analysis useful to understand a threat actor's targets and attack behavior. Currently, CTI is assuming an always more crucial role in identifying and mitigating threats and enabling proactive defense strategies. In this context, NLP, an artificial intelligence branch, has emerged as a powerful tool for enhancing threat intelligence capabilities. This survey paper provides a comprehensive overview of NLP-based techniques applied in the context of threat intelligence. It begins by describing the foundational definitions and principles of CTI as a major tool for safeguarding digital assets. It then undertakes a thorough examination of NLP-based techniques for CTI data crawling from Web sources, CTI data analysis, Relation Extraction from cybersecurity data, CTI sharing and collaboration, and security threats of CTI. Finally, the challenges and limitations of NLP in threat intelligence are exhaustively examined, including data quality issues and ethical considerations. 
This survey draws a complete framework and serves as a valuable resource for security professionals and researchers seeking to understand the state-of-the-art NLP-based threat intelligence techniques and their potential impact on cybersecurity.
\end{abstract}

% keywords can be removed

\section{Introduction}
\label{sec:intro}

Nowadays, Cyber Threat Intelligence~(CTI, hereafter) has gained a paramount role in cybersecurity, providing organizations with valuable insights into the  Tactics, Techniques, and Procedures~(TTPs) employed by cyber adversaries. Indeed, since there has been a significant increase in the variety and number of cyber attacks and malware samples, security practitioners have started to rely on CTI to promptly recognize the indicators
of a cyber attack, collect information about the attack methods, and respond to it accurately and timely. Hence, in broad terms, the CTI pipeline takes as input cybersecurity data and produces valuable insights aimed at enhancing proactive cybersecurity defenses. This includes formulating strategies to minimize the impact of cyber attacks and prevent them in the future.
\par
Due to the intricate and multifaceted nature of the cyber security context, different organizations and experts may emphasize different aspects of CTI based on their specific needs, objectives, and contexts. For this reason, several definitions of CTI exist.
For instance, one of the most popular definitions states that CTI is ``evidence-based knowledge, including context, mechanisms, indicators, implications, and actionable advice about an existing or emerging menace or hazard to assets that can be used to inform decisions regarding the subject's response to that menace or hazard'' \cite{mcmillanDef22}. The authors of \cite{shackleford2015s} state that CTI refers to ``the set of data collected, assessed, and applied regarding security threats, threat actors, exploits, malware, vulnerabilities, and compromise indicators''. Finally, Dalziel \cite{dalziel2014define} describes CTI as ``data that has been refined, analyzed, or processed such that it is relevant, actionable, and valuable''.
\par
Despite the numerous CTI definitions and the various strategies for effectively utilizing them, a common baseline revolves around identifying suitable solutions to collect, manage, and analyze CTI data.
In this context, recent research findings have proved that Artificial Intelligence~(AI), and especially Natural Language Processing~(NLP) techniques, play a crucial for several reasons. Firstly, because much of the information is represented by unstructured text data, such as threat reports, social media posts, news articles, and hacker forums, NLP allows CTI analysts to automatically handle these textual data and to enable the uncovering of relevant threats and trends. The analysis process of such information can be significantly sped up by automatically extracting key details from large volumes of text, such as Indicators of Compromises~(IoCs) and TTP. Moreover, NLP can unravel the relationships between entities 
 and events mentioned in text data, helping contextualize the information and building a more complete view of cyber threats and the actors behind them. Additionally, we can gain information about threat actors by analyzing online communications, forums, and other text sources through NLP techniques. Furthermore, because the CTI landscape is variegated, NLP can help to integrate information from various sources and formats into a coherent view, also assisting cybersecurity practitioners in the reports generation and sharing. Finally, NLP has also been exploited as a powerful tool to identify emerging threats and trends by analyzing large datasets of text-based information, thus proactively averting potential cyber attacks.
\par
Taking all the above reasons into account, we can affirm that NLP plays a crucial role in enhancing the effectiveness and efficiency of CTI by helping analysts cope with vast amounts of textual data, extracting actionable intelligence, and predicting the evolution of cyber threats. 
Nevertheless, a comprehensive survey on NLP-based techniques for CTI is still missing in the current scientific literature. Consequently, with this paper, we aim to fill this gap by providing a clear and updated picture of the CTI landscape, its multiple sources, and its gathering and sharing methods under a main perspective, focusing on NLP techniques and how they can enhance CTI ability to detect, analyze, and respond to cyber threats, effectively.

We list our contributions as follows:
\begin{itemize}
    \item We discuss the notion of Cyber Threat Intelligence and its life cycle.
    \item We undertake a comprehensive examination of crawling techniques to extract data from main CTI Web sources, namely social networks, clear, and Dark Web. 
    \item We analyze NLP models, techniques, and applications within the CTI domain. In particular, we give a detailed picture of the main papers dealing with text classification, similarity, clustering, summarization, cross-lingual, and topic detection in the context of CTI. 
    \item We describe the most recent results of NLP-based techniques for relation extraction and representation.
    \item We address the topics of CTI standardization protocols and results sharing.
    \item We exhaustively examine the challenges and limitations of NLP in threat intelligence, including data quality issues, CTI security, and ethical considerations.
\end{itemize}

The outline of this paper is as follows. In Section \ref{sec:related_work}, we present the related survey studies. Section \ref{sec:methodology} discusses the methodology we adapted to conduct this survey. Section \ref{sec:cti} explains the CTI life cycle and different types of CTI. In Section \ref{sec:nlp}, we make an overview of NLP techniques and their role in CTI. Section \ref{sec:collection} deals with the different Web crawling techniques to collect CTI data from clear, Dark/Deep Web, and social networks. Section \ref{sec:analysis} details classical NLP-based techniques applied to CTI data, whereas NLP-based techniques for Relation Extraction are described in Section \ref{sec:rel_extraction}. Section \ref{platform} deals with papers about CTI sharing and collaboration. Section \ref{sec:attack} focuses on security aspects related to CTI and, in particular, on adversarial attacks. In Section \ref{sec:challenges}, we
discuss several open challenges and insights for further research work. Finally, in Section \ref{sec:conclusion}, we draw our conclusion to the survey.

\section{Related Work}
\label{sec:related_work}

This section provides an overview of the existing literature surveys in the field of CTI.
\par
A survey on CTI mining techniques and the CTI knowledge acquisition taxonomy is presented in \cite{sun2023cyber}. This survey is quite recent and analyzes the methodology that transforms cybersecurity-related information into evidence-based knowledge for proactive cybersecurity defense using CTI mining, proposed by papers published before 2022. 
Cascavilla et al. \cite{cascavilla2021cybercrime} focused on papers that deal with CTI data crawling, incident prediction and avoidance, and standardization of cyber-criminal activities in the Surface, Deep, and Dark Web.
Similarly to our work, Rahaman et al. \cite{rahman2023attackers} analyzed different extraction purposes, namely CTI text classification,
IoCs and TTPs extractions. Moreover, they identified several types of textual sources for CTI extraction (i.e., hacker forums, threat reports, social media posts, and online news articles), and they observed that NLP and Machine Learning~(ML) based techniques, such as supervised classification, Named Entity Recognition~(NER), topic modeling, and dependency parsing, are the primary techniques used for CTI extraction. 
\par
The surveys presented in \cite{montasari2021application,ibrahim2020challenges,samtani2020trailblazing} analyzed the application of AI and ML in producing actionable CTI. While the authors of \cite{montasari2021application} only mentioned a work related to NLP and gave a brief survey on data collection and sharing. The authors in \cite{ibrahim2020challenges} conducted a short discussion on ML and AI tools (such as adversarial learning). The survey presented in \cite{samtani2020trailblazing} summarizes commonly used cyber security data sources with several approaches dealing with adversarial learning.
The paper illustrated in \cite{samtani2020cybersecurity} provides a systematic review of existing CTI
platforms within the industry; they reviewed the major data sources (without detailing the crawling techniques) and gave some insights on visualization, report generation, and intelligence dissemination. Moreover, they mentioned the enhancement of NLP and text mining as potential opportunities in the field of CTI.
\par
The works described in \cite{basheer2021threats,miloshevska2019dark} provide an analysis of the
current role of the Dark Web as an environment that facilitates cybercrime and illicit gain. In particular,  \cite{basheer2021threats} compares state-of-the-art research studies that leverage the dark
Web as an information source for CTI, describing their goals, approaches, tools, case studies, results, and possible limitations. A group of surveys focuses mainly on NER aspects related to the cybersecurity domain \cite{gao2021review,georgescu2021survey}. Specifically, the authors of \cite{gao2021review}
introduced common NER models, methods, and related resources.
\par
The work presented in \cite{tounsi2018survey} focuses on Technical Threat Intelligence (TTI, hereafter) and the major problems related to it, CTI sharing, and evaluates the most known open source tools offering TTI. Surveys \cite{el2023survey,wagner2019cyber,abu2018cyber,sauerwein2021threat} specifically deal with CTI Sharing.
In particular, El-Kosairy et al.\cite{el2023survey} highlight the most recent contributions, which discussed how blockchain is integrated with CTI to solve the issue of CTI sharing.
The survey presented in \cite{wagner2019cyber} gives an insight into diverse problems pertaining to CTI sharing between 2001 and 2018.
Specifically, it deals with the way of establishing a threat-sharing program with decentralized stakeholders, focusing on
what information can be shared, with whom, and how to automate some of the collaboration processes. Moreover, it also covers topics like CTI anonymization, encrypting the data, and presenting privacy risk scores. The work presented in \cite{schlette2021comparative} sheds light on existing standardization approaches for incident responses. Finally, Sun et al. \cite{sun2018data} described 
state-of-the-art cybersecurity incident prediction schemes and methods; moreover, the type of datasets used in each work is identified and referenced in minute detail.
\par
With respect to the previous surveys reported in this section, our paper introduces important contributions to the research community. Indeed, first, we analyze the publications produced in a very recent period (i.e., from 2018 to 2023); this is a crucial aspect because, with the technological improvement of AI solutions, novel NLP-based strategies for CTI are more and more advanced, and effective with respect to the results obtainable only a few years ago.
Moreover, while existing works already consider some overlapping aspects with our survey, our proposal analyzes the recent CTI literature completely from the perspective of the exploitation of NLP techniques, thus providing a comprehensive report of how NLP-based solutions have been successfully adopted in the reference context.
Table \ref{tab:relatedSurveys} provides a summary analysis of the contributions of the existing related surveys compared to ours.

\begin{table*}
\caption{Surveys related to our work}
\footnotesize
\centering
\begin{tabular}{|l|llllllcp{1cm}|}
    \hline
    Paper & \rot{Literature timeline}& \rot{CTI Definition} &\rot{Data Crawling} &\rot{NLP for CTI} &\rot{NLP for Relation}\rot{Extraction} &\rot{Adversarial Attack}&
    \rot{Standardization}\rot{Protocols}&\rot{CTI Sharing} \\
    \hline \hline
    Sun et al. \cite{sun2023cyber}& 2014-2022 &\cmark & \cmark & \cmark& \cmark & & \cmark & \cmark \\
    Rahman et al.\cite{rahman2023attackers}& 2013-2022 &\cmark & \cmark & \cmark& \cmark & &  &  \\
    Cascavilla et al.\cite{cascavilla2021cybercrime}& 2002-2020 &\cmark & \cmark & \cmark& & &  & \cmark \\
    Montasari et al. \cite{montasari2021application}& 2015-2020 &\cmark & \cmark & & & &  & \cmark \\
    Samtani et al.\cite{samtani2020cybersecurity}& 2013-2019 &\cmark & & & & \cmark &  & \cmark \\
    Samtani et al. \cite{samtani2020trailblazing} & 2012-2020 &\cmark & \cmark & & & &  & \cmark \\
    Ibrahim et al. \cite{ibrahim2020challenges}& 2014-2020 & \cmark&  & &  & \cmark &  & \\
    Basheer et al. \cite{basheer2021threats}& 2017-2021 &\cmark & & \cmark & &  &  &  \\
    Miloshevska \cite{miloshevska2019dark}& 2015-2018 & \cmark & &  & & &  &  \\
    Gao et al. \cite{gao2021review}& 2007-2019 & & & & \cmark &  &  &  \\
    Georgescu et al. \cite{georgescu2021survey}& 2007-2019 &  & & & \cmark & &  &  \\
    Tounsi et al.\cite{tounsi2018survey}& 2010-2017 & \cmark & & & & & \cmark & \cmark \\
    Abu et al.\cite{abu2018cyber} & 2011-2017 & \cmark & & & & &\cmark & \cmark \\
    Wagner et al. \cite{wagner2019cyber} & 2001-2018 & \cmark & & & & & \cmark & \cmark \\
    Sauerwein et al.\cite{sauerwein2021threat} & 2012-2021 & & & & & & & \cmark \\
    El-Kosairy et al.\cite{el2023survey} & 2019-2022 & \cmark & & & & &\cmark & \cmark \\ 
    Schlette\cite{schlette2021comparative}& 2001-2020 & & & & & &\cmark & \\ 
    Sun et al. \cite{sun2018data}& 2013-2018 & & \cmark& & & & & \\ 
   \hline
   \textbf{Our Survey} & \textbf{2018-2023} & \cmark & \cmark & \cmark & \cmark& \cmark &\cmark &\cmark  \\
    \hline\hline

\end{tabular}
\label{tab:relatedSurveys}
\end{table*}

\section{Methodology}
\label{sec:methodology}

\begin{figure}[ht]
    \centering
    \includegraphics[width=0.4\linewidth]{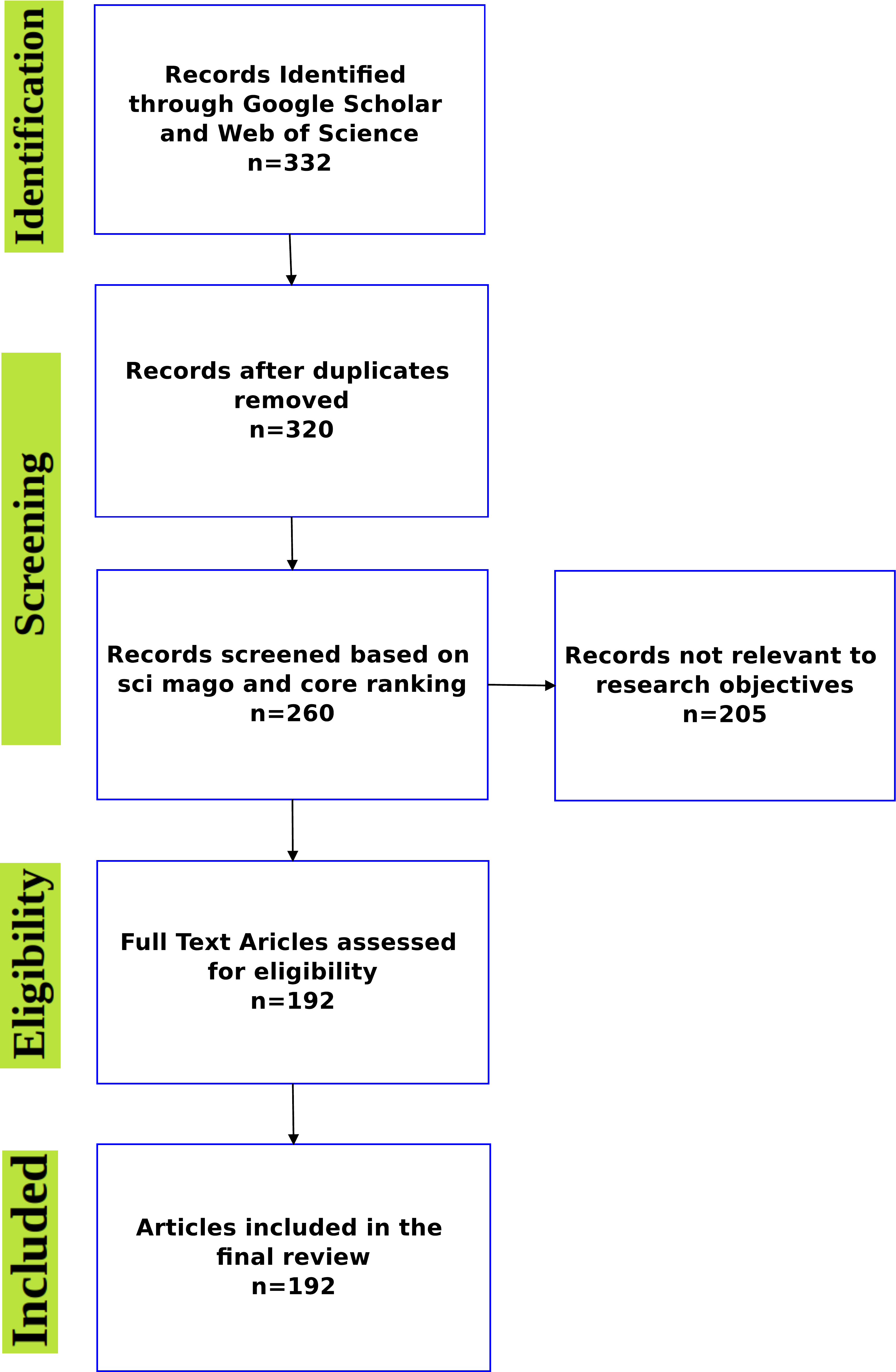}
    \caption{PRISMA Flowchart for paper selection process}
    \label{fig:prisma}
\end{figure}

In this study, we undertake a comprehensive literature review process, which can be outlined as follows: it involves establishing clear search objectives, entails the identification of relevant literature from journals and conference proceedings via a search engine driven by a well-defined search strategy, and encompasses the application of selection criteria to filter the articles during this iterative process. Figure~\ref{fig:prisma}  presents the PRISMA flow diagram, providing a visual representation of the screening process. The diagram highlights the count of research works identified, excluded, and included.
\subsection{Research Objectives}
The primary objective of this comprehensive literature survey is to conduct an in-depth analysis of NLP models, techniques, and their applications within the CTI domain. This entails a detailed exploration of common data sources and the methodologies employed for data collection in the CTI field. The study delves deeply into the relevant literature, with a specific focus on papers that address topics like text classification, similarity, clustering, summarization, cross-lingual analysis, and topic detection, all within the context of CTI. The research extends its purview to encompass recent developments in NLP-based techniques for relation extraction. The survey also examines the literature related to adversarial attacks in the CTI field. Moreover, this research explores CTI standardization protocols and the sharing platforms, contributing to the enhancement of collaboration and knowledge dissemination in the CTI community. Furthermore, the study aims to identify areas in CTI where improvements are needed and to pinpoint potential research directions that can advance the capabilities of CTI. Lastly, this paper undertakes an exhaustive examination of the challenges and limitations that NLP encounters within the context of threat intelligence.
\subsection{Search Strategy}
To gather relevant research focusing on NLP-based techniques for CTI, we designed a search strategy to align with our research objectives. We meticulously conducted searches through Google Scholar and Web of Science, thus accessing a wealth of scholarly resources. In addition, our search strategy encompassed a wide array of esteemed academic databases, such as the Institute of Electrical and Electronics Engineers (IEEE) Xplore, the Association for Computing Machinery (ACM) Digital Library, ScienceDirect, and SpringerLink. The search scope encompasses a broad spectrum of publication years, spanning from 2018 to 2023, ensuring a well-rounded coverage of recent research in the field. Further, we developed the search terms to facilitate our initial exploration. The search terms assist in identifying relevant research and employ a combination of specific keywords and phrases to encompass various aspects of the threat intelligence and cybersecurity domain. The primary search term includes ``threat intelligence", ``cyber intelligence", and ``CTI". In conjunction with the primary term; we integrated additional search terms such as ``indicators of compromise", ``tactics, techniques and procedures", ``hacker forum", ``dark web", ``APT attack", ``cyber attack", ``cyber security", ``named entity recognition", ``knowledge graph", ``cyber security relation extraction", ``adversarial attack", ``CTI platform", and ``CTI standardization".
\subsection{Selection Criteria}

In this section, we list the set of selection criteria we exploited to determine whether a scientific paper, identified through the search queries, is relevant to our study and reaches enough quality to be included in this survey. A paper is eligible for inclusion in the present work if it meets at least one of the inclusion criteria and none of the exclusion criteria applies.
At the end of this screening, 192 papers have been selected based on inclusion/exclusion criteria. 

\subsubsection{Inclusion Criteria}

To evaluate the relevance of a paper and include it in our survey, we followed the subsequent criteria:

\begin{itemize}
    \item we consider the corresponding author or the supervisor's importance in the field under analysis;
    \item we count the number of citations (we refer to Google Scholar \cite{Scholar} and Scopus; \cite{Scopus} Web sites to establish this value);
    \item we privilege more recent works;
    \item we take into account the importance of the journal (or the conference) where the paper has been published (we consider Scimago \cite{Scimago} and Core.edu \cite{Core} as ranking Web sites for journals and conferences respectively to establish this value). 
\end{itemize}

\subsubsection{Exclusion Criteria}

After the inclusion, the exclusion process is followed. A paper is excluded if only one of the following criteria is met:

\begin{itemize}
    \item the paper is not written in English;
    \item the paper is not peer-reviewed or not presented at reputable conferences;
    \item the paper's year of publication is earlier than 2018;
    \item the paper is relevant in the NLP context but it is not specifically focused on CTI;
    \item the paper lacks relevance, it is an incremental refinement of an earlier proposed approach, there is a duplicate publication, or there is a most recent and cited advancement of the work published in a more important journal or conference.
\end{itemize}
\section{Overview of Cyber Threat Intelligence}
\label{sec:cti}

CTI enables organizations to understand the evolving threat landscape, enhance their defenses, and respond effectively to emerging cyber attacks. Organizations adopt various types of CTI and go through various stages. The CTI life cycle and types are explained below
\subsection{Threat Intelligence Life Cycle}
CTI is a data-driven process that helps organizations defend against cyber threats and attacks. It involves several stages, working together to collect, process, and analyze data, providing valuable insights for cyber security. This systematic process transforms data into actionable intelligence, requiring continuous adaptation to evolving needs and real-world events. Different researchers have highlighted various versions of the CTI life cycle, with Irshad et al.~\cite{irshad2023cyber} mentioned six stages, Basheer et al.~\cite{basheer2021threats} and Ainslie et al.~\cite{ainslie2023cyber} mentioned four stages. They all typically follow a cyclical and iterative process, allowing for ongoing review and adaptation. The core principles of intelligence collection, analysis, and dissemination remain crucial in addressing the ever-changing cyber threat landscape. In light of existing literature, the main CTI phases can be grouped into six stages, namely Planning and Direction, Collection, Processing, Analysis, Dissemination, and Feedback. The CTI process begins with \textit{Planning and Direction}, where organizations establish their objectives, priorities, and requirements for threat intelligence. In the \textit{Collection} phase, raw data is gathered from diverse sources, including open-source feeds, commercial threat intelligence providers, and internal logs. This data may include IoCs, malware samples, and network traffic logs collected through automated tools and manual research. Once data is collected, it needs to be refined and organized. \textit{Processing} phase involves the conversion of raw data into a more structured format that can be used for analysis. In the \textit{Analysis} phase, the processed data is analyzed to identify patterns, trends, and potential threats. Once the threat intelligence is analyzed and assessed, the findings are shared with relevant stakeholders during the \textit{Dissemination} phase. Information is typically disseminated in a clear and actionable format, such as reports or alerts. Continuous improvement is a crucial aspect of a mature CTI system, with feedback collected from various stakeholders to evaluate the effectiveness of threat intelligence, identify areas for improvement, and address emerging requirements. The \textit{Feedback} stage supports the refinement and adaptation of the CTI program to changing threats and organizational needs.
\subsection{Types of Cyber Threat Intelligence} 
CTI encompasses a wide range of subcategories, highlighting the diversity of information sources and purposes. These subcategories include open-source intelligence, technical intelligence, human intelligence, measurement and signature intelligence, social media intelligence, geospatial intelligence, signal intelligence, Deep or Dark Web intelligence, and communication intelligence~\cite{basheer2021threats}. However in most of studies~\cite{tounsi2018survey,sauerwein2021threat}, CTI classified into four types:
\begin{itemize}
    \item \textit{Strategic Intelligence}: Strategic threat intelligence aims to provide valuable insights for business decisions and processes, considering the impact of external factors such as geopolitics on adversaries' actions. Strategic threat intelligence relies on expert analysts and authoritative sources to collect and interpret data from various channels, facilitating organizations in gaining insights into past incidents, understanding the intentions of threat actors, recognizing emerging trends, and developing risk mitigation strategies. It encompasses aspects like financial consequences, attribution of attacks, and industry-specific perspectives. By connecting global events and policies with potential cyber threats, strategic threat intelligence empowers decision-makers to assess risks and align cybersecurity investments with their strategic objectives. This type of intelligence is typically conveyed through comprehensive reports. It provides top executives and IT management with the means to identify risks, threat actors, and the origins of breaches while focusing on long-term challenges and real-time alerts for critical assets.
\item \textit{Operational Intelligence}: Operational threat intelligence serves as a vital tool for organizations to gain a deeper understanding of potential threat actors, including their motives, capabilities, and opportunities for launching attacks, while also identifying vulnerable IT assets and assessing the potential consequences of successful attacks. This form of intelligence is typically amassed by government entities and greatly benefits incident response and forensic teams by enabling them to implement security measures for enhanced detection, early recognition of attacks, and safeguarding critical assets. The sources of operational threat intelligence predominantly comprise human intelligence, data from social media platforms, chat rooms, and real-world events linked to cyber attacks. It involves the analysis of human behavior and evaluating threat groups to forecast and prepare for forthcoming attacks. Typically presented in comprehensive reports, operational threat intelligence includes in-depth information about documented malicious activities, suggested courses of action, and alerts about emerging threats.
\item \textit{Tactical Intelligence}: Tactical threat intelligence is a pivotal component in the protection of an organization's assets, offering valuable insights into the TTPs employed by threat actors in cyber attacks. This intelligence category primarily serves the needs of cybersecurity professionals, including IT service managers, security operations managers, administrators, and architects. It empowers these experts to gain an in-depth understanding of adversaries' attack methodologies, anticipate potential data breaches, assess the technical proficiencies and objectives of the attackers, and pinpoint the avenues of attack. Consequently, it enables security personnel to proactively devise strategies for detecting and mitigating threats, which may involve integrating known indicators into security products and rectifying system vulnerabilities. Tactical threat intelligence is sourced from various outlets, such as campaign reports, incident reports, malware analysis, attack group reports, and human intelligence. The collection of this data entails activities such as analyzing technical papers, collaborating with other organizations, and obtaining information from third-party sources. This intelligence encompasses intricate technical details, including the specifics of malware, details of campaigns, the methodologies employed in attacks, and information about the tools used, often documented in forensic reports.
\item \textit{Technical Intelligence}: Technical threat intelligence is centered on understanding the tools and resources used by attackers in their malicious activities, encompassing elements like command and control channels. In contrast to tactical threat intelligence, this form of intelligence has a shorter lifespan but is highly specific, enabling swift distribution and response to threats. While tactical threat intelligence covers the broader malware used in attacks, technical threat intelligence delves into the detailed implementation aspects of that malware, encompassing specifics such as IP addresses, domains, phishing email headers, and malware checksums. This intelligence is typically consumed by Security Operations Center~(SOC) staff and incident response (IR) teams, which is crucial in identifying malicious activities. The indicators used in technical threat intelligence are sourced from active campaigns, attacks on other organizations, or data feeds from external parties and are often gathered during investigations of attacks on different entities. Security professionals leverage this information to bolster the detection capabilities of security systems like IDS/IPS, firewalls, and endpoint security systems. This, in turn, facilitates the identification of malicious traffic and suspected IP addresses involved in malware distribution and spam emails. Technical threat intelligence is directly integrated into digital security devices to block and identify malicious traffic, both incoming and outgoing, within an organization's network.
\end{itemize}

\begin{table}
\caption{Summary of the acronyms used in the paper.}
\scriptsize
\centering
\begin{tabular}{|l|l|}
\hline
    Symbol & Description \\
    \hline \hline
    APT & Advanced Persistent Threat\\\hline
    CTA & Cyber Threat Actor\\\hline
    CTI & Cyber Threat Intelligence \\\hline
    CVE & Common Vulnerabilities and Exposures\\\hline
    DNF & Dark Net Forum\\\hline
    DNM & Dark Net Marketplace\\\hline
    FinTech & Financial Technology\\\hline
    GNN & Graph Neural Network\\\hline
    IDS & Intrusion Detection System \\\hline
    IoC & Indicator of Compromise \\\hline
    IRC & Internet Relay Chat \\\hline
    KG & Knowledge Graph\\\hline
    ML & Machine Learning\\\hline
    NER & Named Entity Recognition\\\hline
    NIST & National Institute of Standards and Technology\\\hline
    NLP & Natural Language Processing\\\hline
    OIE & Open Information Extraction\\\hline
    OSINT &  Open Source Intelligence \\\hline
    OSN &  Online Social Network \\\hline
    SVM & Support Vector Machine \\\hline
    TTI & Tactical Threat Intelligence\\\hline
    TTPs & Tactics, Techniques \& Procedures\\\hline
    CRF & Conditional Random Field \\ \hline
    STIX & Structured Threat Information Expression \\ \hline
    BERT & Bidirectional Encoder Representations from Transformers \\ \hline
    XAI & eXplainable Artificial Intelligence \\ \hline
    NVD & National Vulnerability Database \\
    \hline
\end{tabular}
\label{tab:SystemSymbols}
\end{table}
\section{Natural Language Processing in Cyber Threat Intelligence}
\label{sec:nlp}

This section is devoted to presenting the most common techniques employed in the NLP pipeline and how they can be applied to process CTI reports. In particular, we will describe the fundamental strategies of preprocessing and tokenization to sanitize the text from possible noise to enhance the robustness of the model built upon. Following that, we will introduce different representation techniques, from the most basic to the advanced ones, that make use of Deep Learning~(DL) models to obtain more sophisticated and contextualized representations. Then, we will introduce the most modern NLP models that represent the state-of-the-art in combination with the tool that implements them.  

\subsection{Preprocessing and Tokenization}
\label{sub:processing}

Preprocessing and Tokenization are fundamental preliminary cleaning phases in the pipeline of an NLP  task. Data cleaning is important to reduce the dimension of each text, remove noise, and improve the performance of models that take sanitized text as input. 
Common strategies to preprocess the data \cite{deliu2018collecting}, are listed in the following:

\begin{itemize}
    \item remove HTML tags if the crawled resource is an HTML page;
    \item replace characters not in {A-Za-z0-9(),!?’‘} with whitespace;
    \item convert text to lowercase,
    \item replace leading/trailing white spaces with a single space, 
    \item remove stop-words and duplicates,
    \item filtered out frequent and common words (i.e., appearing in more than 85\% of the documents),
    \item filtered out rare words (i.e., appearing fewer than three times in the entire corpus).
\end{itemize}

A common tool for data preprocessing is Stanford CoreNLP\footnote{ https://stanfordnlp.github.io/CoreNLP/} \cite{zhao2020timiner}.
Also, the studies that leverage social network posts (i.e., tweets) follow similar preprocessing steps. The common ones, listed in \cite{behzadan2018corpus}, are: (1) conversion of all the characters of a tweet to lowercase, (2) tokenizing the text according to white-space separations, (3) removing tokens that are
not encoded in ASCII or that are not comprised
of alpha-numeric characters, (4) removing punctuations from each token, (5) substitute digits with word representations, (6) remove stop words, and (7) stem tokens. For the last two steps, \cite{behzadan2018corpus} utilizes the functionalities available in the Natural Language ToolKit~(NLTK) libraries \cite{loper2002nltk}. 

\par
To disambiguate terms, which are inflected forms of the same word, two techniques are the most common: lemmatization and stemming.
The former consists of substituting the inflected forms in the text with the original word and the latter, instead, reduces the inflected token to their root. These techniques have an important role also in the CTI approaches that involve NLP. An example of this can be seen in \cite{orbinato2022automatic, gao2021enabling, gao2021system, kadoguchi2020deep, behzadan2018corpus, yang2020automated,ji2019feature, chan2022feedref2022} where the authors apply NLP-based strategies to extract CTI from unstructured sources, automatically. In particular, they applied a sanitization phase that includes stemming, lemmatization, and stopword/tag removal, to the descriptions of adversarial techniques. Such techniques are very common and can be applied before any ML algorithm. Models like BERT \cite{devlin2018bert} require as input a sequence of a fixed length. In these situations, the input must preprocessed by either truncating it or adding a padding sequence to reach the required length. In addition, because such a model can interpret only numerical values, tokenizers are used to convert
input tokens into numerical identifiers, encoding representations according to their vocabulary.
For example, this approach is adopted by the authors of \cite{evangelatos2021named, kuehn2023threatcrawl, dasgupta2020comparative, wang2023novel} where they use BERT to perform \ac{NER} and threat classification in CTI.
In particular, in addition to padding and truncation, they took into consideration a case-sensitive tokenizer because capitalized letters could be informative of phrases referring to a ransomware name. The authors of these papers also use tokenizers to split complex words into terms to make them identifiable in a reference vocabulary.
Table \ref{tab:preprocessingTokenization} reports a summarized view of the different techniques adopted by reference papers in the CTI domain.

\begin{table}
\caption{Preprocessing and Tokenization.}
\scriptsize
\centering
\begin{tabular}{|l|l|l|l|}
\hline
    Paper & Lemmatization/ & Stopwords Removal & Tokenization \\
          & Stemming       &                   &              \\
    \hline \hline
    Orbinato et al. \cite{orbinato2022automatic} & \cmark & \cmark & \cmark \\
    Gao et al. \cite{gao2021enabling} & \cmark & \xmark & \xmark \\
    Gao et al. \cite{gao2021system} & \cmark & \xmark & \xmark \\
    Kadoguchi et al. \cite{kadoguchi2020deep} & \cmark & \cmark & \cmark \\
    Evangelatos et al. \cite{evangelatos2021named} & \xmark & \xmark & \cmark \\
    Kuehn et al. \cite{kuehn2023threatcrawl} & \xmark & \xmark & \cmark \\
    Dasgupta et al. \cite{dasgupta2020comparative} & \xmark & \xmark & \cmark \\
    Wang et al. \cite{wang2023novel} & \xmark & \xmark & \cmark \\
    Yang et al. \cite{yang2020automated} & \cmark & \xmark & \cmark \\
    Ji et al. \cite{ji2019feature} & \cmark & \cmark & \cmark \\
    \hline\hline

\end{tabular}
\label{tab:preprocessingTokenization}
\end{table}

\subsection{Text Representation Techniques} 
\label{sub:RepresentationTechinques}

As already mentioned in the previous section, neural networks and statistical algorithms cannot interpret text without converting it into a numerical representation. Representation techniques can leverage different strategies, ranging from statistical approaches, such as the very common Bag-of-Words \cite{zhang2010understanding}, or representation obtained by a neural network, as done by Word2Vec \cite{mikolov2013efficient}.
\par
As for statistical approaches, Bag-of-Words is a well-known strategy that represents text by counting the occurrences of the words in the corpus of the documents. The idea behind this approach is to measure the importance of a word according to its frequency in relation to a specific outcome of the target ML algorithm. With this strategy, word frequencies are considered with the same relevance in the computation of the final result. However, this can be considered the main flaw of the Bag-of-Words approach since some of the words with high frequency, like `IP' or `Port' in the CTI field, are actually noise that negatively affects the performance of the target model. To solve this problem, some enhancements to the Bag-of-Words strategy have been proposed. One of the most common solutions in this sense is TF-IDF \cite{ramos2003using}. TF-IDF follows the same philosophy of Bag-of-Words by encoding the importance of a word and assigning a score to them, but instead of simply counting the word occurrences, it weights their contributions according to how frequent they are in the different documents. In this way, this approach lowers the contribution of too frequent terms.
\par
Another method based on statistical analysis is GloVe \cite{pennington2014glove}. GloVe is an algorithm that generates vector representations of words, building the co-occurrence matrix between words and encoding the information regarding the probability ratio of co-occurrence probabilities of words. As for the approaches using neural network representations, a well-established method is Word2Vec. Unlike previous methods, Word2Vec uses a neural network to create a space where words are represented as vectors, and their positions in this space reflect how they relate to each other. The Word2Vec approach proposes two strategies. The first strategy is called Continuous Bag-of-Words~(CBOW) and is used to generate word representation starting from the representation of the context. The second one, called Skip-Gram, is used to generate the context vector starting from a single word.
\par
The strategies introduced above are fundamental tools to generate representations of text to be fed into NLP models targeted to CTI tasks like, topic modeling or classification \cite{kadoguchi2020deep, nayak2019analyzing, ampel2020labeling, kadoguchi2019exploring, dionisio2019cyberthreat, kim2020automatic, le2019gathering, zhao2020timiner, adewopo2020exploring, bo2019tom, kristiansen2020cti, pantelis2021strengthening, li2020open}.
For instance, the solution proposed by \cite{nayak2019analyzing} applies a TF-IDF strategy to generate representations of malicious URLs that are then fed to a K-means algorithm to obtain the different topics.
\par
In \cite{kim2020automatic} Bag-of-Characters representation, a Bag-of-Word strategy applied at a character level is used to generate the embedding of short-text CTI reports to be processed by a BiLSTM network for the NER task. Similarly, in \cite{dionisio2019cyberthreat}, Word2Vec and GloVe are used in the same way on short text documents about CTI from Twitter.
Another example of this is proposed by \cite{kadoguchi2020deep}, where the authors analyze posts from the Dark Web. To do so, they exploited \emph{doc2Vec}, which works analogously to Word2Vec but instead of generating word representations it produces vector embeddings of the entire document. Then, the obtained vectors are used as input to an unsupervised clustering algorithm. The approach presented in \cite{ampel2020labeling} uses an embedding matrix generated using GloVe as input to a C-BiLSTM network that aims at exploiting source code labeling. Analogously, the work presented in \cite{kadoguchi2019exploring} generates document representations using doc2Vec, as done by \cite{kadoguchi2020deep}, but in this case, it is used to classify cyber attacks using a Multilayer Perceptron~(MLP) network.
\par
A schematic report of the techniques used by the papers analyzed in this section is visible in Table \ref{tab:representationsTech}.

\begin{table}
\caption{Text Representation Techniques in the literature}
\scriptsize
\centering
\begin{tabular}{|l|l|l|l|l|}
\hline
    Paper & Bag-of-Words/ & TF-IDF & GloVe & Word2vec/ \\
          & Bag-of-Characters & &     & doc2Vec             \\
    \hline \hline
    Nayak et al. \cite{nayak2019analyzing} & \xmark & \cmark & \xmark & \xmark \\
    Kim et al. \cite{kim2020automatic}  & \cmark & \xmark & \xmark & \xmark \\
    Dionisio et al. \cite{dionisio2019cyberthreat}  & \xmark & \xmark & \cmark & \cmark \\
    Kadoguchi et al. \cite{kadoguchi2020deep}  & \xmark & \xmark & \xmark & \cmark \\
    Ampel et al. \cite{ampel2020labeling}  & \xmark & \xmark & \cmark & \xmark \\
    Kadoguchi et al. \cite{kadoguchi2019exploring}  & \xmark & \xmark & \xmark & \cmark \\
    Le et al. \cite{le2019gathering} & \xmark & \cmark & \xmark & \xmark \\
    Zhao et al. \cite{zhao2020timiner}  & \xmark & \xmark & \xmark & \cmark \\
    Adewopo et al.\cite{adewopo2020exploring} & \xmark & \cmark & \xmark & \xmark \\
    Bo et al. \cite{bo2019tom} & \xmark & \cmark & \xmark & \xmark \\
    Kristiansen et al. \cite{kristiansen2020cti} & \xmark & \cmark & \xmark & \xmark \\
    Pantelis et al. \cite{pantelis2021strengthening} & \xmark & \cmark & \xmark & \xmark \\
    Li et al. \cite{li2020open} & \xmark & \cmark & \xmark & \xmark \\
    Queiroz et al. \cite{queiroz2019eavesdropping} & \xmark & \xmark & \xmark & \cmark \\
    Kadoguchi et al. \cite{kadoguchi2019exploring} & \xmark & \xmark & \xmark & \cmark \\
    Biswas et al. \cite{biswas2022text} & \xmark & \cmark & \xmark & \xmark \\
    Li et al. \cite{li2019extraction} & \xmark & \cmark & \xmark & \xmark \\
    Ayoade et al. \cite{ayoade2018automated} & \xmark & \cmark & \xmark & \xmark \\
    Goseva et al. \cite{goseva2018identification} & \xmark & \cmark & \xmark & \xmark \\
    \hline\hline

\end{tabular}
\label{tab:representationsTech}
\end{table}

\subsection{NLP Algorithms and Models}
\label{sub:algorithmsTools}

In Section \ref{sub:RepresentationTechinques}, we presented well-known and established representation techniques. Nowadays, more advanced neural models have emerged to build sophisticated techniques of representation and learning. As a matter of fact, one of the main drawbacks of previous approaches is the inability to have different representations of the words in different contexts. For instance, algorithms such as Word2Vec and GloVe generate the same word representations in any context. To overcome this flaw, modern models like ELMo \cite{peters2018deep} or those based on the transformer architecture can generate different embeddings according to the surroundings of the words \cite{devlin2018bert}. ELMo uses a bidirectional LSTM trained to generate different contextualized representations for each word to solve the ambiguity of words with different meanings. In particular, the ELMo model has been trained on a large set of data to predict the next word of a sequence. With such characteristics, this model is capable of generating generally enough representations to be used in different tasks or even improved with accurate fine-tuning.
\par
This aspect is shared also with transformer-based models like BERT \cite{devlin2018bert} and GPT \cite{radford2018improving}. The original architecture of BERT is characterized by an encoder-only model proposed in two versions that differ in the number and dimension of their encoding blocks. The basic one is composed of 12 encoding blocks with 768 hidden units each; the larger model is characterized by 16 blocks with a size of 1024 units. Compared to LSTM models, transformers are unaware of the concept of sequence. To solve this issue, a positional embedding layer is added at the entrance of the BERT model to encode in each word representation information about their position in the sentence. One of the main characteristics of BERT refers to the strategy used to pre-train the model. The training has been performed on two different tasks contemporaneously. In particular, in the first task, the model objective is to predict the original value of a masked word in a sentence. With the second task, instead, the authors train BERT to predict the next sentence. Like ELMo, BERT can be used to produce different representations of a word according to the context. The second example of a transformer-based model is GPT, which recently received huge attention due to its text generation capabilities \cite{gao2023scaling, eloundou2023gpts, openai2023gpt}. Different from BERT, GPT is characterized by a decoder-only architecture. The authors proposed different evaluations of their GPT model over the years. In its latest version, GPT-4, the training is characterized by a reinforcement learning phase that uses rewards given by humans.
This strategy allows the model to achieve extra quality in the produced text compared to the previous versions. 
\par
As said earlier, due to the large amount of data used for the training, the models discussed in this section are general enough to be used as a backbone in NLP pipelines in the CTI field. This is demonstrated by the authors of \cite{zhou2023cdtier}, who test their proposed NER dataset with different combinations of models that exploit representation produced by ELMo and BERT. Another example can be seen in \cite{wang2023novel}, where the authors slightly modify the input of a BERT model to adapt it to a different objective, i.e., restoring the original order of a permuted text. The authors of \cite{alam2022cyner} proposed a Python library for \ac{NER} tasks on CTI documents, which allows the import of pre-trained BERT models. In \cite{yin2020apply}, the authors presented EXBERT, a framework that uses BERT as a backbone to generate the embeddings that are hence provided as input to a classifier. The objective is to predict the exploitability of vulnerabilities from their software descriptions. The approach proposed by \cite{liu2022tricti} uses BERT for data augmentation in the context of cybersecurity. In ~\cite{ranade2021generating}, a GPT-2 fine-tuned model is used to generate fake CTI reports to demonstrate a data poisoning attack on a knowledge extraction system. Instead, the authors of \cite{setianto2021gpt}  proposed to use a GPT-2 fine-tuned model to parse Unix commands from log files in real time. A schematic summary of the NLP tools used by the papers described above is visible in Table \ref{tab:representationsTech1}.

\begin{table}
\caption{NLP solutions used for Text Representation in the reference literature} 
\scriptsize
\centering
\begin{tabular}{|l|l|l|l|l|}
\hline
    Paper & ELMo & BERT & GPT \\
    \hline \hline
    Zhou et al. \cite{zhou2023cdtier} & \cmark & \cmark & \xmark \\
    Wang et al. \cite{wang2023novel} & \xmark & \cmark & \xmark \\
    Alam et al. \cite{alam2022cyner} & \xmark & \cmark & \xmark \\
    Yin et al. \cite{yin2020apply} & \xmark & \cmark & \xmark \\
    Liu et al. \cite{liu2022tricti} & \xmark & \cmark & \xmark \\
    Ranade et al. \cite{ranade2021cybert}  & \xmark & \cmark & \xmark \\
    Chan et al. \cite{chan2022feedref2022} & \xmark & \cmark & \xmark \\
    Ranade et al. \cite{ranade2021generating} & \xmark & \xmark & \cmark \\
    setianto et al. \cite{setianto2021gpt} & \xmark & \xmark & \cmark \\
    \hline\hline

\end{tabular}
\label{tab:representationsTech1}
\end{table}

\subsection{NLP Libraries and Tools}

The processing procedures presented in \ref{sub:processing} are well-known and established techniques implemented in many libraries. Two of the most popular choices are the libraries implemented in Python NLTK\footnote{https://www.nltk.org/} and spaCy\footnote{https://spacy.io/}. NLTK is a very easy-to-use library that implements many different useful tasks for preprocessing and also basic algorithms for clustering, such as K-means and classification, e.g., Naive Bayes. As for the preprocessing tasks, it implements several lemmatizations and stemming approaches and provides lists of stopwords to remove in different languages.
SpaCy, instead, provides more advanced tools like wrappers for neural network models implemented in the main frameworks Tensorflow and Pytorch with GPU support. Interestingly, in the latest version of spaCy at the time of writing this article, developers added support to transformer-based models. Another powerful tool that offers advanced capabilities like tokenization, POS-Tagging, and \ac{NER} is the one proposed by Stanford, called Stanford CoreNLP\footnote{https://stanfordnlp.github.io/CoreNLP/}, available for both Python and Java languages.
When it comes to transformer-based models, such as BERT and GPT described in Section \ref{sub:algorithmsTools}, one of the most well-known and complete ecosystems providing easy and intuitive access to such models in an NLP workflow is Hugginface\footnote{https://huggingface.co/}. Huggingface allows an easy import of pre-trained models and associate tokenizers to be used for many different tasks, including fine-tuning. In addition, the developers offer an open hub where it is possible to upload fine-tuned models freely downloadable by the community.
\par
Most of the papers in the literature that involve NLP \cite{behzadan2018corpus,le2019gathering,pantelis2021strengthening,bose2019novel,orbinato2022automatic,dong2019towards, goseva2018identification} use the NLTK library to exploit implemented algorithms, such as TF-IDF vectorizer, stop-words or stemming/lemmatization. However, many others \cite{liu2022tricti,ji2019feature,gao2021system,gao2021enabling,ranade2021cybert, satvat2021extractor} use spaCy for vector representation and stemming/lemmatization. 
Stanford CoreNLP is typically used as an additional advanced library by many solution-proposing approaches for \ac{NER} and POS-tagging \cite{wang2023novel,satvat2021extractor,dionisio2019cyberthreat,evangelatos2021named,sun2022cyber}.
Finally, as we stated before, transformer-based models are recently becoming very popular, and pre-trained ones are often exploited thanks to the support of the Huggingface community \cite{ranade2021cybert,evangelatos2021named,chan2022feedref2022}. In such papers, pre-trained transformer-based models are used after a fine-tuning step for the specific task of CTI reports analysis.

\section{NLP-Based Techniques for Threat Intelligence Collection}
\label{sec:collection}

The aim of this section is to survey the solutions for CTI data gathering from Online Social Networks~(OSNs, for short), the Clear Web, and the Dark/Deep Web, where malicious actors collaborate and communicate to plan cyber attacks. 
Not all information gets published into standard CTI databases and appliances; CTI is often shared in unstructured ways like blogs, posts, or threat reports from security companies or experts. For this reason, multiple online data sources are used as signals to generate warnings indicative of new potential cyber threats. Information gathering is the first and crucial step for collecting relevant data about new vulnerabilities, exploits, security alerts, threat intelligence reports, and security tool configurations.
From the collection perspective, data can be divided into two categories: {\em(i)} indicator-based data, which mainly provides IoCs for quick attack prevention, and {\em(ii)} document-based data, which may contain richer and more comprehensive threat information than the former category, but, at the same time, may require more complex NLP techniques for the analysis. The first group is, in turn, divided into high-level IoCs (represented by TTPs, malware, and tools) and low-level IoCs (e.g., IP, URL, hash, domain name, source/destination port, timestamp, and infection type). In both cases, preliminary steps for crawling Web pages about malicious content involve {\em(i)} gathering URLs on the internet, {\em(ii)} filtering out benign content, and then {\em(iii)} downloading malicious ones.

\subsection{Crawling from Clear Web}

\begin{figure}[ht]
	\centerline{
        \includegraphics[scale=0.4]{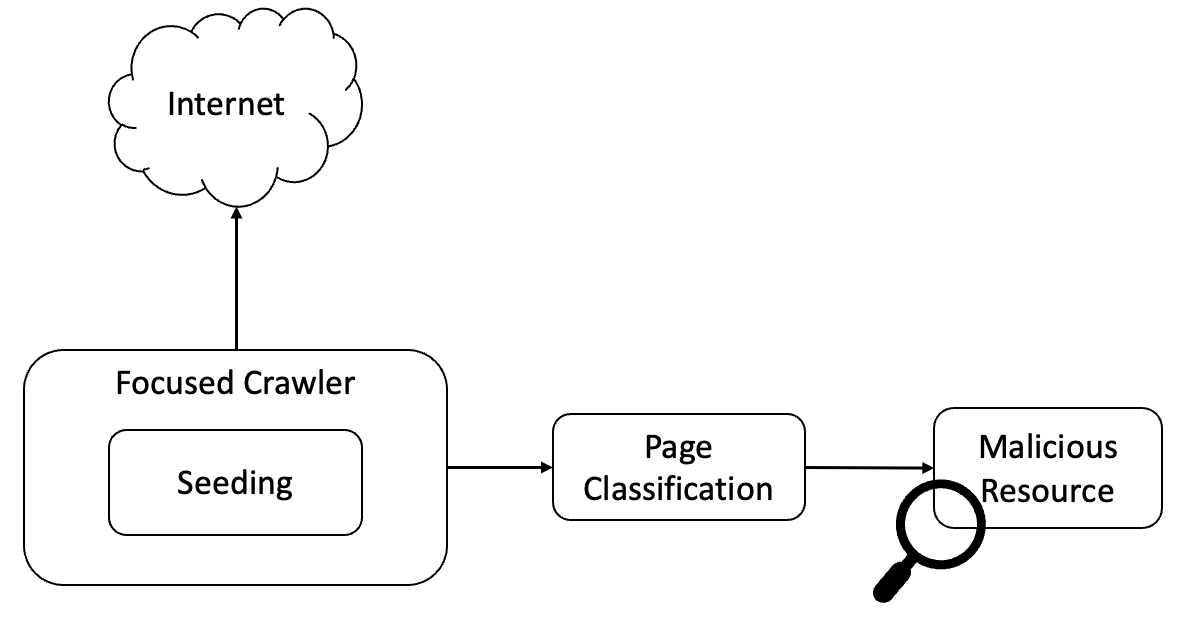}
    }
    \caption{General Architecture for crawling the Clear Web \label{fig:genArchitecture}}
\end{figure}

The clear (also Open or Surface) Web represents the standard, publicly accessible portion of the Internet, where information is openly available, indexed by search engines, and accessible via conventional Web browsers.
Since their introduction, Web crawlers have been intended to gather data from Clear Web.
In Figure \ref{fig:genArchitecture}, a general architecture for crawling this kind of Web is shown.
Focused crawlers scan the Internet starting from some seed URLs. Usually, a {\em Seeding} module is in charge of locating initial URLs to be used by focused crawlers. Since they only search
for relevant information, their performance can go down if different clusters of Web pages are isolated; hence, choosing a high number of appropriate seeds can solve this issue. To improve the performance of the overall architecture, inside the Seeding component, ML models can be used to classify URLs before navigating the Web pages. Finally, a page classification module is launched to assess if the Web page includes malicious content.

In \cite{koloveas2019crawler}, the authors proposed a two-stage 
architecture, including a crawling module and a content ranking module. The first part produces a focused crawler that employs NLP to decide if some target Web sites deal with specific cybersecurity-related vulnerabilities. Specifically, a Support Vector Machine~(SVM) classifier is combined with a user-provided query relevant to the topic. Hence, with respect to the reference architecture in Figure \ref{fig:genArchitecture}, the Seeding module is enriched with this facility to improve seeds' selection. Moreover, the second component is in charge of assessing the relevance and usefulness of the crawled content. In the extended version of the paper \cite{koloveas2021intime}, the authors presented an architecture called inTIME, including a more complex crawling infrastructure built on top of NYU's ACHE crawler\footnote{https://github.com/ViDA-NYU/ache} and enhanced through the use of regex-based filters to direct the crawl to specific parts of a Web site. Moreover, to support focused crawling, inTIME includes the SMILE\footnote{https://haifengl.github.io} page classifier, which is based on an ML-based text classifier (e.g., SVM and random forest), trained by a selection of positive and negative examples of Web pages, to direct the crawl toward topically relevant Web sites.
\par
The issue of discriminating between benign and malicious Web pages before crawling their content is also faced by ``MalCrawler'', which has been designed to crawl and search malicious Web sites efficiently \cite{singh2017malcrawler}. In this system, the set of URLs representing the initial seeds' collection is processed, and only malicious links are navigated. MalCrawler can also prevent redirections and cloaking, generally used to avoid showing pages containing malicious content to search engine crawlers. The libraries used to develop MalCrawler are:
{\em(ii)} JSoup Library\footnote{http://www.jsoup.org/} for parsing Web pages and extracting hyperlinks, document content, and JavaScript tags; {\em(ii)} Rhino\footnote{https://developer.mozilla.org/docs/Mozilla/Projects/Rhino/} a JavaScript Emulation Library to analyze the runtime behavior of JavaScript; and {\em(iii)} HTML Unit\footnote{http://htmlunit.sourceforge.net/} a browser emulation library written in Java to test redirection, and cloaking.
Another recently presented focused crawler, called ThreatCrawl \cite{kuehn2023threatcrawl}, leverages the aforementioned general architecture, and it is based on both a retrieval and an extraction step. Additionally, this system presents a monitor module as well, that stores the state of each
retriever and extractor thread and can check 
for any critical error. Moreover, to extract the main content of an HTML Web Page and return it as parsed text, ThreatCrawl leverages both Trafilatura \cite{barbaresi2021trafilatura} and Beautiful Soup\footnote{https://www.crummy.com/software/BeautifulSoup/}.
Both the works presented in \cite{li2018security} and \cite{sun2022cyber} design a cybersecurity engine, according to which the data collection step is done manually from multiple data sources, which are updated periodically via a monitor. Specifically, the main scope of \cite{sun2022cyber} is to model and visualize security datasets based on users' queries and demands. Three datasets are exploited in this work, namely: {\em(i)} 1000 security news articles \cite{satyapanich2020casie} mentioning security events and annotated by experts; {\em(ii)} some vulnerability archives collected from authoritative vulnerability database; {\em(iii)} a number of tweets related to security keywords and extracted using a Python library called Twitterscraper\footnote{https://pypi.org/project/twitterscraper/0.2.7/}. 
\par
Similarly to the previous paper, also in the work presented in \cite{noor2019machine}, the authors crawl multiple data sources. To do so, a semantic search system is developed to facilitate searching for high-level IoCs in the CTI corpus. They identified 36 Cyber Threat Actors~(CTAs) and collected publicly available documents describing both the attack incidents (APT) and the CTAs. As also done in \cite{chen2022automatic,chen2021threat,perry2019no}, they leveraged a publicly available repository called APTNotes\footnote{https://github.com/kbandla/APTnotesarchived} containing cyber threat reports since 2008. Using the search engine and the repository, they collected 327 unstructured CTI documents. The work described in \cite{perry2019no} collects a set of publicly available reports starting from MITRE ATT\&CK2\footnote{https://attack.mitre.org/} (a globally accessible knowledge base of adversary tactics and techniques based on real-world observations); and APT Groups and Operations\footnote{https://airtable.com/shr3Po3DsZUQZY4we/}, which link threat actors to relevant reports. Through a manual expert review, the authors selected the most important threat actors for which there are a sufficient number of reports. The final dataset contains 249 reports, which describe attacks performed by 12 different actors. 
Similarly, the authors of \cite{wu2021price} leveraged a dataset of corpora collected on the XianZhi platform \footnote{http://service-corp.odps.aliyun-inc.com/api.}
\par
Another important source of CTI data can be found in the authoritative cybersecurity databases available online. As a matter of fact, authors of several papers \cite{dong2019towards,sun2020data,alves2020follow,ji2019feature,vishnu2022deep,sun2022cyber} leveraged data from online databases. Often, they used this data in combination with posts gathered from the Twitter Social Network as a ground truth database. The most popular employed databases are listed in the following:
\begin{itemize}
    \item Common Vulnerabilities and Exposures~(CVE) database\footnote{https://cve.mitre.org/}. Each vulnerability present in this storage is identified by a CVE ID. The Web portal allows the user to make queries by CVE IDs to gather detailed information about the already known threats as well as to provide a Common Vulnerability Score~(CVS) evaluating the impact of the given vulnerability.
    \item Common weakness enumeration (CWE)\footnote{https://cwe.mitre.org/} is a community-driven project that aims at identifying and classifying common software security weaknesses in a standardized and systematic way. Moreover, for each weakness listed in the CWE, a piece of associated information is present about how to identify, mitigate, and prevent the vulnerability.
    \item Common Attack Pattern Enumeration and Classification (CAPEC) database\footnote{https://capec.mitre.org/} provides a broad hold of a wide list of attack patterns, which are widely used by security analysts, developers, and testers. Similarly to CWE, CAPEC categorizes attack patterns into a hierarchical structure. Each attack pattern is assigned to a unique identifier and organized into different categories and families.
    \item  The Web Application Security Consortium~(WASC)\footnote{http://www.webappsec.org/} collects and organizes the threats related to the security of Web sites. Its main focus is to categorize and classify various Web application security threats and vulnerabilities, making it easier for organizations to understand and address them.
    \item NIST's National Vulnerability Database~(NVD)\footnote{https://nvd.nist.gov/} mirrors and complements CVE entries on their database. It is maintained by the National Institute of Standards and Technology~(NIST), which is a federal agency of the United States Department of Commerce. Every hour, NVD contacts CVE to obtain new CVE IDs of recently disclosed vulnerabilities.
    \item PacketStorm\footnote{https://packetstormsecurity.com/news/tags/database/} is an online resource and repository for a variety of current and historical cybersecurity-related information, tools, and databases. 
    \item Privacy Rights Clearinghouse (PRC)\footnote{https://www.privacyrights.org/} is an independently maintained collection of reports about cybersecurity incidents divided according to victim organization type (i.e., government agencies, businesses, medical service providers, educational institutions, etc.).
    \item Hackmageddon\footnote{https://www.hackmageddon.com/} is another reputable collection of public reports about cybersecurity incidents.
\end{itemize}

Table \ref{tab:clearCrawling} summarizes, for each analyzed paper, the technology used to crawl data and the amount of data crawled.

\begin{table*}
\caption{Crawling from Clear Web}
\scriptsize
\centering
\begin{tabular}{|l|l|l|}
\hline
    Paper & Technology used to crawl data & Sample Size \\
    \hline \hline
    Koloveas et al. \cite{koloveas2019crawler} & ACHE open-source focused crawler & 20,000 Web sites\\
    Singh et al. \cite{singh2017malcrawler}&  JSoup Library, Rhino, HTML Unit & a not defined number of Webpages\\
    Perry et al. \cite{perry2019no} & publicly available reports & 249 reports\\
    Kuehn et al. \cite{kuehn2023threatcrawl} &Trafilatura and Beautiful Soup&259 URLs\\
    Sun et al. \cite{sun2022cyber}&Manual Collection&1000 news articles \cite{satyapanich2020casie} and vulnerability archives from authoritative vulnerability database\\
    Noor et al. \cite{noor2019machine}& Customized Search Engine& 327 unstructured documents\\
    Chen et al.\cite{chen2021threat}& Github's APTNotes& 600 articles\\
    Dong et al. \cite{dong2019towards}& CVE and NVD databases& 78,296 CVE, 78,296 NVD entries \\
    Alves et al. \cite{alves2020follow}& CVE, PacketStorm and other CTI databases&455,026 entries\\
    Ji et al. \cite{ji2019feature}& PRC and Hackmageddon databases & 61,957 cybersecurity events\\
    Vishnu et al. \cite{vishnu2022deep}& CVE database & 134,091 vulnerability descriptions\\
    Li et al. \cite{li2018security}&Manual Collection& 610 collected articles\\
    Wu et al. \cite{wu2021price}& XianZhi platform&971 Web sites\\
    \hline\hline

\end{tabular}
\label{tab:clearCrawling}
\end{table*}

\subsection{Crawling from Social Media}

Online social networks provide an open environment where both offensive and defensive practitioners can engage in discussions, report incidents, and promote timely information about vulnerabilities, attacks, and malware. Among the online social networks, Twitter has always been one of the most popular in the scientific research context, mainly because of {\em(i)} its large and international user base (almost $400$ million users), {\em(ii)} its set of available APIs, that allows developers to easily integrate Twitter data into their systems, and {\em(iii)} the tagging functionality that allows users to label their tweets and search them via keywords (i.e., hashtags), thus providing a natural data aggregation capability. For all these reasons, Twitter has become the preferred platform for disseminating up-to-date IoCs and threat intelligence data and conducting threat hunting\cite{huntingTwitter23,arikkat2023can}.

\begin{figure}[ht]
	\centerline{
        \includegraphics[scale=0.35]{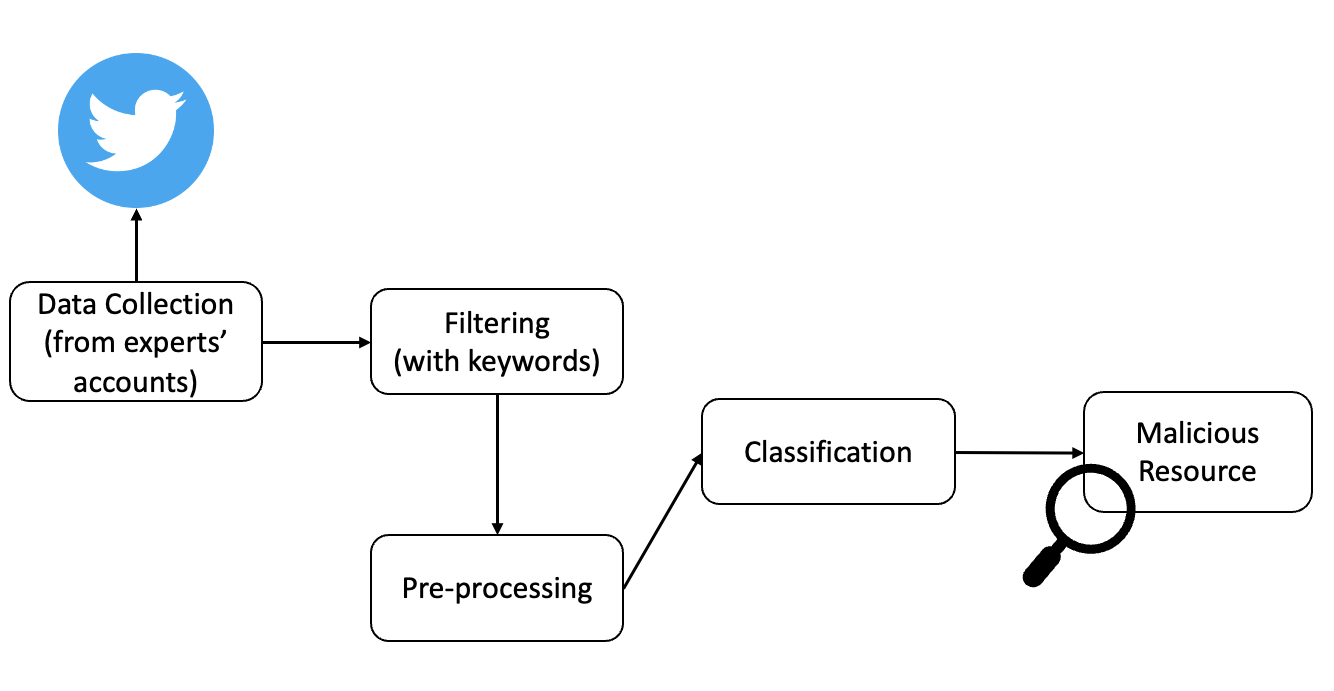}
    }
    \caption{General Architecture for crawling data from Twitter \label{fig:TwitterArch}}
\end{figure}

One of the main issues of gathering data from Twitter is to wisely choose starting keywords to filter the stream listener results and reduce the amount of irrelevant information that is gathered. Nevertheless,  crawling from this social network is quite a standard task for all the papers considered in this section and mainly proceeds with the execution of three tasks: data collection, keyword-based pre-filtering, and text pre-processing.
\begin{itemize}
    \item In the {\em collection} phase, an initial set of Twitter accounts is required. Starting from this list, the crawler usually queries Twitter's streaming API for tweets.
    \item In the {\em pre-filtering} phase, a set of user-defined keywords are used to drop irrelevant tweets.
    \item In the {\em pre-processing} phase, some transformation on data is performed to make representation uniform.
\end{itemize}
Figure \ref{fig:TwitterArch} summarizes the main steps explained above.

The works presented in \cite{dionisio2019cyberthreat,alves2021processing,rodriguez2019generating,sapienza2018discover,adewopo2020exploring} follow this standard crawling architecture to crawl tweets and classify them as potential cyber threats.
As for the collection step, in \cite{rodriguez2019generating}, the authors identified accounts related to Hacktivists, cybersecurity feeds, security researchers, and companies. Then they crawled tweets from their timeline, and after that, they filtered these tweets again based on a security keyword filter and on a keyword list obtained by the Department for Homeland Security \cite{binder2011}.
Also, the framework IoCMiner \cite{niakanlahiji2019iocminer} starts from an initial set of Twitter accounts to collect tweets about CTI. This initial set is built through Twitter user-defined lists of identified CTI experts
with a mechanism of validation that measures the relevancy, popularity, and comprehensiveness of the list in addition to the credibility of its owner. Moreover, it employs a CTI classifier to further filter out non-CTI tweets from the observed data streams. Finally, IoCMiner uses a set of regular expression rules to extract IoCs from the identified tweets.

Similarly, the framework presented in \cite{sapienza2018discover}, called DISCOVER, aims to process data from multiple sources (such as Twitter and security blogs), and by employing data mining techniques, it identifies novel terms related to a potential cyber threat in the form of a warning alarm. To gather data from Twitter, they leveraged some experts' profiles, which it identifies in the timeline of international researchers and security analysts. As for the security blogs, the authors manually chose a list of 290 blogs and extracted data from them via a custom crawler. Data from these two sources are filtered to find words related to cyber threats. 
\par
The works analyzed up to this point collect the set of raw data from experts' profiles and then apply a filter, based on keywords, on their timelines. The framework called \#Twiti \cite{shin2021twiti}, instead, performs data collection in two ways: user tracking and keyword tracking. To perform the latter modality, it extracts the top 100 words that appear in tweets containing IoCs and uses them as starting keywords. Moreover, the presence of external sources is detected and also used as search criteria (such as URLs to security vendor blogs or Pastebin.com).
\par
In \cite{behzadan2018corpus,liu2020event}, the authors presented a framework for the detection and classification of cyber threat indicators in the Twitter stream. Specifically, for \cite{behzadan2018corpus} the data collected consists of a corpus of $21,000$ tweets gathered with a custom stream listener developed through Tweepy\footnote{https://www.tweepy.org/}. This work proceeds only using a list of keywords as
a pre-filter for the stream listener, and after that, it performs two classification steps assessing the relevance of the cyber threats and the threat types. The first step is performed by leveraging the topic modeling API of IBM's Watson
Natural Language Understanding service \cite{chen2016ibm}. The latter, instead, is done manually by simple string matching on the rough
tweet text to find the type of thread. Similarly to \cite{behzadan2018corpus}, the framework CTI-Twitter \cite{kristiansen2020cti} uses a set of manually decided keywords as input for the streaming API, but it also leverages a Python library known as GetOldTweets\footnote{https://github.com/
Jefferson-Henrique/GetOldTweets-java} to enrich this set with a historical collection of tweets. In \cite{liu2020event}, authors presented a work to identify target domain-relevant feeds containing critical patterns, and they preprocessed the whole tweet stream, filtering it via keywords. The chosen keywords are related to traditional cybersecurity events, including account hijacking, data breaches, and denial of service attacks, and also to emerging events, such as ransomware and cryptocurrency mining malware. Interestingly, the authors also applied dynamic query expansion (DQE) to enrich tweet collection with tweets semantically related to cybersecurity. 
\par
The dataset used by \cite{sun2020data} comprises tweets collected using Twitter API and information crawled from vulnerability information databases, such as NVD. They linked the data from the two sources together if they refer to the same CVE ID. In particular, the initial seeds' set is composed by searching on Twitter all the users' profiles who periodically posted tweets containing the keyword ``CVE'' from June to September 2018. By observing the temporal characteristics of tweet activities, the authors found that Twitter message streams can directly reflect cyber threats. Similar works are presented in \cite{le2019gathering,alves2020follow,sauerwein2018tweet,syed2018takes,huang2020monitoring}
that learn the features of CTI from the CVE descriptions and classify each input tweet as either normal or anomalous. In particular, the framework presented in \cite{alves2020follow} aims at comparing some aspects of the information present on vulnerability databases with Twitter data. The authors start searching for tweets mentioning the vulnerabilities indexed on the vepRisk database \cite{andongabo2017veprisk} containing all entries published on NVD, PacketStorm, and other minor security databases. Moreover, they leveraged the GetOldTweets library to have access to tweets at any point in time. The pre-filtering phase is performed manually, and the final database resulted in $3,461,098$ tweets. The authors of \cite{sauerwein2018tweet}, instead, gather their data from Twitter by searching for tweets matching the CVE identifiers and enriching this set with additional information from the NVD database. In addition to the standard gathering of tweets about CVE IDs via the Twitter APIs, the peculiarity of \cite{syed2018takes} is that it queries the vendor sites to collect the patch release dates to determine whether a vulnerability has had immediate or deferred disclosure. Also, the paper presented in \cite{ji2019feature} uses two external sources (i.e., collections of reports, namely PRC and Hackmageddon) as a ground truth database to test the proposed method. Moreover, it leverages a large stream of tweets from GNIP's decahose\footnote{https://www.cabinetm.com/product/gnip/decahose}, which is a real-time trend detection and discovery solution delivering a $10\%$ sample of real-time tweets.
\par
The paper presented in \cite{horawalavithana2019mentions} deals with two social media conversation channels (Reddit and Twitter) and a
collaborative software development platform (GitHub) and aims to compare user-generated content related to security vulnerabilities on these three digital platforms. In particular, Reddit\footnote{https://www.reddit.com/} is also a popular social network among researchers for its openness and for its focused topic-based conversations structured around subreddits, whereas GitHub\footnote{https://github.com/} is one of the most prominent collaborative open-source software development platforms. This work shows that both Twitter and Reddit can be used to accurately predict activity on GitHub. As classically done, the dataset for this work is built by filtering posts from both Twitter and Reddit through keywords containing a vulnerability identifier (i.e., CVE) from the NVD subset. When a CVE identifier appears in a post or a comment, all the related messages (e.g., tweets, retweets, and replies) are also gathered. As for Reddit and Github, a search on subreddits (repositories, respectively) via regular expression is carried out to find the CVE IDs.
\par
Similarly to Github, {\em Paste sites} are Web sites or online platforms that allow users to easily share and store plain text snippets or code snippets for a temporary period. These snippets can include code, configuration files, notes, and other text-based information. The work presented in \cite{vahedi2021identifying}, leverages data from three prevailing {\em paste sites} for collection based on feedback from cybersecurity experts: Pastebin\footnote{https://pastebin.com/}, PasteFS\footnote{https://pastefs.com/}, and Pastelink\footnote{https://pastelink.net/}. The authors developed custom Web crawlers to collect each paste and their associated metadata (e.g., title, author).  

The social Web crawler presented in \cite{koloveas2019crawler} uses links to discussion threads on IoT vulnerabilities to traverse a forum structure and download all relevant discussions on the topic. To filter out parts that are not relevant, part of the forum employs regex-based link filters. Examples of forum crawled are: {\em(i)} Wilders Security Forums\footnote{https://www.wilderssecurity.com/threads/};
{\em(ii)} Oracle Security Blog\footnote{https://blogs.oracle.com/ security/}; and 
{\em(iii)} Security Forum\footnote{https://www.securityforum.org/events/}. In the extended version of this paper, presented in \cite{koloveas2021intime}, the author designed inTIME, a framework based on ML that can crawl data from multiple sources.
As for social network monitoring, the authors show a use case where the system collects live tweets using the Twitter Stream API with keywords related to IoT vulnerabilities and leverages a CVE database as a dictionary of keywords. Finally, the authors of TI\_spider \cite{zhao2020timiner} developed an automated data collection system from different social media, including blogs, hacking forum posts, security news, and security vendor bulletins. In particular, this system exploits $75$ independent distributed crawlers, each of which monitors and collects a specific data source using a breadth-first search to collect threat descriptions. Each crawler starts the collection from a homepage, including links to threat events. For each link, it crawls the HTML source codes and extracts threat event data leveraging Xpath (XML Path language).

Table \ref{tab:socialMediaCrawling} summarizes, for each analyzed paper, the Social Media analyzed, the technology used to crawl contents, the collected sample size, and the type of resource collected (i.e., tweets, users, replies, etc.).

\begin{table*}
\caption{Crawling from Social Media} 
\scriptsize
\centering
\begin{tabular}{|l|l|l|l|}
\hline
    Paper & Social Media & Technology & Sample Size \\
    \hline \hline
    Behzadan et al. \cite{behzadan2018corpus} & Twitter & Tweepy & 21,000 tweets\\
    Adewopo et al. \cite{adewopo2020exploring} & Twitter & Tweepy & 500,000 tweets\\
    Sapienza et al. \cite{sapienza2018discover} & Twitter, & Twitter API,  & tweets of 69 profiles\\
    & Security blogs & HTML parser  & 290 security blogs\\
    Sun et al. \cite{sun2020data} & Twitter & Twitter API  & tweets of 2,917 profiles\\
    Rodriguez et al. \cite{rodriguez2019generating} & Twitter & Twitter API  & 70,475 tweets\\
    Niakanlahiji et al. \cite{niakanlahiji2019iocminer} & Twitter & Twitter API  &  2,300 tweets\\
    Dionisio et al.\cite{dionisio2019cyberthreat}& Twitter & Twitter API  &  5,320 tweets\\
    Liu et al.\cite{liu2020event}& Twitter & Twitter API  & 2,031,766 tweets\\
    Shin et al.\cite{shin2021twiti}& Twitter & Twitter API  & 978,414 tweets\\
    Alves et al. \cite{alves2020follow} & Twitter & GetOldTweets  & 3,461,098 tweets\\
    Ji et al. \cite{ji2019feature}&Twitter & GNIP's decahose& 4,975,992,550 tweets\\
    Sauerwein et al.\cite{sauerwein2018tweet}& Twitter & Twitter API &709,880 tweets\\
    Syed et al.\cite{syed2018takes}& Twitter & Twitter API & 13,277 tweets\\
    Kristiansen et al. \cite{kristiansen2020cti} & Twitter & Twitter API, GetOldTweets & 76,047 tweets\\
    Horawalavithana et al. \cite{horawalavithana2019mentions} & Twitter, & Twitter API & 105,596 tweets, retweets, replies \\
    & Reddit, & Reddit API & 170,486 posts and comments\\
    & GitHub & Public download & 7,240,398 activities\\
    Zhao et al. \cite{zhao2020timiner}& Social Media & Custom crawlers&75 blogs, security forums, etc.\\
    Vahedi et al.\cite{vahedi2021identifying}& Pastebin,PasteFS, and Pastelink & Custom Web crawlers & 4,254,453 posts\\
    \hline\hline

\end{tabular}
\label{tab:socialMediaCrawling}
\end{table*}

\subsection{Crawling from Dark/Deep Web}

With the term Deep Web, we refer to all Web content that is not indexed by search engines (whose content is not necessarily hidden). This includes Web pages and databases that are not accessible through search engine queries. The Dark Web, instead, represents a smaller, hidden portion of the Deep Web that is accessible through specialized software (like TOR -  The Onion Router\footnote{https://www.torproject.org/}, the Invisible Internet Project, or Freenet). It typically makes use of special encryption software to hide users' identities and IP addresses and is often associated with illegal activities. The most frequently used platforms, where hackers gather to exchange malicious tools, information, and other content in the Dark/Deep Web, include hacker forums or Dark Net Forums (DNFs, for short), Dark Net Markets (DNM, hereafter), Internet Relay Chat (IRC, hereafter), and carding shops \cite{du2018identifying,benjamin2019dice}.
\par
Hacker forums are repositories of thousands of readily available exploits enriched with comprehensive metadata, and they primarily revolve around major themes, such as carding and exploits. In contrast, DNMs often feature a substantial amount of non-cybersecurity-related content, such as pornography and illegal drugs, while lacking valuable CTI metadata, as observed by Ebrahimi et al. in their study \cite{ebrahimi2020semi}. Furthermore, researchers often face a significant risk when seeking additional details from DNMs, as products may need to be purchased. IRC and carding shops enable plain-text conversations or the posting of stolen credit card information, but they typically do not allow hackers to share exploits. Due to the analytical challenges posed by DNMs, IRC, and carding shops, cybersecurity researchers frequently prefer to focus their efforts on hacker forums when examining exploit-related content for CTI. Consequently, a significant number of relevant studies focus on the detection of cyber threats within these hacker forums \cite{koloveas2021intime,almukaynizi2018darkmention,deliu2018collecting,williams2018incremental,tavabi2019characterizing}.
While hacker forums on the Dark/Deep Web are built on similar frameworks as traditional forums, they incorporate a variety of anti-crawling measures aimed at impeding large-scale, automated data collection. Common mechanisms include authentication, Turing tests, throttling, CAPTCHA images, IP address blacklists, obfuscation, paywalls, and network traffic analysis \cite{benjamin2019dice}. Moreover, usually, each forum framework
has a unique HTML structure, naming scheme, and different internet software packages. They are accessed through Tor, a network of servers running specialized software and providing anonymity to the user. In addition, many of them are written in different languages depending on their origin (the most frequent are English, Chinese, and Russian). For all these reasons, a traditional Web crawling approach cannot be directly applied to crawling hacker forums in the Dark/Deep Web. 
\par
\begin{figure}[ht]
	\centerline{
        \includegraphics[scale=0.35]{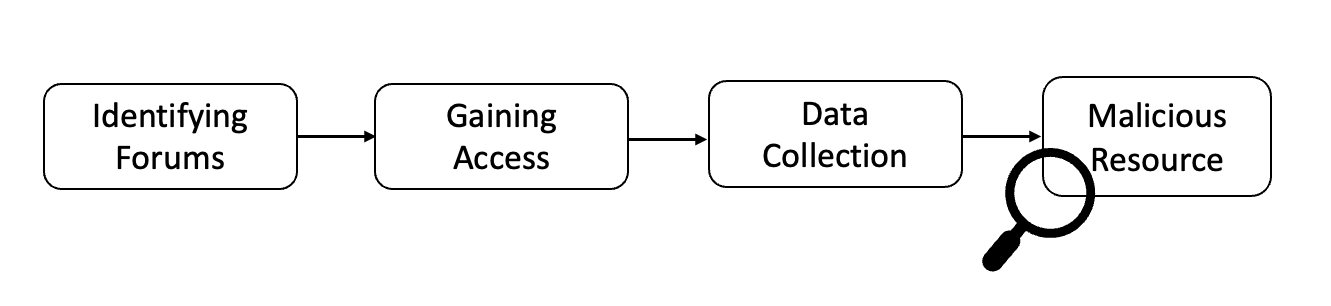}
    }
    \caption{General steps for crawling Dark/Deep Web \label{fig:darkCrawlingSteps}}
\end{figure}

The general steps usually performed by all the cited articles are shown in Figure \ref{fig:darkCrawlingSteps}, and they can be summarized as follows:

\begin{itemize}
    \item Identifying forums. The first challenge is the identification of the relevant target forums, i.e. those containing users and content related to cybersecurity intelligence. Some researchers exploit expert knowledge to identify possible hacker forums\cite{samtani2020proactively,du2018identifying} or leverage an already existing list of possible resources \cite{schafer2019blackwidow,koloveas2019crawler,williams2018incremental,arnold2019dark}. In many Dark Net forum conversations, participants may reference or share hyperlinks to other cyber-criminal communities or underground markets that can, in turn, be crawled\cite{benjamin2019dice,du2018identifying} (this mechanism is referred to as snowball identification). In general, due to the underground nature of the intended targets, obtaining a curated and always updated list is quite challenging; hence, the majority of works realize custom crawlers that traverse the Dark Web to find reachable sites over Tor. For this last set, the first suitable target forums are identified by hand or by a simple keyword search to bootstrap the process \cite{benjamin2019dice}. After obtaining a foothold, the content of these forums is analyzed to obtain further links and addresses to other targets in a more automated fashion in later iterations \cite{schafer2019blackwidow}. 
    \item Gaining access. Since forums often require some sort of authentication to access the site, crawlers need personal accounts to log in on each site. Some sites request new users to provide only a valid email address, and others may work only if invited by other active hackers or even require users to first buy credits.
    \item Data collection. After the previous phases, data collection is usually fully automated. This phase deals with establishing anonymous access to the forums over Tor and the collection of raw data. Usually, a custom Web crawler is developed. The crawler will automatically download the starting seed pages, identified in the first step, and constantly discover new pages by following encountered hyperlinks. Text parser programs, using regular expressions manually identified by researchers, can be used to automatically extract meaningful information from the HTML code of the Dark Net forums (i.e., author names, thread titles, and others).
\end{itemize}

The work presented in \cite{benjamin2019dice} gives useful insights into the steps of Dark Net identification and Data collection. Before beginning to search for data sources, it is crucial for researchers to take precautionary measures to create a secure research environment to safely download, analyze, and archive Dark Net content. Researchers could have to deal with malware or malicious JavaScript code when browsing underground forums. To enhance safety, the following suggestions are given:
\begin{itemize}
    \item Implementing virtual operating systems and networks to collect data on quarantined and isolated computers.
    \item Renting virtual private servers from cloud services to set up research tools on one server and then clone it to scale collection for different forums. This can provide resilience and scalability.
    \item Check for Terms of Service (ToS) violations when identifying potential cloud service providers or other external tools.
    \item Sanitized Collected data can be processed for secure long-term storage.
    \item The database should be placed on a network distinct from the servers used for downloading Dark Net content to prevent potential malware and other security threats from compromising the integrity of the archived data.
\end{itemize}
The work of \cite{benjamin2019dice} presents a complete framework called DICE-E for conducting Dark Net forums identification, collection, and evaluation. Moreover, it carries out an empirical demonstration by collecting information from 4 DNFs located in the United States, China, Russia, and Iran, namely \url{Antichat.ru}, \url{Ashiyane.org}, \url{HackHound.org}, and \url{Unpack.cn}.

The paper \cite{schafer2019blackwidow} presents BlackWidow, a system that monitors Dark Web services in real-time and continuously. Moreover, it adds custom functions that emulate typing and clicking behavior in order to login to forums automatically. Moreover, it employs the node.js headless Chrome browser puppeteer\footnote{https://pptr.dev} as a crawler within the Docker containers to collect forums' content and metadata. A similar framework is presented in \cite{samtani2022linking}, and it is called the Exploit Vulnerability Attention Deep Structured Semantic Model (EVA-DSSM). As for the collection part, the authors identified a large and popular exploit-specific hacker forum containing a variety of malicious tools and used a Web crawler routed through Tor to collect all exploit-category, post-date, author-name, platforms-targeted, and exploit-description data into a relational database. This resulted in $18,052$ exploits across four categories (namely, Web applications, local, remote, and DoS) targeting 31 operating systems, Web applications, and programming languages. 
\par
The custom crawler called HackerRank developed by Huang et al. \cite{huang2021hackerrank} gathers data from 5 forums (i.e., Nulled, HackThisSite, HiddenAnswers, BreachForum, Raidgets) collecting all the threads from the forum first, and then all the posts under the thread, including the username, profile, content, order, and time of the post. In addition, the authors also consider some mechanisms to deal with the anti-crawler mechanisms of the underground forums. Its aim is to realize an automatic method for identifying key hackers, also leveraging social network analysis metrics. In \cite{koloveas2019crawler}, the authors realize an architecture to extract CTI data not only from Clear Web and forums but also from specific Web sites on the Dark Web. This crawler is provided with several onion links that correspond to hacker forums or
marketplaces, where cyber-crime tools are sold, and zero-day vulnerabilities/exploits are monitored. After a first manual authentication, the HTML text of the Web site is extracted along with useful metadata. In the extended version of this paper \cite{koloveas2021intime}, the authors presented a complete framework capable of crawling from multiple and heterogeneous sources and supporting Dark Web crawling. This functionality relies on the use of TOR proxies to visit the user-specified onion links, and all the required actions (i.e., joining the TOR network, using the proxy, initializing the crawler) are automatically carried out via internal API calls. As for the authentication, a manual user login should be performed the first time the crawler encounters an authentication barrier. Then, session cookies are stored and used in all the subsequent crawler visits.
\par
The works presented in \cite{deliu2018collecting,suryotrisongko2022topic} exploit data from a breached Dark Web site called Nulled.IO\footnote{https://www.nulled.to} whose database is publicly available \footnote{http://leakforums.net/thread-719337} and from which the authors extracted hacker forum posts. Starting from this database, the authors of \cite{deliu2018collecting} identified the posts that are most relevant to cybersecurity through the SVM algorithm and clustered the relevant posts into topics using Latent Dirichlet Allocation~(LDA). The paper described in \cite{williams2018incremental} performed an incremental crawling strategy bypassing the anti-crawling measures of hacker forums to collect attachments. Information was collected from 10 hacker forums, namely OpenSC, Garage4hackers, Hacksden, AntiOnline, Crackingzilla, WebCracking, SafeSkyHacks, Ashiyane, Hack, and Haker. They leveraged the internet forum software package vBulletin\footnote{https://www.vbulletin.com/} and allowed users to embed attachments directly into their posts that can be freely accessed. All traffic is routed through Tor, thus ensuring anonymity. After the connection to Tor is successfully established, the Python crawler begins an automatic search for attachments, performing a Depth-First Search (DFS) approach for collection. This means that the crawler starts with one subforum and crawls each topic and posts within that subforum before moving on to the next subforum.

Tavabi et al. \cite{tavabi2019characterizing} examined the dynamic patterns of discussion and identified forums with similar patterns among 80 hacker forums in the Dark and Deep Web. To do so, they exploited an already existing crawler infrastructure presented in \cite{nunes2016darknet}. As classically done by Dark and Deep Web crawlers, it uses anonymization protocols, such as Tor and I2P, and handles authentication to access non-indexed sites. Similarly to the previous research, the works presented in \cite{tavabi2018darkembed,marin2018predicting,marin2018mining,almukaynizi2018darkmention} leverage the same infrastructure for crawling the Dark Web and Deep Web introduced by \cite{nunes2016darknet}. The authors of \cite{tavabi2018darkembed} created a custom crawler and several parsers for over $200$ sites relating to malicious hacking. Instead, the paper presented in \cite{marin2018predicting} deals with an approach to predict the posts on hacking forums. They formulated this problem as a sequential rule-mining task, where the goal is to discover hackers' posting rules through sequences of peers' posts to make future predictions.

The authors of \cite{almukaynizi2018darkmention} implemented DARKMENTION, a system that retrieves information from both marketplaces where users sell information regarding vulnerabilities or exploits, and forums providing discussions about discovered vulnerabilities. They leveraged an already existing architecture introduced in \cite{nunes2016darknet} that includes customized crawlers and parsers built for each site and collects data from more than 400 platforms (both forums and marketplaces). To ensure the collection of cybersecurity-relevant data, they adopted ML models, and the filtering phase is performed by querying the CVE IDs in the NVD database using API calls. Regular expressions are used to identify CVE mentions in the Dark Web pages. The authors of \cite{samtani2020proactively} realized a custom Tor-routed Web spider to crawl and download all HTML
pages of a popular hacker forum. The Web spider leverages a breadth-first search strategy and a specialized Python program using regular expressions to parse all data into a local relational database. The obtained dataset with $32,766$ posts (i.e., threats) made by $8,429$
hackers that span a 23-year period is publicly available online\footnote{https://github.com/HongyiZhu/}. The work presented in \cite{nunes2018risk} uses Dark Web data supplied by a threat intelligence company and accessed via APIs. The data is comprised of forum discussions and marketplace items offered for sale. Moreover, ML models are employed to filter out the data related to drugs, weapons, and so forth and ensure the collection of only cybersecurity-relevant data.
\par
Differently from the papers above, which mainly gather data from hacker forums, the works presented in \cite{ebrahimi2018detecting,ebrahimi2020semi,adewopo2020exploring} focus on threat identification in Dark Net Marketplaces (DNMs).
For all these works, the markets' identification is performed through deepdotweb.com, a Dark Net news Web site. For \cite{ebrahimi2018detecting}, 7 English marketplaces and 1 Russian are identified, whereas for \cite{ebrahimi2020semi}, the selected marketplaces are Valhalla, Dream Market, Hansa, Alphabay, Minerva, SilkRoad3, and Apple Market. A custom Web crawler is developed to traverse the onion links in a breadth-first manner. To avoid anti-crawling measures in dark net marketplaces, it implements several techniques, and it is also capable of waiting for a user's response to access CAPTCHA-protected content. Moreover, Adewopo et al. \cite{adewopo2020exploring} collected data from two Dark Net Markets (Silkroad and Wall Street) extracted from the Arizona State University database\cite{du2018identifying}. The data set contains over $128,000$ posts from different discussion threads. The thread titles are related to Carding, Newbie, Scam, Hacking, and Review threads. 
\par
The authors of \cite{arnold2019dark,ampel2020labeling,du2018identifying} performed a crawling from multiple sources. As for \cite{arnold2019dark}, the different sources are: {\em (i)} 3 major DNFs (providing $204,001$ threads and $14,196$ authors); {\em (ii)} 5 largest DNMs (providing $224,270$ product listings and $7,911$ vendors); and {\em (iii)} the two largest exploit databases, such as Exploit DB and 0day.today (providing $43,678$ exploit listings). The data-gathering process starts by creating an initial list of potential sites based on factors such as the number of listings, discussion threads, threat actors, and the number of listings related to cyber threats found on each site. Subsequently, several Python-based Web crawlers are designed and deployed to collect data from each of these Web sites. The authors' aim is to identify threats across major Dark Net data sources linking assets to threat actors, using text features and Social Network Analysis metrics. The work presented in \cite{ampel2020labeling} employed 11 traditional hacker forums (collected through a crawler routed via Tor), one exploit database specific for DNM (i.e., 0day.today), and several public exploit repositories collected through APIs, such as Seebug, ExploitDB, PacketStorm, Metasploit, Vulnerlab, and Zeroscience. Traditional hacker forums are crawled through a depth-first search strategy implemented for efficiency. This choice makes the process incremental as a growing database of previously crawled links and dates is kept for each Website to ensure links are not visited or scraped twice. The authors of \cite{du2018identifying} collected data from 51 DNFs, 12 DNMs, 13 IRC channels, and 26 carding shops. For the first step of the hacker forums identification, the authors used three approaches, namely: {\em(i)} suggestions from cybersecurity experts; they contacted both the National Cyber-Forensics Training Alliance (NCFTA), a major non-profit organization focusing on the CTI sharing and the Policing in Cyberspace (POLCYB), an internationally recognized law enforcement entity; {\em(ii)} Surface Web and Tor search engines, which are queried based on the platform names suggested by the groups of experts; {\em(iii)} snowball identification, the platforms previously identified are used as seeds for a further in-depth search. Several Web crawlers are developed to collect the raw data in HTML format for forums, DNMs, and carding shops. For IRC data, they employ two bots, emulating fake users, inside each channel.
 \par
Authors of \cite{dong2018new} connect the Dark Net through TOR (using Polipo and Vidalia proxies) to collect information on eight large DNMs in the Dark Net, including Dream Market, Berlusconi Market, etc. The technology used to crawl and parse the Web page is the Scrapy framework\footnote{https://scrapy.org/}. After this step, the authors designed a custom parser for each marketplace to collect important information from cybersecurity-related categories. Moreover, some studies focus on identifying key hackers in Internet Relay Chat (IRC) channels, like the paper in \cite{shao2018autonomic}, which describes an autonomic personality analysis based on author identification in IRC conversations. Using the IRC chat logs collected through autonomic IRC bots in various cybersecurity, underground channels, and general channels (computer and politics), the authors analyze them with the Exploiting IBM Watson Personality Insights to demonstrate that the personality-based solution can work
effectively in user identification. Similarly, the work presented in \cite{marin2018mining} deals with key hackers' identification exploiting a hybrid approach that combines content, Social Network, and seniority analysis. In this work, the authors collected data provided by a commercial version of the system described in \cite{nunes2016darknet}, from which they selected three popular English hacker forums on the Dark Web. The papers presented in \cite{kadoguchi2019exploring,kadoguchi2020deep} rely on an external threat intelligence platform for the Dark Web, called Sixgill\footnote{https://www.cybersixgill.com/}, to collect hacker activity and Social Network information; then, they analyze the organizational hierarchies. 

Table \ref{tab:DarkCrawling} summarizes, for each analyzed paper, the technology used to crawl content in the Dark/Deep Web, the collected sample size, and the type of resource collected (i.e., DNF, DNM, IRC channel, carding shops, and so forth).

\begin{table*}
\caption{Crawling from Dark/Deep Web}
\scriptsize
\centering
\begin{tabular}{|l|l|l|l|}
\hline
    Paper & Technology & Sample Size \\
    \hline \hline
    Almukaynizi et al. \cite{almukaynizi2018darkmention} & Custom crawler by Nunes et al. \cite{nunes2016darknet} & 400 platforms\\
    Tavabi et al. \cite{tavabi2019characterizing} & Custom crawler by Nunes et al. \cite{nunes2016darknet} & 80 DNFs\\ 
    Tavabi et al. \cite{tavabi2018darkembed} & Custom crawler by Nunes et al. \cite{nunes2016darknet} & 200 hacker sites\\ 
    Marin et al. \cite{marin2018mining} & Custom crawler by Nunes et al. \cite{nunes2016darknet} & 3 DNFs\\ 
    Marin et al. \cite{marin2018predicting} & Custom crawler by Nunes et al. \cite{nunes2016darknet} & 1 DNF\\ 
    Deliu et al. \cite{deliu2018collecting} & Database online & Nulled.IO platform\\
    Suryotrisongko et al. \cite{suryotrisongko2022topic} & Database online & Nulled.IO platform\\
    Nunes et al. \cite{nunes2018risk}& Data supplied by a threat intelligence company  & 302 Web site\\
    Williams et al. \cite{williams2018incremental}& Python custom crawler &10 DNFs\\
    Tavabi et al. \cite{tavabi2019characterizing}& Python custom crawler & 80 DNFs\\
    Samtani et al. \cite{samtani2020proactively}& Custom crawler & 1 DNF\\
    Samtani et al. \cite{samtani2022linking}&Custom crawler & 1 DNF\\
    Benjamin et al. \cite{benjamin2019dice}&Custom crawler & 4 DNFs\\
    Huang et al. \cite{huang2021hackerrank}& Custom crawler & 5 DNF\\
    Arnold and Samtani \cite{arnold2019dark} & Python custom crawlers & 3 DNFs, 5 DNMs, 2 exploit databases\\
    Ampel et al. \cite{ampel2020labeling}& Custom crawlers & 11 DNFs, 1 exploit database for DNM, 6 public repositories\\
    Ebrahimi et al. \cite{ebrahimi2018detecting}& Custom crawler & 8 DNMs\\
    Ebrahimi et al. \cite{ebrahimi2020semi}& Custom crawler & 7 DNMs\\
    Dong et al. \cite{dong2018new} & Scrapy framework & 8 DNMs\\
    Adewopo et al. \cite{adewopo2020exploring} & Arizona State University database\cite{du2018identifying} & 2 DNMs\\
    Shao et al. \cite{shao2018autonomic} & Custom IRC bot & 6 IRC channels\\
    Du et al. \cite{du2018identifying}& Custom crawler, Automatic bot & 51 DNFs, 12 DNMs, 13 IRC channels and 26 Carding Shops\\
    Kadoguchi et al. \cite{kadoguchi2019exploring} & Sixgill platform & 3000 posts from DNFs\\
    Kadoguchi et al. \cite{kadoguchi2020deep} & Sixgill platform & 1700 posts from DNFs\\
    \hline\hline

\end{tabular}
\label{tab:DarkCrawling}
\end{table*}

\section{NLP-Based Techniques for Threat Intelligence Analysis}
\label{sec:analysis}

As the manual analysis of all the sources of CTI (i.e., Clear Web, Social Networks, and Dark/Deep Web) is time-consuming, inefficient, and prone to errors, several solutions to, even partially automatize the utilization of CTI information have been proposed in the literature. In this section, we aim to analyze such approaches by focusing on the NLP techniques exploited, such as text classification, text clustering, topic detection, and trend analysis.

\subsection{Text Classification}

Text classification is one of the main tasks in the NLP field, whose objective is to categorize textual documents according to their content.
Another popular application of text classification is sentiment analysis. Sentiment analysis tries to predict whether the sentiment expressed in a target text is either positive or negative. Text classification can be binary, as it happens for sentiment analysis or multi-class. In this case, each class is characterized by specific features present in the corpus of the text. Due to its supervised nature, text classification requires the assignment of a label to each sample in the training set, thus allowing the classifier to learn the characteristics of the different classes.
To perform a classification task, any classifier needs a numerical representation of the text as input. In Table \ref{tab:representationsTech}, we presented the most popular and effective solutions for text representation, spanning from statistical approaches, such as BoW, TF-IDF, and GloVe, to DL models, e.g., Word2Vec. The obtained representation can be used as input to traditional ML algorithms or more advanced neural networks like Convolutional or Recurrent Neural Networks on top of a classification layer.

Recently, transformer-based solutions, like BERT (see Section \ref{sub:algorithmsTools}), have gained popularity due to their impressive results.
Transformer-based approaches exploit an encoder module to generate a context-accurate representation of the text, thanks to the attention system, which can, hence, be provided in input to a classification head. Text classification using the above-mentioned technologies is employed in many CTI tasks. In the following, we analyze the main applications of text classification for CTI to {\em (i)} Social Media and {\em (ii)} CTI Reports.

\par
\textbf{\textit{Social Media:}} In \cite{deliu2018collecting}, the authors build a binary classifier that detects security-relevant or irrelevant posts in a hacker forum. To do so, they used an SVM classifier, achieving an accuracy of 98.82\% on the task. Similarly, the authors of \cite{behzadan2018corpus} presented a Convolutional Neural Network (CNN, for short) model to perform the same binary classification task between security-relevant and irrelevant tweets. Then, if a post is classified as relevant, a second CNN is used to classify the post between 8 classes namely: vulnerability, DDoS, data leak, ransomware, 0-day, and marketing/general. The final performance of the two models is 94.72\% on the binary task and 87.56\% on the multi-class. In \cite{alves2019design, alves2021processing}, the authors proposed a Twitter-based streaming threat monitor on a specific infrastructure that classifies whether the tweet is relevant or not for CTI. In the same way, the strategy proposed by \cite{le2019gathering} uses a TF-IDF representation technique combined with two possible ML algorithms, namely Centroid classifier and SVM, to predict the relevance of the anomaly reported in the post achieving an F1 score of the $64.3\%$. Hackers forums represent a rich source of data; for this reason, in \cite{ampel2020labeling}, the authors use a C-BiLSTM neural network to perform a multi-class classification finalized to predict the label of different types of attack. In \cite{zhao2020timiner}, the authors collected threat description data from different blogs to build a CTI domain recognizer that uses a combination of Word2Vec representation and a CNN classifier. The authors of \cite{adewopo2020exploring} exploited data from Twitter and a hacker forum to build a logistic regression model combined with a TF-IDF representation finalized to predict the relevance of the post for security monitoring. The strategy proposed by the authors of \cite{dionisio2019cyberthreat} is composed of a combination of pre-trained embeddings generated by Word2Vec or GloVe and a CNN to perform a binary classification of tweets related to IT infrastructures, according to the relevance of the security-related information contained. Similarly, in \cite{queiroz2019eavesdropping}, the authors proposed an approach to identify features, classifiers, and practices that provide the best possible detection performance for software-vulnerability-related communication in online social media channels. In particular, the strategy exploits a combination of representations obtained with Word2Vec and an SVM classifier. In \cite{kadoguchi2019exploring}, instead of using the representation of single words, the authors employed doc2Vec for the representation of an entire document. The results are used as input to an MLP network to extract posts that contain important information and to design proper countermeasures to the predicted attack. In \cite{ariffini2019ransomware}, the authors considered informal text about ransomware from forum threads. Hence, they employed well-known algorithms like SVM, Random Forest, and Naive Bayes to detect important entities and conduct analyses. The authors of \cite{dong2018new} developed a framework to classify cyber threats in the context of marketplaces in the darknet. To do so, after testing several strategies, including Naive Bayes and MLP classifier, the authors exploited an SVM classifier in their solution.
The terminology used in forums can be used to estimate the expertise of a hacker. In this sense, the authors of \cite{biswas2022text} proposed a hacker-expertise predictor that exploits features of the text, such as TF-IDF scores of the keywords, in combination with classification algorithms, i.e., KNN and regression trees.
\par
\textbf{\textit{CTI Reports:}}  In \cite{yang2020automated}, the authors classified CTI reports according to the type of threat. To do so, they used different ML algorithms, like KNN, SVM, or decision trees, achieving performance around $80\%$. Another paper that analyzes CTI reports is presented in \cite{liu2022tricti}. Here, the authors proposed a trigger-enhanced CTI~(TriCTI) discovery system, which aims to automatically discover actionable CTI. In particular, they used a fine-tuned BERT with an elaborate design to generate triggers. The trigger vector is further trained according to the similarity with the sentence containing it. Finally, a binary prediction is performed by exploiting similarity information. Analogously, in \cite{alves2022leveraging}, the authors employed a BERT classifier to map TTPs to the MITRE ATT\&CK framework.
In detail, after conducting a brief hyperparameter search, the authors used the MITRE dataset to fine-tune different BERT models. As an additional contribution, they compared the performance of such BERT models with a standard approach using TF-IDF and linear regression. The BERT-based models obtain an accuracy of around 80\% compared to the 61\% accuracy of the traditional strategy. In \cite{noor2019machine}, the authors compared different classifiers to predict the CTI class for an unseen cyber threat incident. In particular, they considered Naïve Bayes, K-Nearest Neighbors~(KNN),  decision tree, and deep learning neural networks. The authors of \cite{li2019extraction} proposed a threat action extraction strategy through multi-label classification. In particular, they selected ATT\&CK as a threat action taxonomy and TF-IDF as a feature. Such information is then used as input to state-of-the-art classifiers. In the same way, \cite{ayoade2018automated} proposed an ML model for threat report categorization. The proposed solution generalizes across reports using a combination of TF-IDF and a variety of classifiers. In \cite{goseva2018identification}, the authors consider the classification of bug reports similar to the descriptions of vulnerability classes from CWEs. In particular, the authors evaluated supervised and unsupervised strategies on data extracted from issue-tracking systems of two NASA missions. The proposed approach concatenates the title, the subject, and the description of each bug report and processes them through TF-IDF and a classification algorithm. The authors of \cite{zhu2018chainsmith} proposed a framework named ChainSmith, which aims at automatically extracting IoCs and their corresponding campaign stages from technical documents. To do so, they consider the most informative words according to their occurrence in the sentence, and then, these features are used to train four neural network binary classifiers that predict topic probabilities. 

\begin{figure}[ht]
	\centerline{
        \includegraphics[scale=0.45]{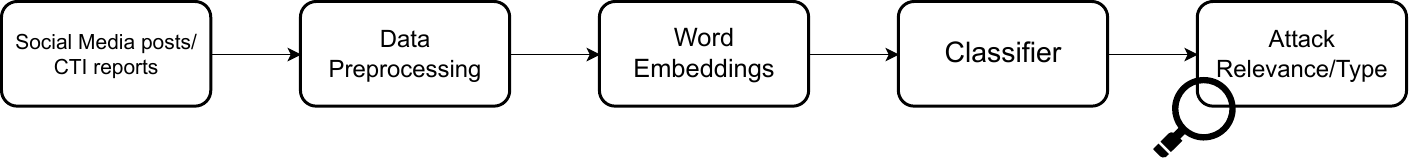}
    }
    \caption{General classification procedure \label{fig:CTIclassifier}}
\end{figure}

%checkpoint

\subsection{Text Similarity and Clustering}

The representation techniques described in Section \ref{sub:RepresentationTechinques}, as we saw in the previous section, illustrate how pre-trained embeddings can be used as input to a classifier. An important characteristic of these representations is the intrinsic relation encoded in the vector representation of similar words. This specific aspect can be exploited to calculate a similarity score between words using distance metrics like cosine similarity. The application of this is particularly useful for dividing a set of data into clusters, giving the obtained vectors to unsupervised clustering algorithms. An example of this is described in the TOM approach presented in \cite{bo2019tom}. In particular, the authors proposed two clustering methods: {\em (i)} a semi-supervised matrix decomposition-based method, and {\em (ii)} an unsupervised term frequency-based method. The first method generates a term-malware mapping matrix from the Threat Expert reports by using the TF-IDF representation technique. The second method determines the name of the clusters based on string distances, and the malware with similar names is considered a cluster using the K-means algorithm. In the strategy presented by \cite{azevedo2019pure}, the authors proposed an approach that improves the OSINT processing by correlating and combining IoCs coming from different feeds. In~\cite{alves2019design, alves2021processing}, authors performed clustering by the k-means algorithm as an aggregator of the outcome of the classifiers of tweets at the previous step of the pipeline of the approach. After the clustering, it is easier to transform unstructured data, like from social media, to standard formats, such as STIX. Similarly, in \cite{kristiansen2020cti}, the authors proposed a framework called CTI-Twitter, which applies a clustering strategy to the outcomes of the classes from a multi-class classifier. The idea is to detect inside each class the groups of semantically similar tweets to help the analysts detect trending keywords. In particular, the authors proposed two text clustering methods to group semantically similar tweets: k-means with TF-IDF and Sentence Transformer embeddings and LDA. Analogously, the authors of \cite{bose2019novel} used the representation of tweets produced by the TF-IDF in combination with the DBSCAN algorithm to cluster tweets in potential novel events. The proposed strategy uses clustering to aggregate all tweet texts in a cluster into a single corpus. Then, they performed the named entity recognition process to detect the keywords. In the proposed paper \cite{pantelis2021strengthening}, the authors exploited the clustering of data from different Dark Web sources to find similar contents. The clustering process has been performed through a combination of TF-IDF and K-means algorithms to find similarities between HTML pages and detect the top-k keywords. In a similar way, the authors of \cite{liu2020event} presented a cybersecurity event discovery and evolution detection framework based on continuous tweet streams called CyberEM.
In particular, the proposed framework consists of three components: pattern clustering, NMF-based event aggregation, and dynamic event inference. Focusing on the pattern recognition component, the objective is to detect the k clusters of the cybersecurity events to remove tweets that talk about cyber attacks in general. To do so, they employed \textit{Kullback Leibler divergence}. The idea behind the approach is to detect the terms with higher KL-divergence following the institution that these terms are more informative in target domain-related tweets. In \cite{kurogome2019eiger}, the authors presented a framework called EIGER with the intent of clustering similar artifacts created by malware of the same family.
The clustering strategy employs Levenshtein distance to calculate the similarity between the abstracts where the ones with the closest distance are merged sequentially.
The authors of \cite{milajerdi2019poirot} formalized the threat-hunting problem from CTI reports and IoC descriptions and, in particular, used a best-effort Similarity search.
The authors of \cite{li2019extraction} used TF-IDF representations to calculate semantic similarities of reports with the considered topics and actions of a threat-related article.
\begin{figure}[ht]
	\centerline{
        \includegraphics[scale=0.45]{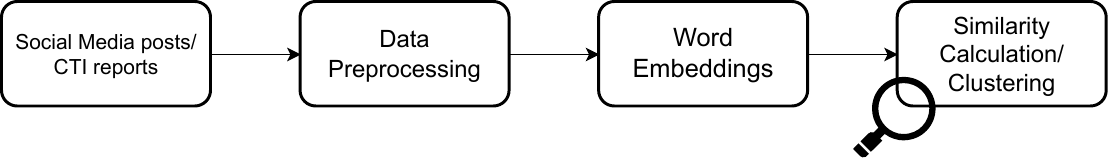}
    }
    \caption{General clustering procedure \label{fig:CTIclustering}}
\end{figure}

\subsection{Text Mining and Open Information Extraction}

Applying Open Information Extraction (OIE, hereafter) to CTI involves using Text Mining and NLP techniques to automatically extract relevant and valuable information about cyber threats, attack patterns, vulnerabilities, and related entities from unstructured text sources without prior knowledge or supervision.
  
In \cite{sarhan2021open}, the authors described Open-CyKG, a framework for extracting information from unstructured Advanced Persistent Threat (APT) reports and representing the retrieved data in a knowledge graph that offers efficient querying and retrieval of threat-related information. One module of this system consists of a neural attention-based OIE model to extract relation triples from unstructured APT reports. This model takes the concatenation of all inputs and passes it to two Bidirectional Gated Recurrent Units (Bi-GRU) layers, followed by an attention layer, two Time Distributed Dense (TDD) layers, and finally, a SoftMax layer for prediction. The work proposed in \cite{vishnu2022deep} develops a system leveraging both a self-attention deep neural network (SA-DNN) model and a text mining approach to identify the vulnerability category from the description text contained within a report. Similarly, the proposal of \cite{irshad2023cyber} aims at extracting features from unstructured CTI reports and attributing them to cyber-threat actors. Detailed feature sets, i.e., TTP, tools, malware, target organization, country, and application, have been used. Moreover, in this work, the authors proposed a novel embedding model called attack2vec to extract features from unstructured CTI reports. This novel model achieves accuracy, precision, recall, and F1-score of 96\%, 96.5\%, 95.58\%, and 95.75\%, respectively. 
\par
The paper presented in \cite{landauer2019framework} deals with the design of software for the automatic extraction of actionable threat intelligence from raw log data. The authors employed anomaly detection to disclose unknown attacks, and they pursued the iterative clustering and enrichment of anomalies with optional human verification with the purpose of transforming low-level log events into complex attack patterns. In this framework, every anomaly is primarily defined by the feature that triggered the detection mechanism. Then, the anomaly is transformed into an alert with enriched attributes storing contextual data as well as event and execution information. Finally, they eventually generate clusters of alarms that frequently occur together.

\subsection{Cross-Lingual Threat Intelligence}

Cyber threat actors operate on a global scale, and many of them communicate in languages other than English. To comprehensively understand and combat cyber threats, it is essential to monitor and analyze threats originating from communities with different languages. However, while text classification techniques have been mainly used for cyber threat detection in English platforms, this task is hindered in non-English ones due to the language barrier and lack of ground-truth data. Recent approaches usually filter out non-English data or make an automatic English translation to leverage monolingual models and overcome this issue. For instance, in all the works based on a Twitter crawler, the authors usually start the search with an English keyword or set a condition on the preferred language of the post or its country of origin (i.e., ``lang: en''). In a popular work of Samtani et al. \cite{samtani2017exploring}, Russian data was machine-translated to English using Google Translate. However, translation errors can deteriorate the classification results, and on the other hand, training separate monolingual models on low-resource non-English languages is impracticable.
\par
For the above reasons, the authors of \cite{ebrahimi2018adversarial} demonstrated that their deep cross-lingual model can jointly learn the common language representation from two languages (namely, English and Russian), and it outperforms a monolingual model trained through machine-translated data for identifying cyber threats in non-English DNMs. They randomly sampled the product descriptions in each language by preserving the ratio of cyber to non-cyber products, and then the manual labeling step as cyber threats or non-threats was performed by cybersecurity experts and a Russian speaker. Moreover, the CL-LSTM (CrossLingual LSTM) algorithm leverages BiLSTM to jointly learn the common hacker language representation from English and Russian DNMs. The authors of \cite{ebrahimi2020detecting} collected data from an English forum, a Russian forum, and two French forums and designed A-CLKT, an adversarial learning procedure that learns language invariant representations across two languages automatically. In the first phase
of automated text representation, A-CLKT reserves an LSTM for each language to automatically create an embedding of hacker forum text. In the second phase, the authors leveraged a GAN to devise a novel adversarial learning strategy to operate on the English and non-English representations. Finally, a classification step is performed to classify the text as a threat or non-threat.

The work presented in \cite{ranade2018using} collects data from Twitter both in English and in Russian, exploiting Twitter API language capabilities to detect language through the related flag (en=English, ru=Russian). Through two lexical databases, Wordnet\footnote{https://wordnet.princeton.edu/} and Russnet\footnote{http://www.russnet.org/}, the authors created relationships between the two languages' cybersecurity words and align them in vector embeddings. Then, an LSTM-based neural machine translation architecture is used to translate cybersecurity text from Russian to English. Finally, the work by Wu et al. \cite{wu2021price} adopted a Language Technology Platform (LTP \cite{che2010ltp}) that is an integrated Chinese language processing platform that includes a suite of high-performance NLP modules and relevant Chinese corpora. 

\subsection{Topic Detection and Trend Analysis}

Topic modeling is an unsupervised ML technique used to automatically identify topics in a collection of text documents and discover patterns in groups of words or documents that are semantically similar. Two of the most commonly used algorithms for topic modeling are Non-negative Matrix Factorization (NMF) and LDA \cite{blei2003latent}. NMF is a statistical method that aims to reduce the dimension of the input corpora, using the factor analysis method to provide comparatively less importance to the words with less coherence. LDA, instead, computes the posterior probability of the distributions of topics per document and the distribution of words per topic. These estimations are presented as matrices and used to infer two outputs, {\em(i)} the dominant topic per document and {\em(i)} an ordered list of words that constitute that topic. The works that base their topic modeling approach in LDA are the following \cite{deliu2018collecting,vishnu2022deep}. Specifically, the authors of \cite{deliu2018collecting} applied LDA after using SVM to remove irrelevant posts to extract ten topics. The interpretation of each topic's meaning is based on reviewing its top documents(posts) by human operators. The extracted topics deal with leaked credentials, malicious proxies, undetected malware, and asset-specific CTI. The work by Vishnu et al. \cite{vishnu2022deep} performs topic modeling on a 13-class categorization problem using the LDA method to discover the topic trends within the dataset for the various CVE categories. In particular, summarized vulnerability descriptions are converted to a document-term matrix and given to the LDA model to create the topic model. The model produced ten topics, from which the authors manually selected the most suitable to describe each vulnerability category.

As seen before, the topic generated by LDA may not be meaningful to a user. Hence, the authors of \cite{nagai2019understanding} apply to CTI the algorithm SeededLDA proposed by Jagarlamudi et al. \cite{jagarlamudi2012incorporating}, which allows a user to give additional information to the
topic model in order to learn topics of specific interest to a user.
In SeededLDA, a user provides seed sets according to the state
and the environment of the organization (i.e., IoT or Financial Industry) to guide the topics together with security blog posts. 

The paper by Hossen et al. \cite{hossen2021generating} utilizes both LDA and another popular algorithm for topic modeling: Non-negative Matrix Factorization (NMF). They collect and analyze data from a well-known hacker forum for the purpose of identifying
and classifying possible CTI.
The paper described in \cite{behzadan2018corpus} leverages topic modeling as the preliminary step of manual annotation of
tweets to speed up and increase the accuracy of the whole process. The topic modeling API of IBM's Watson Natural Language Understating service \cite{chen2016ibm} is used to recollect text classification into five categories connected to cybersecurity for the textual contents of each tweet. The text category
assignment was restricted to the top three categories with the highest confidence score.

The approach proposed in \cite{li2020open} describes an improved keyword feature extraction method, namely the ITFIDF-LP (Incremental TF-IDF method considering word location and part of speech) method. In particular, this strategy considers keyword identification tools, whether a word is a stop word, its position in the article, and which part of speech this word represents.

Given the limitations of prevailing topic models, several works developed custom approaches based on LDA \cite{suryotrisongko2022topic,vahedi2021identifying}.
More recent topic modeling approaches (namely, BERTopic and Top2Vec) are leveraged by \cite{suryotrisongko2022topic}. BERTopic relies on pre-trained transformer-based language models to compute document embedding, cluster these embeddings, and generate topics with the class-based TF-IDF (Term Frequency–Inverse Document Frequency) procedure \cite{grootendorst2022bertopic}. Top2Vec is a new topic modeling and semantic search algorithm that creates embedded document/word vectors, makes lower dimensional embedding, finds dense areas of documents, calculates the centroid, and finally finds the closest word vectors \cite{angelov2020top2vec}.

To categorize long and contiguous text (e.g., pastes), the authors of \cite{vahedi2021identifying} incorporate BERT into LDA, proposing the BERT-LDA framework. It consists of three main components, namely {\em(i)} BERT's encoder that tokenizes each word in each sequence (sentence) for every input paste; {\em(ii)} BoL model (produced by BERT) replacing the traditional BoW in the conventional LDA model captures information at about each paste's semantics at the sequence-level, rather than at the word-level; and {\em(iii)} Topic Generation using LDA produces topics based on each paste's BoL. The proposed BERT-LDA model was applied to all pastes gathered from Pastebin, PasteFS, and Pastelink platforms. The authors manually assigned names to five prevailing topics extracted by the model (namely, hackers, malware, networks, Web sites, and PII) and checked the results by comparing them with the selected keywords.

The authors of \cite{sleeman2021understanding} used the Dynamic Topic Model (DTM, hereafter) to model multiple document collections over time. They exploit Wikipedia concepts related to cybersecurity as a context model for training the DTM to understand how the concepts found among documents are changing over time. The leveraged corpus belongs to arXiv Cryptography and Security research papers. An important contribution of this work is the automatic domain concept extraction using Wikipedia concepts. They captured 3,836 concepts from Wikipedia, exploiting them to establish the context for the topic modeling portion.

Table \ref{tab:topicmodeling} summarizes, for each analyzed paper, the tool and/or the technique employed by the authors to perform topic modeling.

\begin{table}
\caption{Tool/Technique for Topic modeling}
\scriptsize
\centering
\begin{tabular}{|l|l|}
\hline
    Paper & Tool/Technique for Topic Modeling \\
    \hline \hline
    Deliu et al. \cite{deliu2018collecting}&LDA\\
    Vishnu et al. \cite{vishnu2022deep}&LDA\\
    Hossen et al. \cite{hossen2021generating}& LDA and NMF\\
    Nagai et al. \cite{nagai2019understanding}&SeededLDA\\
    Behzadan et al. \cite{behzadan2018corpus} & Watson Natural Language Understating service \cite{chen2016ibm}\\
    Li et al. \cite{li2020open} &  ITFIDF-LP \\
    Suryotrisongko et al. \cite{suryotrisongko2022topic}&  BERTopic and Top2Vec \\
    Vahedi et al. \cite{vahedi2021identifying}& BERT-LDA\\
    Sleemans et al. \cite{sleeman2021understanding} &DTM\\
    \hline \hline
\end{tabular}
\label{tab:topicmodeling}
\end{table}

\subsection{Text Summarization}

Text Summarization is often used to reduce the verbosity of a document or Web content and obtain a concise description of the attack behavior that can be directly used to detect the attack itself.
Indeed, threat reports are usually characterized by a significant amount of complex and irrelevant text, and only a small portion of the report describes attack behavior. Although topic classification has been used to identify topic-related context among out-of-domain contexts (e.g., advertisement text versus technical text), they are not good enough to provide a description of observable attack actions discriminating among technical concepts. Moreover, inside each sentence, some parts of the speech, such as adverbial and adjectival, do not contribute to the behavior description of the attack, and that can be safely removed. To solve this issue and provide a good summary of CTI content, authors of \cite{satvat2021extractor} designed a two-step approach that consists of a BERT classifier, which deals with sentence verbosity, and a BiLSTM network, which deals with word verbosity. In particular, for the first step of sentence verbosity, they use a 12-hidden layer BERT to discern the productive sentences from the non-productive ones. To train the model, they used 8,000 labeled sentences. For removing the word verbosity, instead, understanding the words' roles in the text summary is crucial, and for this reason, they used a re-implementation of a deep BiLSTM model \cite{he2017deep}. Since this model was not fine-tuned to handle cybersecurity sentences, they trained the model using 3,000 manually labeled sentences. Similarly, the paper presented in \cite{chen2022automatic} performed text summarization, adopting the BERT transformer model to perform word embedding and BiLSTM to extract threat entities.

The authors of \cite{russo2019summarizing} proposed a lightweight scheme, called CVErizer, for summarizing CVE's content, which consists of two steps: {\em(i)} information extraction and {\em(i)} classification of vulnerabilities. They observed that the most relevant information pieces usually contained in a CVE description are: the vulnerability name, the name and the version of the software affected by the vulnerability, the kind of attacker who could exploit the vulnerability, the mechanism used, and the effect. This recurrent pattern is then formalized in an XML file, and through a lightweight taxonomy, these vulnerabilities are categorized into different types, reaching an accuracy score of 81\%. To monitor the cybersecurity events, Li et al.~\cite{li2018security} summarised CTI text using \textit{TextRank}, a graph-based ranking algorithm. This approach entails the transformation of natural language text into a graph, where individual sentences are considered as vertices, and connections are established based on sentence similarity. The similarity is determined by calculating the count of common tokens found in the lexical representations of two sentences and then dividing this count by a normalization factor. Sentences exhibiting greater connectivity to other vertices are accorded higher importance and are subsequently ranked, resulting in the extraction of pivotal sentences. Arranging these summaries chronologically using timestamps reveals the development of cybersecurity events.

\section{NLP-Based Techniques for Relation Extraction}
\label{sec:rel_extraction}
In this section, we delve into research dedicated to extracting relationships from cybersecurity data. To perform relationship extraction, multiple researchers have pre-identified cybersecurity entities. Furthermore, certain studies have successfully identified cybersecurity events from these recognized entities and extracted valuable relationships. Additionally, we have reviewed various research works that center on the development of graphs using this extracted knowledge. The process of relationship extraction and constructing a Knowledge Graph~(KG) from security text can significantly contribute to generating actionable CTI.
\subsection{Named Entity Recognition} 

 Named Entity Recognition~(NER) assists in recognizing entities(proper nouns) from the unstructured text and serves as a fundamental technique for tasks such as information retrieval, constructing knowledge bases, question answering, generating automatic text summaries, and providing semantic annotations. The named entities include real-world objects like persons, organizations, locations, dates, and other entities with proper names. NER holds a pivotal role in NLP by focusing on identifying and categorizing named entities within the text. %The initiation of the NER task occurred during the Sixth Message Understanding Conference in 1996, organized by Grishman and Sundheim. 
\par
\begin{figure}[ht]
    \centering
    \includegraphics[scale=0.23]{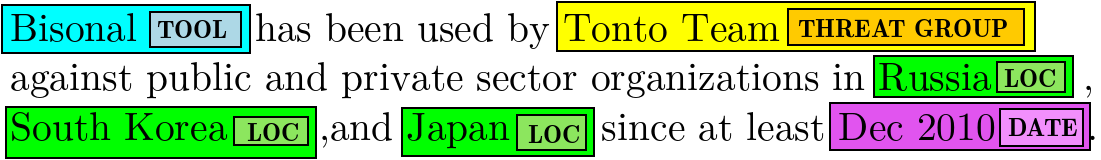}
    \caption{Cybersecurity named entity}
    \label{fig:nerdemo}
\end{figure}

There are general tools available, such as Spacy, Stanford NER, NLTK, Flair, etc., that can recognize generic entity types like locations, persons, organizations, and dates. However, these entities might not fulfill the specific demands of specialized domains like cybersecurity~\cite{hanks2022recognizing}. Cybersecurity requires the identification and classification of entities unique to the field, including malware types, operating systems, attack methods, and other terms specific to cybersecurity. Figure~\ref{fig:nerdemo} displays a sample of cybersecurity text sourced from the MITRE Web site\footnote{https://attack.mitre.org/software/S0268/}, along with its corresponding entities. The significance of identifying and labeling these domain-specific entities becomes evident in various cybersecurity tasks. For instance, in the context of malware analysis, NER can pinpoint the types of malware within a text document~\cite{piplai2020creating}, thus aiding analysts in comprehending the threat's nature. In the classification of attacks and vulnerabilities, NER plays a role in identifying the precise types of attacks or vulnerabilities~\cite{dong2019towards} mentioned in security reports or incident logs. Additionally, NER contributes to the creation of cybersecurity KG and structures that organize and present cybersecurity-related information systematically.

\subsubsection{Annotation Strategy}
Recently, NER has been predominantly executed through supervised learning classifiers that require labeled text as input. Since the rapidly evolving threat landscape is characterized by emerging attack mechanisms and the convergence of various attack and malware types, the imperative arises to develop customized NER datasets. In the NER task, each word of unstructured text is assigned a specific tag instead of assigning a label to an entire paragraph. Researchers have employed various tagging schemes and annotation tools for annotating this task.

\par
\textbf{\textit{Tagging Schemes:}} Researchers explored tagging schemes, including BIO, IO, IOE, BIOES, BILOU, etc. Among these schemes, BIO and BIOES are extensively utilized in CTI.
\begin{itemize}
    \item \textit{BIO or IOB}: BIO stands for ``Begin, Inside, Outside," also referred to as the IOB format. The B (Beginning) and I (Inside) tags are primarily employed to label relevant entities. The B index indicates the start of a specific entity instance, while the I index signifies the continuation of the particular entity. Additionally, an O tag is used to denote terms that exist outside of any entity context. For instance, consider the sentence extracted from the Bleeping Computer site: \texttt{Major US energy organization targeted in QR code Phishing attack.} In this example, \texttt{US} is tagged as B-Location, \texttt{Energy} is assigned with the B-Industry tag, \texttt{QR} is labeled as B-attack type, \texttt{code} as I-attack type, \texttt{Phishing} as I-attack type, and all other tokens are assigned the O tag. In studies~\cite{kim2020automatic, wang2022cyber, alam2022cyner,zhou2023cdtier,guo2021cyberrel,zhao2020cyber,sarhan2021open,ren2022cskg4apt,gao2021data,luo2021framework}, researchers employed the BIO/ IOB tagging scheme for annotating the tokens as cybersecurity entities.
\item BIOES: The BIOES scheme also employs BIO tags akin to the BIO approach but contains two additional tags: E and S. The E tag indicates the end of a specific entity, while the S tag is utilized for single-word entities. In the sentence, \texttt{Major US energy organization targeted in QR code Phishing attack}, \texttt{US} is designated as S-Location, \texttt{Energy} is assigned the S-Industry tag, \texttt{QR} is marked as B-attack type, \texttt{code} as I-attack type, \texttt{Phishing} as E-attack type, and all other tokens are labeled with the O-tag. Few researchers have adopted the BIOES tagging scheme in their studies~\cite{wang2022aptner, wang2023novel,yang2023enriching}.

\item BIOFR: Traditional BIO tagging typically concentrates solely on identifying entities and doesn't consider the sentence's structure. In contrast, BIOFR~\cite{wei2023automated}, which stands for Begin (B), Inside (I), Outside (O), Front (F), and Rear (R), incorporates attention to words surrounding the entities. However, it's worth noting that BIOFR does not regard nouns as responsible for threats or vulnerabilities; instead, it summarizes them using verbs. Take the sentence, ``A stack out-of-bound error was flagged in $\cdots$." When we apply BIOER, we get tagged sentences like this: ``A [F-B-Tri] stack [B-Tri] out-of-bound [I-Tri] error [I-Tri] was [R-B-Tri] flagged [R-I-Tri] $\cdots$."

\end{itemize}
\textbf{\textit{Annotation Tools:}} For expediting the manual annotation process, the researchers employed various annotation tools, including \textit{Prodigy, BRAT, YEDDA, Doccano, LightTag}, etc. The BRAT Rapid Annotation Tool (BRAT)~\cite{stenetorp2012brat} serves as a Web-based text annotation mechanism primarily designed for NLP tasks, encompassing both entity and relation annotation. The authors~\cite{wang2022cyber, alam2022cyner, wang2022aptner,luo2021framework} opted for BRAT in their work. Prodigy~\cite{prodigy}, a commercial tool, supports not only text-based tasks like NER, dependency parsing, and part-of-speech tagging but also tasks for audio and video annotation and computer vision. Few works~\cite{hanks2022recognizing,zacharis2023aicef} used Prodigy for annotating the text. YEDDA~\cite{yang2017yedda}, on the other hand, is an open-source lightweight tool supporting command-line annotation. It supports languages such as  English and Chinese and works with symbols and even emojis. YEDDA is adopted by Ren et al.~\cite{ren2022cskg4apt} in their study. Further, Zhou et al.~\cite{zhou2023cdtier} employed YEDDA for annotating the Chinese security-related text.

\subsubsection{NER Approaches}
\textbf{\textit{Conventional Methods:}}
Early research on NER relied on rule-based and statistical machine-learning models. Rule-based strategies hinged on rules crafted by domain experts, amalgamating gazetteers(curated lists of known entities), and syntactic-lexical patterns~\cite{li2020survey}. Hanks et al.~\cite{hanks2022recognizing} employed gazetteers to identify a range of entities, including protocols, malware types, programming languages, attack types, file extensions, and operating systems. These gazetteers were constructed by querying data from Wikidata. Additionally, the authors employed regular expressions to detect indicators such as CVEs, ports, hashes, and IP addresses. For recognizing URLs and Email addresses, they utilized the Spacy tool. In ~\cite{kim2020automatic}, the authors adopted hybrid techniques, incorporating rule-based and DL-based methodologies to identify cyber entities effectively. Specifically, they extracted CVE IDs, commonly employed to denote vulnerabilities within software or products, using regular expressions and replacing them with the CVE tag.
Further, in \cite{alam2022cyner}, Alam et al. adopted regular expressions to recognize indicators such as IPv4, CVE, SHA1, SHA256, Email, and File Path. On the other hand, statistical methods harnessed the power of ML algorithms such as Perceptron, Support Vector Machine~(SVM), Hidden Markov Model (HMM), and Conditional Random Field~(CRF). The most effective methods for NER often involve the utilization of CRFs. CRFs address the NER task by implementing a sequence labeling model, where the assignment of a label to an entity is intricately linked to the labels assigned to neighboring entities within a predetermined contextual window. However, nowadays, researchers are using DL classifiers instead of ML classifiers.

\textbf{\textit{Deep Learning for NER:}}
The traditional approaches involved meticulous manual intervention in creating rules and undertaking intricate feature engineering. Following the rise of deep learning, many researchers have embraced DL-driven techniques in their research on NER to circumvent the laborious task of feature engineering. Significantly, the transformer framework showcases remarkable efficacy across diverse NLP tasks with models like BERT. In~\cite{dionisio2019cyberthreat,ma2020cybersecurity}, authors explored BiLSTM with CRF for identifying cybersecurity entities. In the study~\cite{gasmi2018lstm}, the authors employed LSTM recurrent neural networks, a domain-agnostic approach, in combination with a CRF method to extract entities within the cybersecurity domain. VIEM~\cite{dong2019towards} used a NER mechanism to identify vulnerable software names and versions. The NER model developed by Bidirectional Recurrent Neural Network achieves 97.8\% and 99.1\% precision and recall, respectively. Furthermore, they employed a heuristic approach that incorporated dictionary lookups to rectify inaccuracies in the NER model. With the introduction of a gazetteer, the overall accuracy increased to 99.69\%. In~\cite{georgescu2019named}, Georgescu et al. developed an automated system aimed at serving as a semantic indexing solution to identify pre-existing vulnerabilities within IoT environments. They created a domain ontology and crafted a NER model to conduct semantic text analysis. To train the NER model, they employed the Watson Knowledge Studio tool, which leverages a semi-supervised learning approach. Pingchuan et al.~\cite{ma2020cybersecurity} generated word embedding vectors by simultaneously training the model and the word vectors with the help of one hot encoding. The study by Kim et al.~\cite{kim2020automatic} integrated character-level attributes through Bag-Of-Character representations and incorporated GloVe for word embedding. Their strategy encompassed using a BiLSTM model along with CRF to extract cybersecurity entities. Similarly, Gao et al.~\cite{gao2021data} adopted a BiLSTM-CRF architecture, which incorporates domain dictionary embedding and a multi-attention mechanism for enhanced performance. They constructed a list of 15,357 words for domain embedding by collecting information from security blogs, CVE, NVD, and Wikipedia. In~\cite{qin2019network}, the authors employed a novel approach to identify entities in both English and Chinese text. This method leveraged a neural network, CNN BiLSTM CRF model, in conjunction with a Feature Template~(FT). The FT serves to extract local contextual features, while CNN was employed to capture character-level feature sequences, and BiLSTM was used to create the global feature vector. 
\par
The study in~\cite{dasgupta2020comparative} assessed various DL-based NER algorithms across a wide spectrum of sentence types, encompassing concise and lengthy structures. They investigated both domain-independent Word2Vec embeddings and domain-specific Word2Vec embeddings (trained on a cybersecurity corpus). Their findings demonstrated that the utilization of domain-specific embeddings resulted in superior performance. Further, they pinpointed that the optimal choice for cyber NERs entails the fusion of BERT embeddings with Bidirectional LSTMs. In~\cite{wang2022aptner}, Wang et al. introduced APTNER, a NER dataset comprising 21 entities derived from STIX 2.1. The authors evaluated multiple NER models and discovered that the BERT embedding-based BiLSTM model with CRF achieved the highest F1 score of 82.31\%. In another study, Wang et al.~\cite{wang2022cyber} leveraged knowledge engineering to boost the performance of their NER model, which relied on a combination of BERT-embedded BiLSTM and CRF. The model built using knowledge engineering achieved an F1-score of 91.62\%.  Wang et al.~\cite{wang2023novel} developed a NER model using a novel Neural Network Cell called GARU, which combines a Graph Neural Network and a Recurrent Neural Network. They concatenated diverse embedding dimensions such as character level, word level, POS, and dependency relation in the embedding layer. For word embedding, they used two methods: Word2Vec, trained on cybersecurity datasets~\cite{roy2017learning}, and PERT~\cite{cui2022pert}, a BERT-inspired auto-encoding model. They incorporated character-level features from character-based CNNs. Additionally, they created a module for detecting entity boundaries and predicting entity heads and tails. Sun et al.~\cite{sun2022cyber} employed supervised neural network approaches to extract and categorize 15 cybersecurity entities. Their approach commenced with the use of Stanford CoreNLP to segment sentences in articles based on punctuation marks. Subsequently, token segmentation was performed within those sentences. They combined word embeddings with feature embeddings and inputted them into a BiLSTM RNN. They conducted an evaluation and performance comparison of their entity recognition method with two approaches: one using LSTM with ELMo embeddings and the other with fine-tuned BERT embeddings. The neural network incorporating ELMo and BERT embeddings achieved exceptionally high accuracy levels, reaching 99.8\% and 99.2\%, respectively. In~\cite{chan2022feedref2022}, various fine-tuned models, including BERT, RoBERTa, ELECTRA, and DeBERTa, were utilized to perform NER and extract IoCs. Token annotation followed the CoNLL-2003\footnote{https://paperswithcode.com/dataset/conll-2003} structure, and IoCs were extracted from sentences based on predefined rules for 16 entities using the pyparsing\footnote{https://pypi.org/project/pyparsing/} module. When evaluating model performance, the fine-tuned DeBERTa model demonstrated superior results compared to the other models. However, in practical case studies, it was observed that the ELECTRA model effectively identified new IoCs.
\par

The enhancement of the identification and categorization of cybersecurity-related terms is the aim of Liu et al. who proposed a semantic augmentation approach in ~\cite{liu2022multi}. Their approach consists of three primary elements: internal domain augmentation, external general augmentation, and mixed linguistic features. Internal domain augmentation enriches the meaning of input words by incorporating insights from the K most similar words within the cybersecurity domain. Additionally, they fine-tuned the BERT model and utilized the resultant token representations for external general augmentation. Moreover, they generated a composite set of linguistic features, encompassing elements such as POS, morphology, and other component-related features. In~\cite{yang2023enriching}, Yang et al. underlined the significance of improving word information in the context of Chinese cybersecurity NER. They have introduced a novel model called Star-HGCN, in which the hybrid embedding layer enriches word information by integrating character-level data and POS embeddings. In this model, the Star-Transformer layer adeptly captures implicit semantic connections among words in a sentence, while the GCN layer is applied to model explicit syntactic dependencies, drawing from dependency tree structures. In the decoding phase, a CRF is employed to carry out sequence labeling for sentences.
 
Alam et al.~\cite{alam2022cyner} developed an open-source toolkit named CyNER to extract cybersecurity-related entities from unstructured text. To accomplish this, they combined various methods: transformer-based models for identifying cyber entities, heuristic techniques like regular expressions for indicators, and established models like Spacy and Flair for general entities such as locations and persons. They gathered 60 threat reports from MITRE ATT\&CK to generate a benchmark dataset. Using the BRAT annotation tool, they labeled the dataset with different types of entities, including vulnerabilities, organizations, systems, indicators, and malware. The transformer-based model, specifically XLM using RoBERTa large, achieved an F1 score of 76.66\% on this task. Most research in NER relies on labeled data, but in cybersecurity, such data is scarce. So, In~\cite{zhang2019multifeature}, the authors introduced an approach that combines a Generative Adversarial Network~(GAN) with the BiLSTM Attention CRF model to acquire labeled data from crowd annotations. The GAN is utilized to identify shared features within crowd annotations, which are then integrated, with domain dictionary features and sentence dependency features, into the BiLSTM-Attention-CRF model. This integration is aimed at enhancing the quality of crowdsourced annotations. Similarly, In~\cite{li2021adversarial}, the authors proposed an adversarial active learning approach for cybersecurity NER. They employed a BiLSTM layer to encode word embeddings and then utilized an additional LSTM layer to decode hidden representations obtained from the dynamic attention layer. More specific information about the representative work on neural network-based NER is included in Table~\ref{tab:ner}. The table lists the different data sources the researchers employed as well as the embedding technique they applied. 

\begin{table*}[htbp]
 \caption{Research work on neural network-based named entity recognition}
  \fontsize{6.2}{8}\selectfont % Adjust the numbers as needed
    \centering
    \begin{tabular}{|l|p{2cm}|p{2.5cm}|p{.8cm}|p{1.5cm}|p{2.5cm}|p{3.5cm}|}
    \hline
        \textbf{Paper} & \textbf{Data Source} & \textbf{Entities} & \textbf{Tagging Scheme}  &    \textbf{NLP Technique} & \textbf{Result} & \textbf{Remarks}\\
        \hline \hline
        Dionísio et al.~\cite{dionisio2019cyberthreat} 
         & Twitter & Organization, product or asset, version number, vulnerability, ID 
         & Entity label +O tag  &  GloVe, Word2Vec 
    & BiLSTM - 92\% (Average F1 score)  & Number of entity type is less\\
    \hline
Pingchuan et al.~\cite{ma2020cybersecurity} & Lal et al. dataset~\cite{lal2013information} & Software, modifier, OS, consequences, attack, means, file name, network, hardware & BIO & NM & BiLSTM + CRF - 89.38\%(F1 score) & Low f1 score for hardware and network types \\
 \hline
   Kim et al.~\cite{kim2020automatic}  & PDF documents &  Main entities: Malware, IP, Hash, Domain/URL, Total 20 sub-entities   & BIO & Bag-Of-Character &  BOC+BiLSTM+CRF 75.05(F1 score)   & CVE achieves 100\% performance and .url.normal, url.unknown, malware.unknown, exhibit high differences in precision and recall.  
   \\
    \hline
Gao et al.~\cite{gao2021data}& NER: Bridges et al. dataset~\cite{bridges2013automatic}, Domain dictionary: Wikipedia, CVE\& NVD, blog  & Application, version, hardware, OS, file, vendor, edition  & NM  & Domain dictionary embedding(N-gram feature) & BiLSTM+Dict+Att+CRF 88.36\%(F1 score)  & Lowest performance on hardware and edition\\
 \hline
     Dasgupta et al.~\cite{dasgupta2020comparative}  & Microsoft security bulletin, Adobe security updates, NVD Twitter, Technical reports, CVE & STIX 2.0 + Exploit-Target & NM &  Word2Vec, BERT  & BERT+BiLSTM+CRF  88.60\%(F1 score) & Trained on dataset with varying text lengths and complexity. 
   \\
    \hline
   Wang et al.~\cite{wang2022aptner}  & APT reports &  APT, SECTEAM, IDTY, OS, EMAIL, LOC, TIME, IP, DOM, URL, PROT, FILE, TOOL, MD5, SHA1, SHA2, MAL, ENCR, VULNAME, VULID, ACTION &BIOES & ELMo, BERT & BERT+BiLSTM+CRF 82.31\%(F1 score) &  Generated a publicly available NER dataset \\
    \hline
 Wang et al.~\cite{wang2022cyber} &  Twitter, Blogs, Forums, Cybersecurity company sites & Hacker groups, sample files, malicious samples, security teams, attack time, tool, country, industry, organization, user,  methods, vulnerability, mode of transmission, file type, alias  & BIO &   Word2Vec, GloVe, ELMO, GPT, BERT & BERT+ BiLSTM+ CRF+ Knowledge engineering 91.62\%(F1 score)& 
Performance of vulnerability recognition increased by the Knowledge Engineering \\
 \hline
Wang et al.~\cite{wang2023novel} & Bridges et al. dataset~\cite{bridges2013automatic}, OpenCyber: APT reports, CVE, security bulletins &OpenCyber: Threat actor, Attack pattern, Vulnerability, Software, Malware, Campaign, Indicator, Course of action, Tool & BIOES & Word2Vec, PERT &  PERT based FIEBD: Bridges et al.- 94.56\% 
OpenCyber - 87.34\%(F1 score) & Predicted entity boundaries(start and end points) \\
 \hline

  Alam et al.~\cite{alam2022cyner}  & MITRE ATT \& CK &  Malware, indicator, system, organization, vulnerability, email, URL, file path, hash, CVE, generic entity types of spacy and flair & BIO & BERT & XLM+RoBERTa - Large 76.66\%(F1 score)   & Developed CyNER python library \\
   \hline
  Sun et al.~\cite{sun2022cyber} & Security news article, vulnerability database, Twitter & organization, person, device, product, Web site, capability, file, malware, money, number, purpose geopolitical entity (GPE), time, CVE, vulnerability, version & BIO & POS, ELMo, BERT & ELMO + BiLSTM RNN -  99.8\%(F1 score) &  ELMO embedding better than BERT\\
   \hline
  Li et al.~\cite{li2021adversarial} & APT report, CVE, security news, blogs & organization, location, software, malware, indicator, vulnerability, course-of-action, tool, attack-pattern, campaign & NM & Word2Vec & Dynamic-att+ BiLSTM+ LSTM -45\% sample achieves 88.27\% (F1 score) & Adversarial learning to reduce labeling effort \\
   \hline
  Dong et al~\cite{dong2019towards} & CVE, vulnerability report & vulnerable software name and version & Entity label+ O tag & Skip-gram  & Bidirectional RNN - 0.978(precision) and 0.991(recall) & Extracted vulnerable software name and version \\
   \hline
  Qin et al.~\cite{qin2019network} & WooYun vulnerability database and Freebuf Web site & VUL ID, network relevant term, software, organization, location, person & BIO & Word2Vec & FT+CNN+BiLSTM+CRF - 86\%(F1 score) & Detection of entities in both English and Chinese languages\\
   \hline
  Chan et al.~\cite{chan2022feedref2022} & AlienVault OTX & attack technique, bitcoin address, CVE, Microsoft defender threat, domain, email, MD5, SHA-1, SHA-256, file path, hostname, IPV4, IPV6, sslcert fingerprint, URI and URL & BIO & BERT & DeBERTa-97.51\%(F1 score) & The fine-tuned ELECTRA model demonstrated the ability to extract novel IoCs accurately.\\
   \hline
  Gasmi et al.~\cite{gasmi2018lstm} & Bridges et al. dataset~\cite{bridges2013automatic} & vendor, application, version, file, OS, hardware, edition & BIO & Word2vec & LSTM + CRF - 83.37\%(F1 score) & Poor performance on the edition and hardware \\
   \hline
Yang et al.~\cite{yang2023enriching} & FreeBuf portal and blogs & Vulnerability ID, Computer terminology, Software, Organization, Location & BIOES & Word, Character level, POS embedding & Star-HGCN -90.98\%(F1 score) & Performance of organization and software improved by Star-HCN method \\
 
  \hline 
\end{tabular}
\\
\vspace{.1cm}
\begin{flushleft}
* NM indicates Not Mentioned
\end{flushleft}
    \label{tab:ner}
\end{table*}

\subsubsection{NER Corpus}
Generating actionable threat intelligence through DL-based NER necessitates labeled data. Based on different tagging schemes and annotation tools, researchers invest significant effort in annotating security reports to create an NER corpus. Researchers defined different entity types according to their specific threat intelligence objectives. Few research works explain how they conducted and validated the annotation. In the study by Kim et al.~\cite{kim2020automatic}, the annotation was carried out by five cybersecurity experts who were assigned text segments ranging from 1500 to 5000 lines. In another study~\cite{wang2022aptner}, the authors enlisted the assistance of 30 undergraduates and four graduate students to label security-related text and verify the labeling process. Another study~\cite{wang2022cyber} involved three graduate students for cross-verification of the annotated data. 
\par
To evaluate the quality of the annotation, Hanks et al.~\cite{hanks2022recognizing} conducted an inter-annotator agreement assessment. Notably, during the initial round of annotation, the authors found that less than 50\% of the annotations reached a consensus. The primary source of disagreement stemmed from varying interpretations of entities, such as distinguishing between software names and tools or products. For instance, when encountering a term like \texttt{Microsoft Word}, some annotators treated \texttt{Microsoft} and \texttt{Word} as separate entities, classifying \texttt{Microsoft} as an organization and \texttt{Word} as a software name. In contrast, other annotators regarded \texttt{Microsoft Word} as a single entity and categorized it as a software name. To address these disagreements, the authors conducted a subsequent round of annotation. They provided additional training to annotators and redefined entity categories, merging certain entities into a single type.
\par
 In the majority of NER studies, researchers create NER corpora, but few of them choose to share these valuable resources with the public. Wang et al.~\cite{wang2020dnrti} have contributed a cybersecurity NER dataset in which they gathered threat reports from various sources such as GitHub, government agencies, and security companies' Web sites. In the DNRTI dataset, they meticulously annotated 13 categories, including features, vulnerabilities, methods, organizations,  industries, geographical areas, motives, timestamps, tools, security teams, sample files,  attack types, and hacker organizations. Their annotation process employed the BRAT and utilized the widely accepted BIO labeling mode. However, the description of how they validated the quality of the annotations is not provided in sufficient detail. Table~\ref{tab:ner_corpus} lists the publicly accessible NER corpus.

%% Dont use tools to create table
%%%% https://en.wikibooks.org/wiki/LaTeX/Tables
\begin{table}[ht]
 \caption{Publicly available NER corpus}
\scriptsize
    \centering
    \begin{tabular}{|l|p{1.5cm}|p{6cm}|}
    \hline
       \textbf{Paper}  & \textbf{Data Instance} & \textbf{Link of the corpus}\\
     \hline \hline
 Bridges et al.~\cite{bridges2013automatic}  &   853560 tokens  73964 tags & \url{https://github.com/stucco/auto-labeled-corpus}     \\ 
Kim et al.~\cite{kim2020automatic}     & 498000 tokens 15720 tags   & \url{http://github.com/nlpai-lab/CTI-reports-dataset}         \\ 
 Alam et al.~\cite{alam2022cyner}    & 106991 tokens  4350 tags  & \url{https://github.com/aiforsec/CyNER}    \\ 

 Wang et al.~\cite{wang2022aptner}   & 260134  tokens  39565 tags & \url{https://github.com/wangxuren/APTNER}                      \\ 
  Wang et al.~\cite{wang2020dnrti} & 6574 sentences 36412 tags & \parbox[t]{4.2cm}{\url{https://github.com/SCreaMxp/}\\ \url{DNRTI-A-Large-scale-Dataset-for-} \\ \url{Named-Entity-Recognition-in-} \\ \url{Threat-Intelligence}}\\
\hline \hline
   \end{tabular}
 
   \label{tab:ner_corpus}
\end{table}
\subsection{Event Identification}

To respond rapidly to potential cyber threats, security analysts and IT personnel must stay informed about critical security events, regardless of their frequency of reporting. Given the rising number of cybersecurity incidents reported in articles, the capability to identify these events becomes a critical necessity. In NLP, event detection revolves around the discovery of details concerning entities within events, their respective functions, and the temporal and spatial attributes associated with these events. Event extraction is the process of extracting semantic and structural information from text, represented through typed phrases comprising trigger words and associated arguments~\cite{li2023event}. It generally encompasses two primary sub-tasks: type classification, responsible for identifying specific event types within sentences, and element extraction, which captures trigger words and related arguments following diverse role patterns~(schemas) to ensure comprehensive semantic understanding. When extracting events from the text, such as in the sentence ``\textit{Russian cybercriminals \textbf{breached} the International Criminal Court’s IT systems amid an ongoing probe into Russian war crimes committed in Ukraine,}" we can analyze both the event type and trigger words. In this context, the term ``\textit{breached}" serves as a trigger word, indicating an unauthorized intrusion or access to the IT systems of the International Criminal Court. This event is classified as a ``\textit{Attack.Databreach}" due to its involvement in compromising or accessing sensitive data, which is a defining characteristic of data breach attack. 

\par
Trong et al.~\cite{trong2020introducing} generated a corpus for cybersecurity event detection, encompassing 8,014 event triggers spanning 30 event types and drawn from 300 articles. These events types were classified into four categories related to cyber vulnerabilities/attacks: IMPACT, ATTACK, PATCH, and DISCOVER. They also evaluated various models for event detection using embedding techniques such as Word2Vec and BERT. The Document Embedding Enhanced Bidirectional RNN achieved 68.4\% F1-score in the event detection task. In~\cite{bose2019novel}, the authors explored social media to extract cyber events and devised a method to prioritize potential cyber threats, considering criteria like user influence scores, named entity confidence, and keyword relevance. They mimicked the process of identifying named entities and keywords using TextRazor\footnote{https://www.textrazor.com/}(NER API), and the TextRank~\cite{mihalcea2004textrank} algorithm from Gensim. In addition, TextRazor furnished confidence scores for named entities, while TextRank assigned scores to keywords based on word graphs. In~\cite{luo2021framework}, Luo et al. addressed the extraction of cybersecurity event details from Chinese text, focusing on four event types: IncidentsOnVulnerability, RansomwareAttack, Phishing, and Data Breach. Their approach involved converting the event extraction task into a sequence labeling task. This process began by embedding characters and incorporating word information into character representations. They introduced a k-window-size BiLSTM to capture contextual information spanning sentences. 
\par
TCEDCL~\cite{tang2023trigger} is a novel cybersecurity event detection method that doesn't rely on event trigger word labeling but instead incorporates sentence semantics. It employs unsupervised comparison learning to find the most suitable candidate instances by using a contrastive learning model. They defined two main event types, further categorized into nine subtypes: VulnerabilityPatch,  VulnerabilityDiscover, VulnerabilityImpact, SupplyChain, Malware, DDoSAttack, Ransom, Phishing, and Data Breach. A key benefit of TCEDCL is that it doesn't rely on detecting trigger words, simplifying the data sorting process during model training and lessening trigger word accuracy's influence on the model's performance. Instead, the system only needs to determine the event type, making it exceptionally flexible in adapting to new event types with minimal labeled data needed to train a new classifier. W2E(Words to Events)~\cite{shin2020cybersecurity} identifies emerging cyber threats with low false positives and high coverage by monitoring individual words instead of using semantic clustering methods. They used NLP techniques, such as POS tagging, lemmatization, and NER, to reduce false positives effectively. Additionally, it filters and categorizes tweets based on keywords associated with five event types: Data Breach, DDoS attack, Vulnerability, Exploit, and Malware, while also monitoring CVE-related events separately. In~\cite{mohammad2023deep} Mohammad et al. introduced a dual-level approach to cyber-event detection involving medium-level detection utilizing LDA and high-level detection based on Google Trends data. Their research encompassed comprehensive preprocessing steps, which included NER, POS tagging, symbol removal, stop-word elimination, and lemmatization. Also, they leveraged Word2Vec for feature vectorization and t-distributed stochastic neighbor embedding (t-SNE) for dimension reduction. This method achieved 95.96\% accuracy in cyber event detection.
\par
While existing methods can identify emerging cyber events, they often fall short of providing detailed threat characteristics. To overcome these limitations, Associated EE~(AEE)~\cite{li2023event} is introduced, which is a capable of performing event extraction for every sentence in a document. AEE consists of two components: event type classification and element extraction, both utilizing BERT to grasp sentence-level semantic understanding. To classify event types, it constructs a document-level graph that links sentences of diverse types and words within each document, utilizing a Document-Aware Graph Attention Network~(DGAT) to represent sentences and grasp contextual information. AEE also introduces a novel schema to facilitate associations among argument roles and incorporates type-aware parameter inheritance to enhance extraction precision and enrich contextual knowledge across different event types and roles. Satyapanich et al.~\cite{satyapanich2019extracting} introduced a framework aimed at classifying cybersecurity incidents through the use of semantic schemas, focusing on five distinct event categories: Attack.Databreach, Attack.Phishing, Attack.Ransom, Discover.Vulnerability, and Patch.Vulnerability. Within this schema, there are 20 argument types designed to represent various elements and characteristics within these events. The study utilized two distinct word embedding methods, namely context-free (Word2Vec) and context-dependent (BERT). Word2Vec was trained on a set of 5,000 cybersecurity news articles, generating domain-specific embeddings with a size of 100. Conversely, BERT word vectors were derived by averaging Word-Piece embeddings from the fourth-to-last layer. The paper introduces Context2Vector~\cite{liu2022context2vector}, which employs representation learning for detecting event contexts to alleviate triage pressure. This enables security analysts to concentrate on resources requiring urgent attention. In essence, context represents a sequence of events generated from various perspectives, indicating specific security events. Context2event comprises four primary components: context extraction, context embedding, deviation detection, and human-in-the-loop annotation. To achieve this, Context2Vec utilizes LDA2Vec for creating event, context, and topic representations within the same vector space. Additionally, it employs the Local Outlier Factor algorithm to identify behavioral deviations and anomalies in dynamic contexts. Ultimately, this proposed approach generates informative and interpretable labels for human inspection, enhancing high-risk event detection models in dynamic environments. In~\cite{perera2018cyberattack}, a method was introduced to predict future cybersecurity events employing probabilistic soft logic rules. This approach involved detecting explicit event references within a cyber attack kill chain by combining lexical matching and NER.

\subsection{Relation Extraction}

\label{sec:relationextraction}
Cyber Threat Intelligence (CTI) systems must aim to provide security experts with actionable information for detecting and mitigating cyber attacks. Traditionally, security experts have summarized this information in the form of IoCs, also known as low-level indicators, which include IP addresses, domain names, and malicious file hashes used to create detection rules. However, low-level IoCs have a short lifespan and lack the semantic context necessary for real-time attack detection. Additionally, attackers can easily modify such IoCs through obfuscation techniques, alterations to malicious code, changes in attack IP addresses, and the updation of phishing domains.

\par Collecting a substantial volume of valuable information from sources like social media, hacker forums, APT reports, threat reports shared by security companies, technical blogs, and more can transform this data into meaningful CTI. However, enterprises and organizations encounter three significant challenges when dealing with the vast, unregulated landscape of cyber threat information online. Firstly, CTI resources are complex, unstructured, and continually evolving. This complexity makes it challenging to extract actionable insights from the data. Secondly, there is a need for automated systems capable of synthesizing information about threat entities, attack semantics, and malicious actions. Automating this process can help rapidly identify and understand emerging threats. Lastly, organizations face the challenge of integrating threat data from multiple sources. They must harmonize data from various origins and understand the interactions between different threat elements, all while discerning and filtering out ambiguous or duplicate information.

\par To effectively detect cyber attacks, trace the source of attackers, and understand their behavior, the fundamental requirement is extracting cyber objects such as attackers, tools, infrastructure, files, and more. This can be accomplished by extracting named entities and subsequently inferring relationships between these entities. Specifically, a significant body of research work can be categorized into two approaches: (i)\textit{Pipeline Approach}: In this method, entities are extracted first, and ML or DL algorithms are employed in a subsequent phase to discern relationships between these entities. The information extracted includes entity heads, relations, and entity tails, providing crucial insights into various aspects such as the details of the attack incident, attacker behavior, threat actions, exploitation of vulnerabilities, and threat hunting. However, a critical drawback of this approach is that errors introduced during the entity extraction phase can propagate and influence the accuracy of relationship extraction, (ii) \textit{Joint-Entity Relation Model}: In contrast to the pipeline approach, the joint-entity relation model involves learning entities and their relationships simultaneously. This approach aims to capture associations between entities and their corresponding relationships in a more integrated manner, potentially mitigating the impact of errors in entity extraction on the subsequent relationship extraction stage. 
\par
\begin{figure*}[ht]
    \centering
    \includegraphics[scale = 0.23]{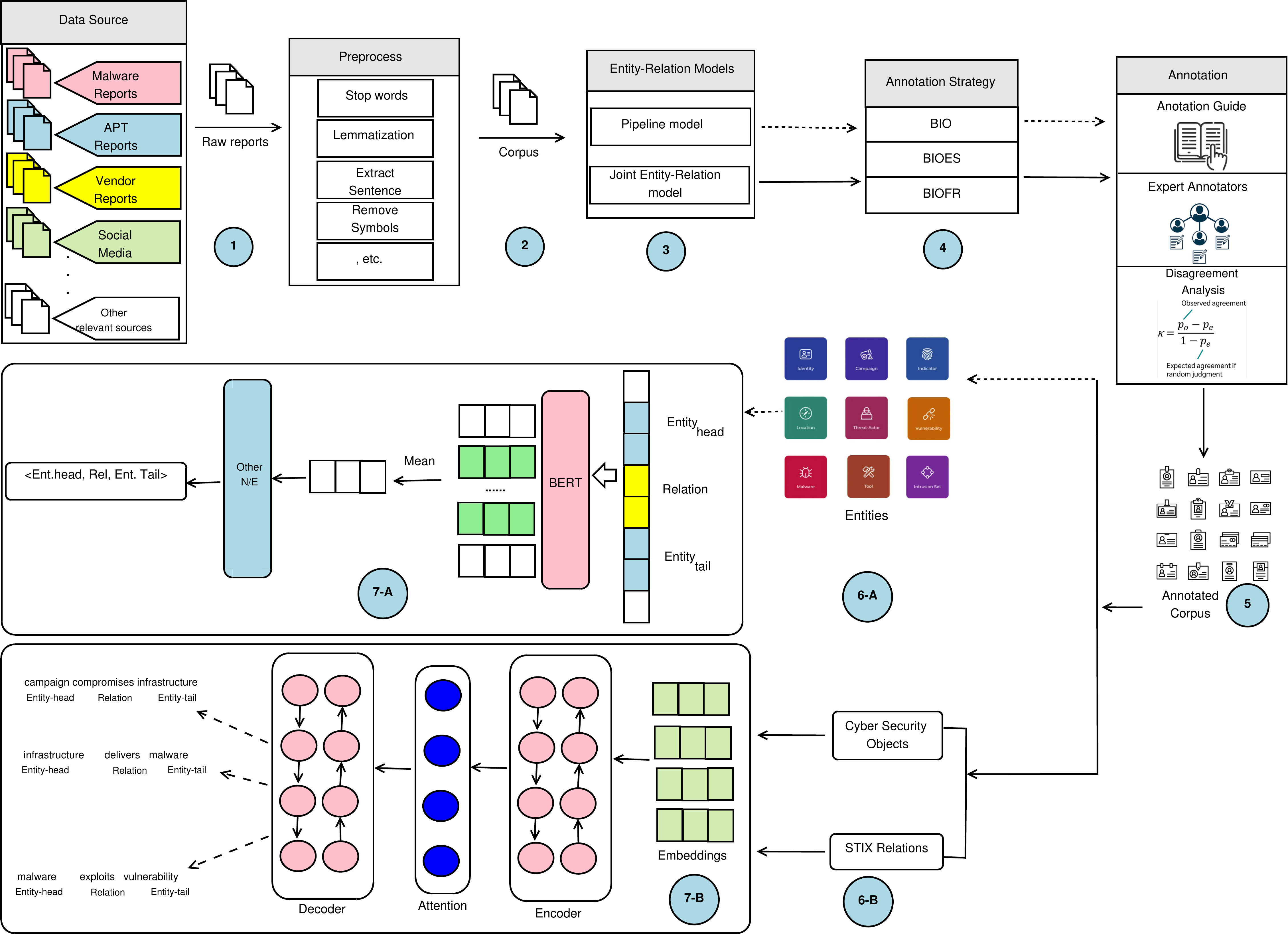}
    \caption{Relationship Extraction Pipeline}
    \label{fig:relationextraction}
\end{figure*}

 In general, we can summarize the task of extracting relationship triplets as follows (refer to Figure~\ref{fig:relationextraction}): collecting data, preprocessing it, preparing annotation guidelines, annotating the text, and performing entity-relation extraction. In the following paragraphs, we will explain how to extract triples from text and why it is essential to understand the relationships between words for CTI tasks, summarizing the work of essential articles.

\textbf{Tactics Techniques \& Procedures~(TTP):} SeqMask~\cite{ge2022seqmask} uses \textit{Multi-Instance Learning (MIL)} to find TTPs in CTI. It predicts TTP labels using behavioral keywords and conditional probabilities. SeqMask has two mechanisms to ensure credibility: expert validation and blocking existing keywords to assess their effect on accuracy. Experiments show SeqMask achieves a notable 86.07\% F1 score for TTP classification and improves TTP extraction from extensive CTI and malware data. The \textit{Attention-based} Transformer Hierarchical Recurrent Neural Network (ATHRNN)~\cite{liu2022threat} performs TTP extraction from unstructured CTI. This involves several key steps. First, they design a Transformer Embedding Architecture (TEA) to capture high-level semantic representations of both CTI and the ATT\&CK taxonomy. Then, they develop an Attention Recurrent Structure (ARS) to model the relationships between tactical and technical labels within ATT\&CK. Lastly, they create a joint Hierarchical Classification (HC) approach to understanding the interdependencies between tactics and techniques, enabling the prediction of the final TTPs. ATHRNN demonstrates superior performance in extracting TTPs, surpassing benchmarks like TF-IDF, Word2Vec, SVM, Deep Neural Networks (DNNs), BERT, and Text-CNN. Recently, the authors proposed TTPpredictor~\cite{aghaei2023cve}, which maps CVE descriptions to threat actions mentioned in the MITRE knowledge base. This work addresses two critical issues: (1) the scarcity of datasets containing labels for threat descriptions and (2) the semantic disparity between threat descriptions and MITRE techniques. To tackle these issues, the authors created a large corpus of threat reports mentioning CVEs correlating with MITRE descriptions. They estimated the correlation by matching the verb-object pairs of the corpus with the MITRE text using semantic role labeling. This resulted in a labeled dataset, which the authors used to fine-tune BERT to predict actions. Additionally, they compared the performance of TTPpredictor to ChatGPT by injecting 20 CVEs and prompting both tools to predict MITRE techniques for a vulnerability description. They found that ChatGPT failed to predict the techniques for vulnerability reports 80\% of the time.
\par 
\textbf{Threat entity extraction and relation extraction:} CDTier \cite{zhou2023cdtier} identified five types of threat entities and 11 types of relationships between these entities based on standards like STIX and real-world CTI usage. The process began with manually annotating a text corpus to create a labeled dataset. This dataset was then used to train a word2Vec model, which represents word meanings using vectors for word semantics, position, and grammatical relationships. These vectors form the basis for a neural network. After further feature extraction, the significance of semantic features was enhanced using Softmax, resulting in the extraction of entity-relationship pairs. To extract semantic information from cybersecurity documents, the authors~\cite{wang2019automatic} used Java Annotation Patterns Engine~(JAPE) with General Architecture for Text Engineering~(GATE) framework. The approach focused on noun and verb tokens that describe the relationships between entities, but the accuracy was only 32.69\% due to the limited number of samples. 

\par The CARE~(Cyber Attack Relation Extraction)~\cite{chen2023useful} system extracts entities and relations from Chinese cybersecurity Web sites. CARE comprises four important phases: (i) A crawler collects text from several Chinese cybersecurity Web sites, (ii) Google Translate converts each article to English, (iii) Entity and relation extraction of nine STIX and ATT\&CK and five relationships, (iv) Finally, a pre-trained BERT model fine-tunes with labeled data to capture relationships between entities in the articles. The evaluation of 150 articles using BERT, RoBERTa, and ALBERT demonstrates that BERT outperforms its counterparts, showing superior performance. The paper ~\cite{wang2021method} proposed an information extraction process that involves two aspects. First, it employs the Dictionary Template-BERT-BiLSTM-CRF (DT-BERT-BiLSTM-CRF) entity extraction model with a CRF to obtain entity labels. It creates a comprehensive dictionary template containing 3857 entries of entities, such as hacker groups, malware names, attack types, and security teams. The dictionary template is used to enhance the accuracy of the labels predicted by CRF. Second, the relationship extraction model is based on distant supervision and reinforcement learning, with one of its essential components being the Piecewise CNN classifier. To improve the quality of instance/sentence selection, reinforcement learning assigns rewards to the sentences based on the F1 score. This rewards function also improves sentence selection.

The author in~\cite{guo2021cyberrel} introduces CyberRel, a \textit{joint entity and relation extraction model} for cybersecurity concepts. CyberRel tackles the problem by performing sequence labeling, generating unique labels for relations that capture crucial information about the entities and their roles. CyberRel uses a pre-trained BERT model to create word embeddings, subsequently leveraging a BiGRU neural network with an attention mechanism to extract semantic features. Lastly, combining BiGRU and CRF constructs cybersecurity triples. CyberRel achieves an F1 score of 80.98\%, surpassing the performance of the conventional pipeline model.

\par \textbf{Vulnerability detection, and exploitation:}
Predicting the exploit potential of vulnerabilities is vital for decision-makers to prioritize their actions and address critical flaws. However, the limited size of the vulnerability descriptions makes it challenging to train a comprehensive NLP model. ExBERT \cite{yin2020apply} is an enhanced BERT model tailored for \textit{exploitability prediction}. The approach involves word piece tokenization for embedding extraction, addressing infrequent cybersecurity terms and polysemy. Furthermore, pre-trained BERT is fine-tuned using a domain-specific dataset. Additional layers, including a Pooling Layer and Classification Layer, are added to the fine-tuned BERT model to capture sentence-level semantic features and predict vulnerability exploitability using an LSTM model. ExBERT achieves an impressive 91\% accuracy in exploitability prediction. 

\par The author proposed VE-Extractor~\cite{wei2023automated}, a tool for \textit{vulnerability analysis} and \textit{classification} that uses dynamic information from text. VE-Extractor was trained on 4638 vulnerability reports, and it can extract vulnerability events, which consist of vulnerability triggers (causes) and arguments (such as attacker, consequence, operation, location, and version). The triggers and arguments were first manually extracted using a tagging scheme called BIOFR. Then, the tool used BERT-BiLSTM-CRF to map the descriptions to triggers. Next, FastText was used to classify the triggers into different types. Finally, BERT Q\&A was used to design question templates to extract the vulnerability arguments, depending on the trigger.

\par The paper~\cite{yitagesu2021unsupervised} proposes an unsupervised approach to label and extract important \textit{vulnerability concepts} (e.g., root cause, attacker type, impact, etc.) from vulnerability descriptions. The system first constructs a vocabulary of absolute paths and relative paths, which captures the syntactic differences between sentence structures and phrase expressions. Then, a Categorical Variational Autoencoder (CaVAE) is used to learn embeddings for each path. CaVAE places identical paths in the same cluster, representing the same type of concept. Finally, vulnerability concepts are extracted based on the human labels and considering noun phrases.

\par The paper~\cite{dong2019towards} proposes a Vulnerability Information Extraction Model (VIEM) to detect inconsistencies in vulnerabilities reported in NVD and CVE, particularly related to software names and versions. Security analysts primarily rely on vulnerability reports to protect their organizations from ongoing and future cyber attacks. However, inconsistencies in the information published by different sources can be a concern, as specific versions of unpatched software expose stakeholders to more significant security risks. VIEM utilizes the Named Entity Recognition (NER) and Relationship Extraction (RE) models to address this limitation by recognizing software names, versions, and associated relationships between them. VIEM's NER model encodes the text sequence into word and character-level vectors. It leverages bi-directional GRU to predict one of the following labels for each word: (i) software name, (ii) software version, or (iii) others. Subsequently, the RE model takes the entity labels predicted from the NER model and creates all possible combinations of software names and versions. It encodes this combination as one-hot vectors. Finally, the RE model takes a sequence of vectors as input and, using a hierarchical attention mechanism predicts if the pairing of the software name and version is appropriate.

\par Cyberhunters and security analysts rely on diverse sources to comprehend threats and formulate mitigation strategies. However, vulnerabilities originate from one document, while exploitation and countermeasures exist in others. Thus, manually establishing connections between documents poses a challenging and intricate task. The work~\cite{hemberg2022sourcing} identifies available entity information in one document and deduces missing data from other documents using NLP techniques. It accomplishes this by analyzing free-form text to infer concealed relationships that illustrate threat actions and mitigation strategies. The investigation was conducted on the BRON dataset~\footnote{http://bron.alfa.csail.mit.edu/info.html}. Text is subsequently represented in free form using four Language Embedding Models (LEM): Bag-of-Words, GloVe, BERT, and Fine-tuned BERT. Following this, a Random Forest classifier was trained on a dataset containing labels denoted as links and no-links. GloVe and BERT excelled in illustrating relationships, while Bag-of-Words yielded unsatisfactory results. However, GloVe yielded mixed conclusions.

\par \textbf{Action/Event extraction and detection:} CASIE~\cite{satyapanich2020casie}, a cybersecurity event extraction system, detected events in 1000 English news and CTI texts by using a combination of the Attention Mechanism, BiLSTM model, and linguistic features. CASIE defined five event types, their corresponding semantic roles, and the arguments that fulfill these roles. However, CASIE defined five event types but only focused on event arguments near the event triggers. This approach overlooked event arguments located further away from the event triggers despite events often spanning multiple sentences within a document. Husari et al. \cite{husari2018using} developed ActionMiner, which utilizes entropy and mutual information for threat extraction. Specifically, these information-theoretic measures are employed to identify object-verb pairs representing threat actions, such as ``execute-file", ``run-registry", ``delete-file", and so on. ActionMiner reported improved performance compared to alternate solutions employing the Stanford dependency parser solution. Also, ActionMiner faced limitations in identifying actions from CTI text containing syntactic blocks composed of verb, verb-direct object, verb-indirect object, verb-subject complement, etc. 
\par CyEvent2vec \cite{ma2022cyevent2vec} is an event-embedding framework based on Heterogeneous Information Network (HIN). The framework aims to predict events using three datasets comprising three actions: malware, misuse, and physical. The framework creates an event matrix to depict the event with attributes, where the matrix elements denote the relationship between the event and the object with attributes. Subsequently, using an autoencoder, intricate relationships between the heterogeneous objects are obtained, which produces low-dimensional embeddings. This autoencoder comprises many sub-encoders, each designed to generate embeddings for the object or attribute. Finally, to map the event matrices to the embedding space, hidden layer embedding of sub-autoencoders is combined. 
\par Ex-Action~\cite{zhang2021ex} extracts threat actions from CTI reports by applying rule-based matching and multimodal learning. It identifies each threat action as a triple (subject, object, verb) obtained by matching syntactic rules. The subject and object correspond to security entities, and the verb describes the semantic relationship between the entity pair. The TF-IDF and BM25 algorithms calculate the similarity between candidate actions and a CTI report. Ex-Action also identifies threats using multimodal learning algorithms. To evaluate the completeness of the threat action obtained using the proposed method and manual extraction, Ex-Action utilizes Normalized Mutual Information (NMI). Experiments on 243 CTI documents from ATT\&CK demonstrate the efficacy of the proposed solution, which provides 79.09\% accuracy and 85.26\% NMI.

\par \textbf{Malware behavior :} Huang et al.~\cite{huang2021open} used the MITRE website to develop MAMBA, a Malicious Behavior Analysis system. They employed DL and ATT\&CK knowledge to uncover TTPs and create behavior profiles for Windows malware. MAMBA has three phases (a) Extraction: Knowledge from ATT\&CK is combined with malware traces from sandbox execution, (b) Fusion: ATT\&CK info and execution traces are merged using PV-DM embedding and (c) Threat ID: Malicious behaviors (TTPs) in malware traces, involving API calls, are identified. The model achieved over 90\% precision and recall. 

\par CTDroid~\cite{fan2019ctdroid} employs NLP technologies to comprehend the behavior of Android malware as documented in security blogs, thereby enhancing the detection and classification of such malware. The proposed approach specifically addresses the semantic gap between natural language and programming language. Initially, textual data is gathered from technical blogs published between 2011 and 2017, and sensitive behaviors, represented as subject-object pairs, are extracted through Stanford dependency parsing. Subsequently, the frequency of these sensitive behaviors is estimated. Next, each malware sample undergoes reverse engineering using an open-source tool, and permissions, API calls, and intents are extracted from the AndroidManifest.XML file. Descriptions corresponding to these attributes are collected from Android's official documentation. The similarity between the frequently occurring behaviors and the behaviors extracted from the malware samples is then calculated. To assess the similarity of the textual descriptions with the Android system, words are transformed into word2Vec representations. Furthermore, the matching of sensitive behaviors extends to the code level, where keywords within function definitions are extracted, and similarity is computed as previously described. Finally, each sample is represented as a vector, utilizing sensitive behavior frequencies and a boolean vector to train five ML models. In conclusion, CTDroid achieves an impressive true positive rate of 95.8\% with only a 1\% false positive rate.

\par \textbf{Analyzing IoC Relationships:} In their study~\cite{liu2022tricti}, the authors developed a trigger-enhanced system (TriCTI) to determine the \textit{relationship between IoCs and campaign stages} from unstructured cybersecurity reports spanning 21 years (2000-2021). They successfully extracted campaign trigger phrases indicating six different campaign stages. The authors employed the BERT model to train embeddings for sentences, trigger vectors, and IoCs simultaneously. This approach was used to classify sentences into their respective campaign stages. The scarcity of data during the training stage was addressed by augmenting sentences using CBERT. The classification model reported an accuracy of 86.99\%. However, the model misclassified some examples that contained multiple campaign stages. In~\cite{kumar2023leveraging}, authors proposed a system to analyze risk in an Additive Manufacturing (AM) system. The frameworks recognizer module identifies domain-specific IoCs, which include organization\_domain\_specific, regional\_source\_specific, and regional\_target\_specific from multiple sources. Subsequently, IoCs are modeled using a Heterogenous Information Network (HIN), which consists of meta-paths and meta-graphs. HIN captures eleven semantic relations amongst IoCs. The proposed method uses similarity measures on meta-paths and meta-graphs to determine access to the threat. The system also provides a ranking of IoCs based on diverse factors such as lifetime, number of vulnerabilities exploited, and the attack frequency. 

\par \textbf{APT Analysis:} 
SECCMiner~\cite{niakanlahiji2018natural} analyzed the trend for the 13 most-mentioned techniques in its data from 445 APT reports collected from 2008 until 2017. The system used Context Free Grammar to extract all unique noun phrases and compute the TF-IDF score for each noun phrase. These TF-IDF vectors were used to determine the most identical document that matches the 13 most common APT techniques listed in the MITRE ATT\&CK knowledge base. Furthermore, SECCMiner conducted an APT trend analysis to report that PowerShell scripts are one of the rising techniques, using waterhole together with spearphishing and correlating zero-day exploits with Adobe Flash software. Recently, authors proposed ContextMiner~\cite{gutierrez2022contextminer}, a framework that extracts contextual features using a combination of part-of-speech tagging, dependency parsing, and syntactic grammar. It was evaluated for two aspects: retrieval of contextual phrases from labeled and unlabeled APT corpora, and its effectiveness as a pre-processing step for named entity recognition (NER). ContextMiner reduces the size of the original dataset by 70\%, reducing computational and storage requirements while producing the highest performance in NER.

\subsection{Knowledge Representation using Graphs}

\begin{figure}[ht]
    \centering
    \includegraphics [scale = 0.25]{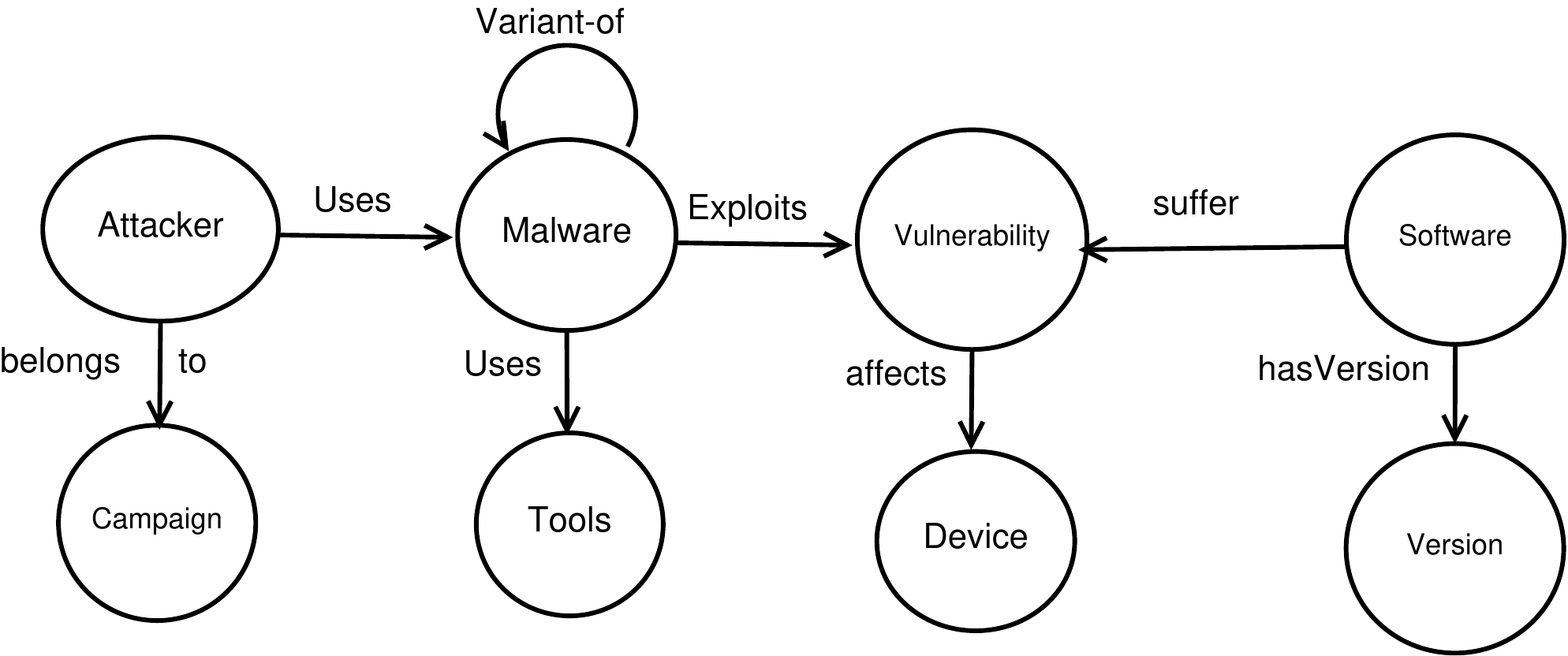}
    \caption{An example cybersecurity knowledge graph}
    \label{fig:examplekg}
\end{figure}

Cybersecurity knowledge graphs are collections of nodes and edges that carry semantic information about vulnerabilities, threat actions, attack patterns, etc. Entities such as attackers, product names, organizations, tools, and malware are represented as nodes, and the relationships between entities are represented through edges (see Figure~\ref{fig:examplekg}). Modeling textual information from different security articles as a KG offers three advantages: (1)security information from heterogeneous sources can be integrated into the KG to provide casual inference, entity correlation analysis, or semantic reasoning between the concepts, (ii)knowledge can be depicted in a structured form and visualized in a graphical format, and (iii) security analysts can query, reason on existing facts to validate consistency in data, derive novel patterns, or refine the KG by inputting new relations to improve semantic modeling.

\par
\begin{figure*}[ht]
    \centering
    \includegraphics[scale=0.26]{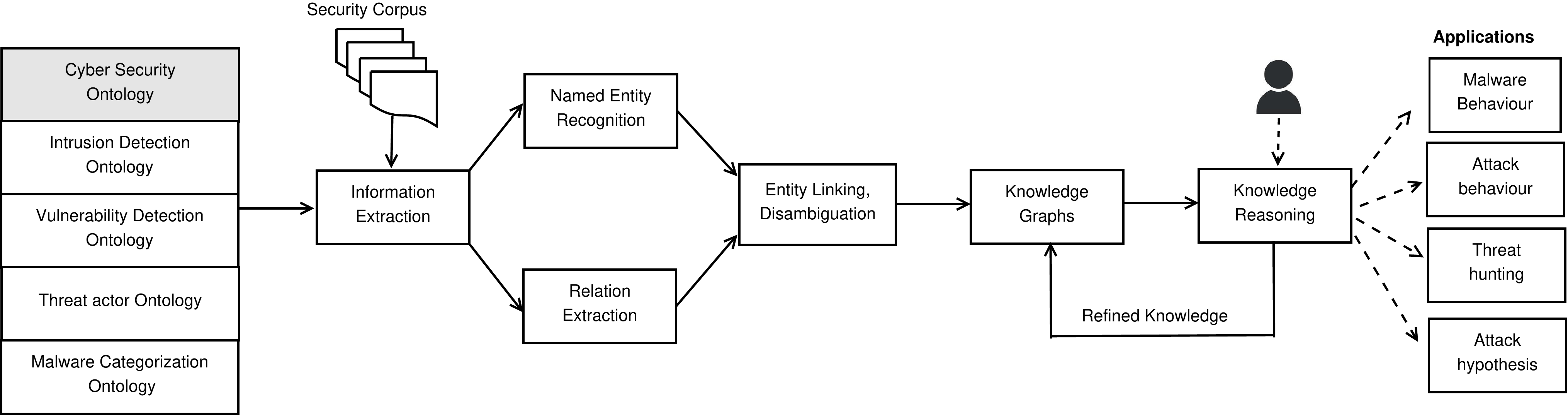}
    \caption{Building Knowledge Graphs}
    \label{fig:buildingKG}
\end{figure*}

 To construct a robust cybersecurity KG~(see Figure~\ref{fig:buildingKG}), we can begin by leveraging pre-designed cybersecurity ontologies developed by esteemed security organizations or cybersecurity experts. These ontologies are tailored for specific applications such as intrusion detection, vulnerability exploitation, and malware categorization. They comprehensively define the concepts and intricate relationships between various attributes. The subsequent step involves Information Extraction (IE), wherein the objective is to meticulously extract cybersecurity-named entities and their corresponding relationships. This extracted data then forms the foundation for constructing the KG, wherein entities are represented as nodes, and relationships are depicted as edges. Following the creation of the initial KG, it is imperative to execute entity linking and disambiguation processes to meticulously eliminate the propagation of errors and enhance the graph's precision. Moreover, to continually enhance the quality of the KG, we should actively seek timely feedback from security experts. This iterative feedback loop enables us to refine and enrich the graph with the latest insights and developments in the field of cybersecurity.
\par The authors in the paper~\cite{sills2020cybersecurity} \textit{generated a repository} of CTI concerning various medical devices and their known vulnerabilities. They augmented this repository by incorporating data from sources such as Wikidata and information regarding insecure medical devices from Global Unique Device Identification Database (GUDID). Using this augmented repository, they created a Cybersecurity Knowledge Graph (CKG). The proposed methodology enhanced the knowledge of entities and relationships derived from the Unified Cybersecurity Ontology (UCO) to represent CTI within the CKG. To devise a means of expressing node features that proved valuable in predicting their behavior within specific neighborhoods of given nodes, the authors employed the breadth-first search (BFS) algorithm. They chose BFS because it provides a local, microscopic graph perspective. Once they defined the n-neighborhood for each node, they used the node2vec algorithm to generate embeddings. As a result of their efforts, they reported a notable 31\% increase in the Mean Average Precision (MAP) value when evaluating an information retrieval task. To enable collaboration in detecting and preventing attacks, organizations must store and represent information in a structured format. The authors~\cite{li2023cybersecurity}, in their research, curated 440 publicly available cybersecurity articles to create a new CKG dataset named CS13K. Subsequently, they expanded the UCO ontology to include 16 categories. This expansion led to the creation of a KG using Neo4j, which represented $12$ types of relationships among security entities, intending to provide early warnings about real threats. KAVAS (Knowledge-Assisted Visual analytics) \cite{bohm2018graph} represents STIX Domain Objects and their relationships in a graph database, specifically Neo4j, and connects this database to the visual interface. This interface aids security analysts in \textit{comprehending and analyzing threat incidents}. KAVAS also facilitates externalization, allowing security analysts to enhance their knowledge by modifying existing threat descriptions.

The AI system Cyber-All-Intel \cite{mittal2019cyber} aids security analysts in knowledge extraction, representation, and analytics. It utilizes a `\textit{VKG structure}' that merges word embedding and KG into a vector representation. To extract information from the KG, it employs SPARQL. LADDER~\cite{alam2022looking} is an automated framework that extracts \textit{TTPs} from CTI sources related to malware and APTs. This information is organized into a KG using an ontology and TTPClassifier. The framework enables predictive analysis, threat hunting, and attribution of APT groups. TTPClassifier employs a novel ML algorithm to extract TTPs from CTI reports, including IoCs and TTPs, and categorizes them using standardized MITRE ATT\&CK pattern IDs. Satvat et al.~\cite{satvat2021extractor} developed a methodology to extract comprehensive insights into \textit{attack behavior} from CTI reports. Then, this information was organized for identifying APTs and conducting threat hunting. They actively transformed intricate descriptions in CTI reports into a provenance graph, with nodes representing entities and edges denoting system calls. They actively employed various NLP techniques such as co-reference resolution, conversion from passive to active voice, text summarization, and word standardization to achieve this transformation.
Furthermore, they actively used Semantic Role Labeling to extract attack behavior elements, including subjects, objects, and actions, ultimately facilitating the conversion of unstructured text into structured provenance graphs. Pingle et al.~\cite{pingle2019relext} proposed a solution called RelExt, a Feed Forward Neural Network (FFNN) classifier, to extract semantic triplets from cybersecurity text and develop KG \textit{to detect cyber threats}. RelExt first removes semantic triples that do not follow the semantics of STIX2.0 and entities that are not present nearby. It then predicts the relationship between the remaining entities using the FFNN classifier. RelExt achieved an accuracy of 96.61\% in identifying relationships between entities and successfully predicted 700 relationships from Dark Caracal and CrossRat malware threat reports. Modeling CTI is challenging due to the heterogeneity of cyber-threat infrastructure nodes (i.e., domain name, IP address, malware hash, and email address). To address this challenge, the authors proposed HinCTI~\cite{gao2020hincti}, a system that models CTI on a heterogeneous information network (HIN). HinCTI integrates infrastructure as nodes and the relationships between them as edges. HinCTI uses meta-schema, meta-paths, and meta-graphs to capture the semantic interdependence between infrastructure nodes. \textit{To identify the threat types} of infrastructure nodes involved in CTI, the authors define a measure of similarity between threat infrastructure nodes based on meta-paths and meta-graph instances. HinCTI also presents a heterogeneous Graph Convolutional Network (GCN) approach that uses this measure to identify the threat types of infrastructure nodes.

\par The authors introduced CSKG4APT~\cite{ren2022cskg4apt} as an APT KG for attributing cyber threat attacks. CSKG4APT's knowledge graph finds application in \textit{threat hunting} and \textit{intrusion analysis} following the diamond model. It defines 12 knowledge types, such as attackers, attack types, tools, assets, and more. The knowledge types/entities from the bilingual threat intelligence corpus are extracted using BERT-BiLSTM-GRU-CRF. Semantic relations between these entities are established by drawing from the STIX framework and NVD.

\par AttacKG \cite{liattackg}, which automatically extracts attack behavior from multiple reports to reduce human effort and speed up response time. Specifically, the authors \textit{summarize techniques} into Technique Knowledge Graphs (TKGs). They first parse CTI reports from different sources to extract attack-relevant entities (using NER) and their interdependencies. Next, they construct technique templates from the technique procedure examples extracted from the MITRE ATT\&CK knowledge base. These technique templates are also represented as graphs. Finally, they propose aligning the multiple subgraphs built from the CTI sources with the technique template graphs. They ascertain the alignment of graphs by computing (i) node alignment score and (ii) graph alignment, which determines the overall alignment score of CTI graphs with a technique template graph. The alignment of the attack graph and corresponding technique template graph enhances the attack graph with knowledge from templates and simultaneously updates the template with rich information from multiple CTIs. AttacKG accurately constructs attack graphs from reports, with F1 scores of 0.887 and 0.896 for entities and dependencies extraction. Recently, the authors presented the Open-CyKG framework in \cite{sarhan2021open}, comprising three modules: the extraction of relation triples from APT reports, NER module for labeling cybersecurity entities, and the construction of a KG. The process begins with the initial parsing of textual documents. Embeddings corresponding to words, Parts of Speech (POS), and predicates are inputted into a Bi-GRU for performing sequence labeling. The output is then passed to an additive attention layer, which assigns a higher weight to essential words in the text. Subsequently, a softmax layer outputs individual probability distributions covering different tags. The NER module further employs Bi-GRU followed by Dense layers and Conditional Random Fields (CRF) to label each entity in the cybersecurity corpus. Finally, the KG is created using the relation triples, with each node augmented with the named entity tags. The KG is disambiguated using hierarchical agglomerative clustering, and ultimately, the graph is visualized using Neo4j. 

\par The paper presents an automated method, K-CTIAA~\cite{li2023k}, for analyzing CTI. K-CTIAA retrieves cybersecurity terminology information from the KG and inputs it into the pre-trained model. Additionally, K-CTIAA offers countermeasures to aid security experts in defending against incoming cyber attacks. K-CTIAA consists of three modules: pre-processing, CTI analysis, and countermeasure modules. It converts CTI articles from various sources into a KG. It extracts named entities from query sentences to generate triples, matching pairs with the KG is also called knowledge query. The knowledge query is subsequently transformed into a sentence tree, augmenting each token with additional information extracted from semantic triples in the KG. The sentence tree is further converted into a linear structure using Depth-First (hard coding) and Branch-First (soft coding) approaches. Encoding is accomplished using BERT, considering position index via soft coding and tokens extracted through hard coding. The visible matrix in this work serves as self-attention, offering precise contextual information about the sentence. Finally, the countermeasure module suggests countermeasures based on the KG created from the ATT\&CK knowledge base.

\par To help security analysts formulate \textit{attack hypotheses} about the system's state, understand attackers' objectives, and prepare an action plan for responding quickly to an attack, the authors in~\cite{kaiser2023attack} focused on detecting higher-level IoCs. Specifically, it aims to detect techniques and then deduce IoCs from observable artifacts. They introduce a tool called the Attack Hypothesis Generator (AHG), which extracts techniques from low-level telemetry data and refines these hypotheses using link prediction. The proposed approach initiates hypothesis generation, traversing a KG if the hypotheses involve techniques closely resembling the currently observable data (COD). It generates initial hypotheses using TF-IDF (utilized for scoring techniques), Naive Bayes (to assess the relevance of techniques to the COD), and Multi-layer Naive Bayes inference (to determine connected techniques). Subsequently, the hypotheses refinement module elevates the scores of techniques frequently employed together in launching an attack. Notably, this method relies on link prediction techniques and similarity measurement metrics. Consequently, when given a KG, the link prediction identifies missing edges (relationships) that are likely to occur.

\par Vulcan~\cite{jo2022vulcan} serves as a CTI system designed specifically for ransomware attacks. It identifies cybersecurity entities through a custom NER process utilizing the Threat Entity Identifier (TEI). The TEI employs BERT, BiLSTM, and CRF models to transform the embeddings of input tokens into six labels that denote ransomware-related entities. Following the entity recognition step, the Entity Linker module performs two crucial tasks: (i) entity segmentation and (ii) entity integration. These tasks address potential issues where (i) the same entity might refer to different things or (ii) multiple variants of entities refer to the same thing. The relationships between entities are established using a pre-trained BERT model. Before training BERT, the input sentence is processed by adding special markers at the beginning and end positions of the entities. To determine the type of relation between two given entities, the relation extraction module utilizes the final hidden state vectors of the BERT model for the [CLS] token and the two entity types. Vulcan stores the extracted CTI data in a graph database and offers security practitioners search APIs to discover the relationships of a given entity with other entities. Recently, the authors of~\cite{guo2023framework} have developed a joint entity and relation extraction model for \textit{threat intelligence extraction and fusion}. They fused entities extracted from various data sources by enhancing the Levenshtein distance algorithm, imposing an additional penalty on edit operations involving numbers rather than characters, to create a preliminary CKG.

\par Zhao et al.~\cite{zhao2020cyber} proposed a HINTI framework to \textit{model the intricate relationship between IoCs} applicable for real-world applications such as (i) profiling and ranking CTIs, (ii) model attack, and (iii) vulnerability analysis. The proposed method enables the creation of a CKG, which involves the following tasks: IoC extraction, relation extraction, and cyber event detection. They propose a multi-granular approach for the extraction of IoCs. The HINTI framework, using BiLSTM+CRF, transforms the words in textual documents into different n-gram types, including character n-grams, 1-grams, 2-grams, and 3-grams, and identifies IoCs of varying lengths and types. The authors also created heterogeneous graphs to obtain granular relationships between IoCs. These graphs were able to summarize nine relationships between six IoCc. Finally, the authors proposed a threat intelligence computing framework using a Graph Convolutional Network (GCN) for effective knowledge discovery.

\par Extracting IoCs is a crucial step in thwarting cyber attacks, and it's equally vital for security operators to assess the ``\textit{maliciousness of IoCs}'' to enhance decision-making and cut down on security operation expenses. To gauge the malevolence of IoCs, Kazato et al.~\cite{kazato2020improving} employed GCN, which exploits two aspects related to IoCs: (i) the distinct features of individual IoCs, encompassing statistical, network-based, and OSINT-based attributes, and (ii) the IoC graph. Specifically, GCN utilizes individual IoC attributes, adjacency data, and label information to determine the malignancy of an IoC. Their approach successfully identified 88.3\% of malicious domains and achieved performance of 96.3\%.

\par SecKG~\cite{kriaa2021seckg} utilizes a KG to analyze event logs within a Windows system. The KG represents entities mentioned in the logs, such as processes, files, and registry entries, and the edges between the nodes depict actions between these entities. The detection module within SecKG employs rules expressed in first-order predicate logic to extract attack techniques from the KG. Subsequently, the system employs a Graph Convolutional Network~(GCN) to create graph representations of the training data, and the prediction module aims to extract subgraphs that match specific attack techniques.

\par Poirot~\cite{milajerdi2019poirot} is a system designed to emulate kernel audit log records to encompass information flows within a system, serving the purpose of \textit{threat hunting}. It comprises three crucial components: (a) the creation of a provenance graph, (b) the generation of a query graph, and (c) the alignment of these graphs. The kernel audit reports are represented as a provenance graph, with nodes representing entities such as files, processes, etc., and edges indicating the flow of information between these entities. Conversely, CTI reports sourced from various origins are transformed into query graphs, with nodes representing IoCs, files, processes, and more, and the relationships between nodes being derived from standard CTI-sharing sources. The graph alignment component employs an inexact graph pattern-matching algorithm to identify matches. Specifically, this component scrutinizes the presence of the query graph within the provenance graph. Poirot does not provide a comprehensive list of aligned paths; instead, it halts when the query identifies the first match that exceeds a predefined threshold. Additionally, constructing Poirot's query graph is a complex task that demands the expertise of a human operator.

\par CTP-DHGL~\cite{zhao2022cyber} leverages multi-source data and a heterogeneous graph to predict \textit{evolutionary cyber threats} by forecasting links within the graph. The graph is structured to encompass nodes categorized into six types, with the potential for up to 14 relations between node pairs. CTP-DHGL predicts newly emerging links within the graph, which can signify the emergence of malware variants, shifts in attacker tactics, or the evolution of vulnerabilities. The proposed CTP-DHGL framework comprises five components:~(I) Heterogeneous Graph Construction: This step involves defining nodes as cyber objects and specifying the relationships between these objects; (ii) Heterogeneous Graph Embedding: It utilizes a hierarchical attention mechanism to maintain representations at both the node-level and meta-path level. These embeddings capture the state of the graph at a specific timestamp; (iii) Transforming Embedding Graph into Temporal Sequence: The embeddings from the previous step are transformed into a temporal sequence; (iv) LSTM Encoder: It learn patterns of evolution within the graph based on the embedded sequence, and (v) Decoder: The decoder dynamically deciphers emerging patterns, revealing novel threat indicators. CTP-DHGL enhances threat prediction by applying a novel loss function that introduces a learnable penalty matrix. This ensures that the model effectively captures the dynamic evolutionary patterns in the data. The paper introduces the \textit{evolving cybersecurity knowledge
graph}, SEPSES~\cite{kiesling2019sepses}, that enables organizations to conduct vulnerability assessments effectively, helping them understand the impact of vulnerabilities on the organization and improving incident response by providing contextualized information. Specifically, this solution uses an Extraction, Transformation, and Loading (ETL) pipeline to integrate security data from various sources, both public and local, and to prepare and update a KG. SEPSES consists of four key modules: (i) Data Acquisition Module: regularly polls diverse sources to ingest fresh information into the KG; (ii) Resource Transformation Component: developed in RDF Mapping Language (RML), converts data collected from different sources in diverse formats into a common ontology, (iii) Entity Linking and Validation Module: links data from different sources that carry identical identifiers. Furthermore, it validates the generated RDF using SHACL, and (iv) the Data Storage Module periodically executes a script to record the data in triple form, enhancing the knowledge generation process.

\par Predicting cyber \textit{attack preferences} offers tremendous value, enabling security operators to formulate defensive strategies for mitigating attacks. The paper introduces a novel framework named HinAp~\cite{zhao2021automatically}, which utilizes an attributed heterogeneous attention network and transductive learning. The primary goal of this investigation is to understand the characteristics of attackers, such as their intent to infiltrate systems, their use of intrusion tools, and their affiliation with malicious groups. Specifically, the authors create an Attribute Heterogeneous Information Network (AHIN) for an attack to model the attacker, vulnerabilities, exploits, compromised devices, and invaded platforms. They use BERT to convert the attributes of nodes (vulnerability nodes and attack script nodes) into 256-dimensional vectors. Afterward, they apply node-level attention within the AHIN to determine the weights of different nodes that characterize attack preferences. The authors then formulate 20 meta-paths and five types of meta-graphs to help explore the relationships between cyber objects. They then use transductive learning to build a meta-graph-based preference model. Finally, they stack these two models together to determine attack preference by investigating node-level and structural-level features comprising meta-paths and meta-graphs.

\par Piplai et al.~\cite{piplai2020creating} extract knowledge from malware After Action Reports (AARs) and transform them into a KG. Their proposed knowledge extraction pipeline comprises three components: Malware entity extraction, relationship extraction, and KG construction based on STIX. They integrate multiple AARs that refer to identical malware and similar attacks into the KG. Suppose the entities in the current document differ from those in previously processed AARs. In that case, the system calculates term similarity using the TF-IDF score for the current document and the AARs seen before. If a close match is found, the system fuses the graph created from the recent document with the existing CKG.

\par The author~\cite{li2022novel} proposes a hybrid model for \textit{information extraction} by combining four important components: entity extraction, coreference resolution, relation extraction, and KG construction. To process a threat article, tokenization is performed, and BERT-based model encodings are obtained for each token. Additionally, POS embeddings are obtained for each token. POS-enhanced word embeddings are created by merging the word embeddings with the POS embeddings. For entity extraction, the POS-enhanced word embeddings are used as input for an attention layer to obtain contextual embeddings. Furthermore, embeddings are also generated for POS-enhanced word representations using Bi-LSTM. A linear classifier combines both the contextual embeddings and Bi-LSTM feature vector to produce entity labels. Coreference resolution is treated as a binary-class problem. POS-enhanced representations are obtained for each mention, and a CNN is employed to determine whether mentions refer to the same entity. Document-level relation extraction is approached as a multilabel classification problem. This module utilizes a type embedding matrix and a distance embedding matrix to enhance entity relation extraction, resulting in an improved document representation. Finally, Neo4j is used to visualize relationships between entities mentioned in multiple documents. Experimental results indicate a 10.56\% improvement in relationship extraction compared to state-of-the-art techniques.

\par More recently, in~\cite{dasgupta2021cybersecurity}, the authors identify incorrect assertions (\textit{outdated data/wrong information}) within the CKG by estimating scores for semantic triples (entity-head, relation, entity-tail). The conventional method for considering a triple as trustworthy involves recording temporal information about the collected data in the CKG. However, since time-related information is often absent in diverse CTI sources, the authors leverage GCN to process semantic triples without time-specific data. During the training phase of GCN, they adopt a supervised approach where each semantic triple is associated with labels. The GCN then generates a probabilistic value between 0 and 1, indicating the relation between a pair of entities. A higher score is indicative of a more credible relationship. The proposed approach periodically updates these scores to filter out outdated triples or data originating from incorrect sources.
\par To address the issue of the potential inclusion of fake CTI within unstructured open-source information, which could lead to incorrect decisions and organizational harm, the authors in~\cite{mitra2021combating} have taken steps to enhance the existing CKG. This enhancement involves augmenting the CKG with \textit{provenance graphs to ensure data authenticity}. Specifically, every node within the CKG is enriched with additional information that indicates its provenance. The process begins by creating a malware CKG, where entities related to malware are extracted using Conditional Random Fields (CRF) and regular expressions. Each entity within this graph is then subjected to a neural network classifier, which establishes links between each entity and one of six predefined relations. Next, the provenance system focuses on each node and constructs the provenance graph by augmenting it with associated information, which includes URL, Publisher, organization name, Author, Country, origin type, creation-date, and provenance score. This comprehensive approach ensures the authenticity of the collected data. Subsequently, the provenance relation encoder is employed to derive relations within the provenance graph. Finally, the provenance fusion system merges the CKG with the provenance graph, thereby incorporating a trust dimension into the KG. This augmentation helps mitigate the risk of fake CTI and supports more reliable decision-making within the organization.

%\subsection{Information Fusion and Integration(graph alignment for kg)}
%\subsection{Pattern Recognition for Visualization}
%\subsection{Temporal Analysis}
\subsection{Explainability in CTI}

Explainability plays a pivotal role in AI, serving as a vital bridge between the complexities of AI models and the understanding of end-users. While there is no universal definition for Explainable Artificial Intelligence~(XAI), its core purpose remains consistent: to render AI outcomes comprehensible and transparent. In essence, XAI encompasses a range of methodologies aimed at empowering researchers to comprehend and trust ML outcomes~\cite{srivastava2022xai}. It evaluates not only the accuracy of AI decisions but also their transparency and underlying rationale. This emphasis on understanding and trust is especially critical in the context of cybersecurity, where the adoption of AI has the potential to enhance defense mechanisms but must also maintain transparency to address evolving and intricate cyber threats effectively~\cite{zhang2022explainable}. Incorporating XAI into cybersecurity models is an essential step towards achieving this balance, ensuring that AI-driven security systems are not only accurate but also comprehensible, enabling users to make informed decisions and effectively manage the complex landscape of cyber defense. In NLP, various XAI techniques are employed to enhance the transparency and interpretability of models. Local Interpretable Model-agnostic Explanations~(LIME) generates localized explanations by perturbing input data, providing insights into why a model made a specific prediction. SHapley Additive exPlanations~(SHAP) employs cooperative game theory to assign values to features, attributing importance to words or tokens in text. Attention mechanisms in transformer models like BERT and GPT reveal which words receive the model's focus. Counterfactual explanations generate alternative texts that would lead to a different model prediction, helping users understand how slight changes in input affect the output.
\par 
In~\cite{moraliyage2022multimodal}, the authors employed explainable DL techniques to categorize onion services based on their content, including images and text. The core components of this approach consist of a CNN integrated with Gradient-weighted Class Activation Mapping~(Grad-CAM) for image analysis and a pre-trained word embedding with Bahdanau additive attention for text analysis. They underscored the importance of explainability in facilitating well-informed decision-making and discussed misclassified samples and observations of false positive predictions. Additionally, the paper highlights how certain false positives yield intriguing explanations, which were explained using Grad-CAM and attention visualization techniques. Li et al.~\cite{li2021explainable} introduced a system that integrates threat intelligence and XAI to detect APTs. This intelligence not only offers valuable insights for refining defense strategies and resource allocation but also ensures the transparency and reliability of AI-driven detection outcomes. Their evaluation of the system involved the use of datasets, including KDD'99 and UNSW-NB15, and employed LIME for providing model explanations. In[~\cite{wang2021explainable}, authors presented an approach for APT attribution using paragraph vectors and bag-of-words vectors to represent function semantics and behavior reports, respectively. They used a Random Forest Classifier~(RFC) and LIME to provide insights into the model's results.
\par
XAI is also applied to explain the TTP classification task, as demonstrated in the work by Ge et al.~\cite{ge2023explainable}. Their method introduces a self-adversarial framework, consisting of an evidence generator and a TTPs classifier discriminator. This framework extracts evidence from CTI text through a topic prototype-based keyword importance filtering technique, resulting in both an evidence set and a perturbation set. A siamese discriminator is utilized to assess these sets' impact on TTP classification, ensuring that only elements from the evidence set are accurately categorized as TTP information. They compared various explanation methods and found that LIME's local semantic approach is less effective when contrasted with methods that consider global semantic contributions. Techniques relying on attention mechanisms or topic prototypes demonstrate enhanced prediction stability compared to SHAP. Furthermore, their approach, which integrates self-adversarial training and $\theta$-sigmoid activation, proficiently eliminates irrelevant data and assigns greater significance to technical keywords. Within this range of methods, the Percent-M model, featuring a multi-topic prototype mechanism, showcases the capability to establish precise and unwavering decision boundaries while offering coherent textual explanations for classifying TTPs.

\section{Cyber Threat Intelligence Sharing and Collaboration}
\label{platform}
This section explores the diverse landscape of CTI-sharing platforms, highlighting the various options available for organizations to collaborate and exchange critical threat information. Furthermore, we delve into the adoption of standards within CTI sharing, underlining the industry's concerted efforts to establish uniform practices for enhanced information sharing. Moreover, our exploration encompasses research works aimed at assessing the quality of CTI sharing and exploring the dimensions proposed to evaluate the effectiveness and reliability of shared intelligence data. Additionally, we reviewed the strategies and mechanisms employed to achieve privacy in CTI sharing, emphasizing the importance of safeguarding sensitive information while fostering collaborative efforts to combat cyber threats.
\subsection{CTI Information-sharing platforms}

Traditionally, organizations relied on informal means like phone calls and emails for sharing threat intelligence. However, there is a growing trend towards establishing connected communities that employ dedicated platforms to facilitate the automated or semi-automated sharing of threat intelligence. Threat Intelligence Platforms(TIPs) are specialized software that assists organizations by gathering, correlating, and analyzing real-time threat information from various sources to strengthen their defensive strategies. While numerous TIPs are available in the market, most are offered under commercial licenses. VirusTotal\footnote{https://www.virustotal.com/} is the leading intelligence service, employing over 80 antivirus engines, sandbox systems, and blacklists to analyze URLs and files for a wide range of malicious content and produce threat reports. MISP~\cite{wagner2016misp}, known as the Malware Information Sharing Platform, functions as an open-source solution created to gather, store, distribute, and exchange cybersecurity indicators and threat data. It streamlines the efficient exchange of information both within the security community and beyond, offering a range of features, including an indicator database, automated correlation, adaptable data modeling, sharing capabilities, user-friendly interfaces, and compatibility with various data formats and standards. In~\cite{stojkovski2021s}, Stojkovski et al. evaluated security information sharing on the MISP platform and the importance of considering user experience to meet the varied needs of security information professionals. The investigation employed a combination of methodologies, such as the user experience questionnaire and sentence completion, to assess MISP's advantages and drawbacks. The findings indicate that the platform leans toward a technical orientation with a steep learning curve, which may pose challenges for newcomers. In another study~\cite{jolles2022building}, Joll{\`e}s et al. explored three collective intelligence dynamics on the ThreatFox\footnote{https://threatfox.abuse.ch/} platform, emphasizing the significance of trust networks in participant onboarding, highlighting the platform's growth patterns in IoC publication, and discussing the potential of a credit system as an incentive for information sharing in the cybersecurity community. ThreatFox, managed by abuse.ch, is a unique open-data TIP that promotes IoC sharing through accessibility, user-friendliness, and a public good approach. Similar to ThreatFox, abuse.ch has created additional initiatives focused on monitoring and disseminating information regarding cybersecurity threats. These projects encompass URLHaus\footnote{https://urlhaus.abuse.ch/}, designed for the exchange of malicious URLs, MalwareBazaar\footnote{https://bazaar.abuse.ch/}, which collects and distributes malware samples, and YARAify\footnote{https://yaraify.abuse.ch/}, a tool for scanning suspicious files using YARA rules. Sauerwein et al.~\cite{sauerwein2021threat} assessed nine TIPs such as ThreatStream\footnote{https://api.threatstream.com/}, ThreatQ\footnote{https://www.threatq.com/}, ThreatConnect\footnote{https://threatconnect.com/}, Open Threat Exchange~(OTX)\footnote{https://otx.alienvault.com/dashboard/new}, MISP, IBM X-Force Exchange\footnote{https://exchange.xforce.ibmcloud.com/}, Falcon X Intelligence Crowdstrike\footnote{https://www.cosive.com/capabilities/crowdstrike-falcon-x}, Collective Intelligence Framework~(CIF)\footnote{https://csirtgadgets.com/collective-intelligence-framework}, and Collaborative Research into Threats~(CRITs)\footnote{https://www.mitre.org/our-impact/intellectual-property/collaborative-research-threats-crits} by examining how they align with the CTI life cycle. Their investigative case studies uncovered that the current focus of these platforms is primarily on the pre-processing and dissemination stages. A similar study by De et al.~\cite{de2020methodology} assessed Anomali STAXX\footnote{https://www.anomali.com/resources/staxx}, OpenCTI\footnote{https://demo.opencti.io/}, CRITs, CIF, and MISP  based on architectural and CTI process. Their findings indicated that among these TIPs, \textit{MISP} and \textit{OpenCTI} were regarded as the most comprehensive and adaptable solutions. OpenCTI is an open-source platform that adheres to STIX2 standards, providing a modern Web application with integration capabilities, enabling the seamless connection and processing of both technical and non-technical threat information. Additionally, it supports data export in various formats, including CSV and STIX2 bundles. The authors also highlighted that while advanced solutions exist for CTI, none of them cover the complete CTI process. Despite the presence of outstanding options, the search for a comprehensive solution for threat intelligence-based defense remains challenging due to the diverse focuses of these platforms, which address only specific stages. Wagner et al.~\cite{wagner2018towards} conducted an analysis of 30 TIPs; they found that AlienVault offers anonymity but lacks content masking, HP Threat Central allows preprocessing and content masking, and Comilion provides architecture-level anonymity but could be susceptible to Personally Identifiable Information~(PII) analysis. Their approach involves storing CTI locally, utilizing Java regular expressions and exceptions for anonymization, and enabling sharing through an MISP instance with TOR network anonymization, permitting customization following the Traffic Light Protocol~(TLP). However, the evaluation revealed vulnerabilities to background knowledge attacks.

\par 
%Existing TIPs, whether commercial or open-source, struggle to effectively address the challenges posed by new and intricate attack patterns.
Haque et al.~\cite{haque2021toward} emphasized the importance of adopting a controlled and automated strategy for sharing CTI while highlighting the effectiveness of Relationship-Based Access Control for facilitating this sharing. They introduced an automated method for identifying, generating, and disseminating structured CTI and demonstrated these concepts through a prototype Automated Cyber Defense System in a cloud-based environment.
ETIP~\cite{gonzalez2021etip}, a platform designed for the purpose, is geared towards the collection and processing of structured data sourced from external sources, encompassing OSINT feeds, along with information originating from an organization's network infrastructure. ETIP comprises three core modules: the input module, responsible for the collection and standardization of IoCs from OSINT feeds and the monitoring infrastructure; the operational module, which correlates the data to generate enriched IoCs and evaluates threat data using a threat score; and the output module, tasked with visually presenting outcomes and sharing them with external entities to bolster cybersecurity defenses. IoC normalization and representation in MISP format are facilitated through the use of the MISP. The platform employs a deduplication technique employing \textit{contained similarity} to eliminate duplicate IoCs, consolidates related IoCs into composed IoCs and assigns a threat score to each IoC. This threat score enables Security Operations Center~(SOC) analysts to prioritize security incident investigations based on the threat score value, which factors in heuristic weights for individual attributes and overall completeness.
\par
Sharing threat intelligence has the potential to improve IT security, but it faces obstacles such as expenses, risks, and legal requirements. To overcome these challenges, Menges et al.~\cite{menges2021dealer} presented DEALER, a platform that promotes secure threat intelligence sharing by providing incentives and addressing compliance concerns. DEALER, which operates on the EOS Blockchain and InterPlanetary File System~(IPFS) distributed hash table, facilitates the anonymous sharing of well-structured incident data. It utilizes impartial quality metrics for reputation assessment and offers safeguards for both buyers and sellers through dispute resolution and cryptocurrency rewards. However, the authors recommended limiting the platform's use to sharing noncritical data, as it may not be suitable for sharing highly sensitive and critical CTI. In response to the challenges related to trust in the source and integrity of threat intelligence data, Preuveneers et al.~\cite{preuveneers2020distributed} improved the security framework TATIS~\cite{preuveneers2020tatis}. This framework guarantees that only individuals with authorization can access sensitive data when it is being transferred between various threat intelligence systems. The practical implementation of this approach involves a distributed system built on top of the MISP, where distributed ledger technology~(DLT) is employed to manage access to CTI data. To safeguard the shared data, encryption is applied using the Ciphertext-Policy Attribute-Based Encryption~(CP-ABE) cryptographic scheme.

\subsection{Standardization efforts}

\begin{figure} [htbp]
    \centering
    \includegraphics[scale = 0.12]{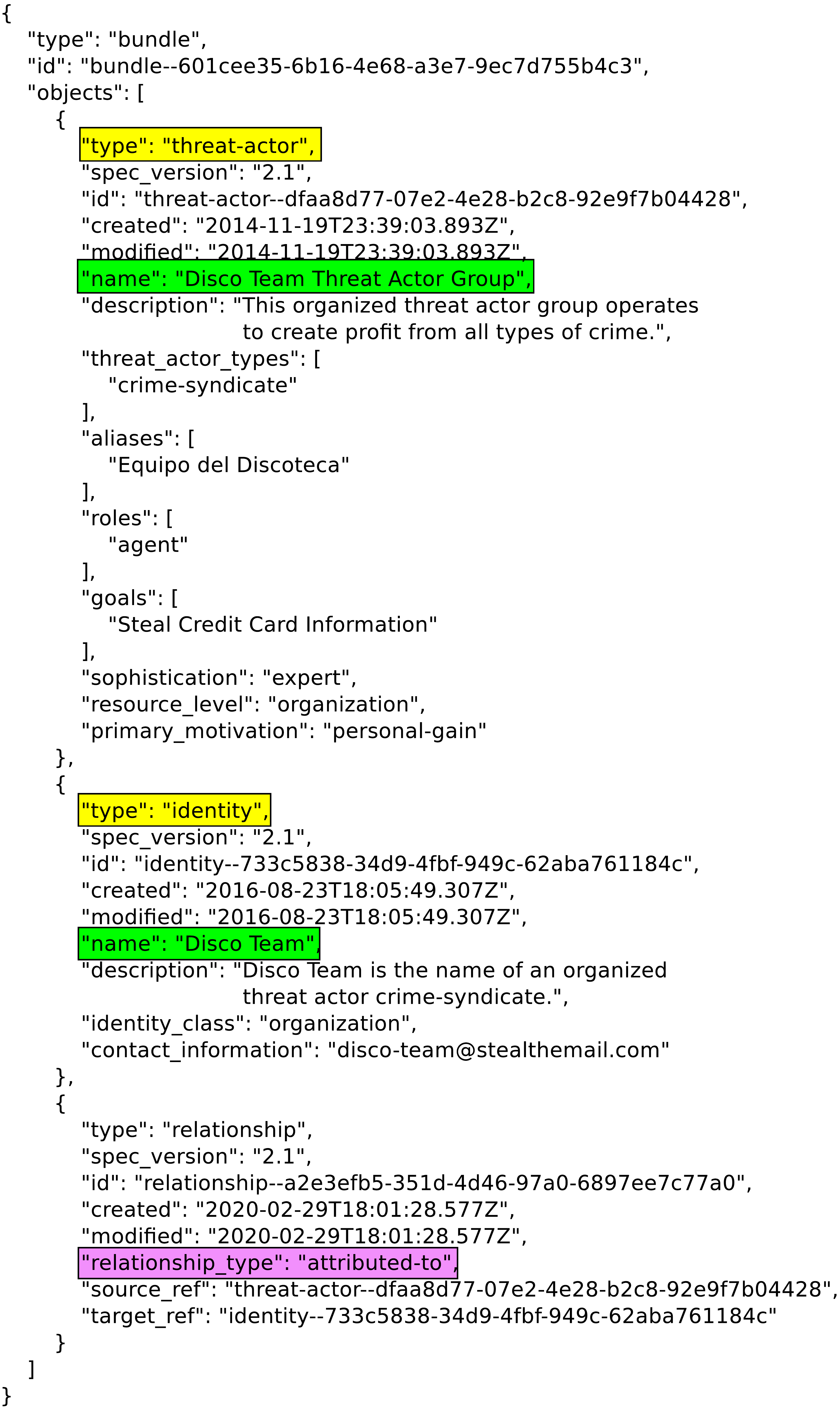}
    \caption{STIX 2.1 example representing entity relationship}
    \label{fig:stix_example}
\end{figure}

The structured representation of CTI data is beneficial, rather than sharing simple plain text. The primary aim of CTI standards is to provide a thorough methodology for describing the details of threats, attacks, and security incidents. These standards can also have specialized purposes beyond just representing security information. They may, in fact, be designed for specific operational scenarios in the cybersecurity field. One widely recognized standard for sharing CTI data is Structured Threat Information Expression~(STIX), launched by the US government and MITRE~\cite{barnum2012standardizing}. STIX acts as a format for data exchange rather than a model for data storage. This standardization, developed in collaboration with a wide range of stakeholders, specifies, captures, characterizes, and communicates standardized information regarding cyber threats. It fulfills various use cases, including analyzing cyber threats, defining indicator patterns for cyber threats, managing cyber threat response activities, and sharing CTI within communities in a standardized format. STIX utilizes JSON format to represent CTI  objects and relationships. STIX 2.1 comprises 18 objects such as Attack Pattern, Campaign, Course of Action, Grouping, Identity, Indicator, Infrastructure, Intrusion Set, Location, Malware, Malware Analysis, Note, Observed Data, Opinion, Report, Threat Actor, Tool, Vulnerability. In~\cite{piplai2020creating,ren2022cskg4apt,pingle2019relext,bohm2018graph,chen2023useful,zhou2023cdtier,dasgupta2020comparative,wang2022aptner}, STIX is used as the standard for defining entities and relationships in entity-relationship tasks. Figure~\ref{fig:stix_example} illustrates the representation of threat information in the STIX format, sourced from the STIX Web site\footnote{https://oasis-open.github.io/cti-documentation/examples/identifying-a-threat-actor-profile}. In this example, there are two distinct objects: the first is a \textit{threat actor} for ''Disco Team Threat Actor Group", and the second is an \textit{identity} type for ''Disco Team". These two objects are interconnected through an \textit{attributed\_to} relationship. Also, the illustration provides additional information, including alias names, contact details, a description, and more, to provide a comprehensive overview of these objects. In ~\cite{riesco2019leveraging}, Riesco et al. incorporated an integrated framework for conducting Dynamic Risk Assessment and Management~(DRA/DRM), employing the Web Ontology Language~(OWL), a semantic reasoner, and the Semantic Web Rule Language~(SWRL). They developed a semantic adaptation of STIX™v2.0 by integrating OWL and SWRL, facilitating the automatic generation of dependencies and the detection of threats. Data formatted in STIX can be transmitted using various communication protocols. One such protocol is Trusted Automated eXchange of Indicator Information~(TAXII)\footnote{https://oasis-open.github.io/cti-documentation/taxii/intro}, which operates at the application layer and is specifically designed to exchange threat information structured in the STIX format. TAXII, initiated by MITRE, supports various sharing models, encompassing publish-subscribe, peer-to-peer~(P2P), and hub-and-spoke approaches. Another structured language, Cyber Observable eXpression~(CybOX)\footnote{https://cybox.mitre.org/about/}, was developed for representing cyber observable events. With the introduction of STIX 2.1, CybOX has been seamlessly integrated and is now an integral part of the STIX standard. Furthermore, MITRE provides MAEC~(Malware Attribute Enumeration and Characterization)\footnote{http://maecproject.github.io/about-maec/}, a highly versatile language for sharing information about malware. MAEC enables the encoding and transmission of comprehensive information regarding malware, including its characteristics, such as attack patterns, artifacts, and behaviors.
\par
VERIS\footnote{https://verisframework.org/} serves as a standardized metrics framework aimed at establishing a universal language for systematically explaining security incidents. It tackles a prominent issue in the security sector(the scarcity of high-quality information) by aiding organizations in the collection of valuable incident-related data and its responsible and anonymous sharing with others. VERIS is organized around four fundamental components: Actor, Action, Asset, and Attribute, with a primary emphasis on documenting and evaluating internal incidents. It includes incident narratives, local incident identifiers, data on affected users, and response actions taken. Currently, VERIS is in version 1.3.1, utilizing the JSON format. Incident Object Description Exchange Format~(IODEF) serves as a structured data framework created to facilitate the exchange of information concerning computer security incidents. It is frequently employed by Computer Security Incident Response Teams~(CSIRTs) and includes both an information model and a corresponding data model that are specified through XML Schema.
 Common Vulnerability Reporting Format~(CVRF) stands out as another standard in cybersecurity, designed with a machine-readable format to facilitate structured creation, updating, and exchange of security advisories. It encompasses data related to products, vulnerabilities, and the current state of impact and remediation, thereby promoting efficient communication within the cybersecurity community. The embrace of CVRF by MITRE's CVE repository, the central hub for vulnerabilities and exposures. CVE is designed to recognize and label vulnerabilities in hardware and software products, with each CVE entry including a distinct identifier, a vulnerability description, and links to supplementary information. Open Vulnerability and Assessment Language(OVAL)\footnote{https://oval.mitre.org/} is a globally recognized initiative, freely available to the public, aimed at standardizing the assessment and reporting of computer system states in the field of information security. Its primary purpose is to provide a language and framework for standardizing the assessment and reporting of computer system states, with a particular emphasis on system vulnerabilities and configuration concerns. While OVAL is not explicitly classified as a CTI standard, it can be effectively combined with CTI standards and tools to bolster cybersecurity practices and elevate the sophistication of threat analysis.
\par 
The study~\cite{schlette2021comparative} emphasizes the significance of a systematic approach and cultivating shared understanding in the development of incident response standards. Also observing a growing shift toward structured incident response formats due to the increasing prevalence of Security Orchestration, Automation, and Response~(SOAR) products. In~\cite{menges2018comparative}, Menges et al. assessed and compared various incident reporting formats, which included STIX, X-ARF, VERIS, and IODEF, with different versions. Their primary focus lies in evaluating general criteria such as aggregability, extensibility, interoperability, and readability and additional aspects like licensing and documentation costs. Also, this paper provides a more detailed examination of structural evaluation criteria, encompassing indicators, attackers, attacks, and defenders, and illustrates how these criteria relate to typical security use cases. Remarkably, their findings identified STIX as the most comprehensive and practical choice for reporting cybersecurity incidents. The study~\cite{ramsdale2020comparative} evaluated specific CTI standards like MISP and STIX; they also included common formats such as RSS and CSV in their examination. The research revealed that even though there are established standards available, many key supporters and CTI producers still opt for creating their own simplified JSON formats, regardless of the time and formal processes required for standardization. Further, the study suggested that for most security use cases, a recommended approach might rely on the IDEA format, supplemented by MISP or STIX, as it proved effective in meeting CTI requirements. The study~\cite{jesus2023sharing} examined distinct attributes of different sharing standards, such as  STIX, IDEA, VERIS, IODEF, OpenIOC, and MISP's Format. They evaluated three categories of risks related to attribute disclosure: direct disclosure, indirect or inference risk, and accidental disclosure. Direct disclosure pertains to situations where an attribute immediately reveals sensitive information, while indirect risk comes into play under specific conditions, like having a sufficient number of related events or being susceptible to reidentification attacks. Accidental disclosure is a concern for attributes situated in fields with less stringent formatting. Excluding VERIS, the analysis showed that the risk of attribute disclosure had an upper limit of approximately 20\%, with direct disclosure typically being below 10\%, especially in the case of the two most widely used formats, STIX and MISP. VERIS displayed a high risk of accidental disclosure due to its emphasis on detailed documentation, whereas IODEF presented lower risks due to its structured approach. STIX carried higher risks of accidental disclosure or inference due to its uncontrolled input, while OpenIOC benefited from well-defined fields, mitigating potential accidental disclosure. MISP utilized secure identifiers but featured free-text fields with accidental disclosure concerns.

\subsection{Quality and Privacy in CTI sharing}

Evaluating the quality of CTI is essential for its effective sharing and utilization, as it directly impacts incident response times. Research findings emphasize the need to maintain CTI quality during collaboration~\cite{sillaber2016data}. They have pointed out that assessing the quality of CTI poses a major challenge in contemporary cybersecurity information-sharing contexts. When considering integrating a CTI source into an organization's security tools, one must consider various attributes, including the quality of the data or information it provides. In~\cite{schlette2021measuring}, the authors proposed a relevant set of quality dimensions and established corresponding metrics for evaluating CTI. They also differentiated which metric could be computed automatically and which needed security experts to assess. They categorized the quality dimensions into three levels: the Attribute Level, which focuses on specific attributes of STIX objects (e.g., evaluating timeliness using attributes like modified and created); the Object Level, which is not restricted to predefined attributes and can be assessed based on variable attribute sets (Representational Consistency) or the entire object itself (Reputation); and finally, the Report Level, which addresses experts' need to determine if a report contains an appropriate amount of data. Also, Zibak et al.~\cite{zibak2022threat} identified different quality dimensions:\textit{ timeliness, reliability, relevance, provenance, interoperability, actionability, and accuracy} through a systematic review of existing CTI literature, subsequently refining and validating them via a modified \textit{delphi} study. In another study, Qiang et al.~\cite{qiang2018quality} conducted a multidimensional quantitative evaluation of threat intelligence services with a user-centric perspective. Experts assess the criteria manually through a questionnaire and adjust their significance using a multi-objective algorithm to derive final ratings. Although this method primarily focuses on commercial CTI providers, it does not explore the granular aspects of threat intelligence. Empirical findings suggest that this approach excels in evaluating the comprehensiveness and depth of threat intelligence data feeds as well as providing an overall situational assessment. 
\par
In~\cite{schaberreiter2019quantitative}, authors assessed the credibility of CTI sources through quantitative parameters and employed a customizable weighted evaluation approach to enhance trust in these sources. The mechanism operates within a closed world assumption, evaluating each source's trustworthiness relative to others using parameters such as \textit{similarity, completeness, timeliness, compliance, interoperability, intelligence, verifiability, positives, false, maintenance,} and \textit{extensiveness}. Their approach can dynamically reassess parameters when new threat intelligence is shared, allowing for real-time trust adjustments without needing expert intervention in source validation. Griffioen et al.~\cite{griffioen2020quality} evaluated the quality of 17 open-source CTI feeds over a 14-month period and seven more feeds over seven months. They evaluated based on four quality dimensions such as\textit{ originality, impact, sensitivity,} and \textit{timeliness.} The findings reveal significant variations in performance among these feeds, providing timely and effective indicators while others exhibit delays in reporting and minimal influence on malicious activities.

\par AttTucker~\cite{li2023cybersecurity}, a graph quality assessment model based on transformers, employs multiple attention heads to capture information about entities and their relations in a low-dimensional space. In the encoding phase, the model constructs query, key, and value vectors using the embeddings of entity heads and relations, with the aid of multiple attention heads, to generate the output representation. In the decoding phase, it calculates the predicted probability for each triple. The probability vector is then concatenated with path-level information and fed into a multi-level perceptron to derive the final score for the triple.
\par
The significance of trust in threat information sharing arises from the presence of private and sensitive data in CTI. Many organizations are cautious about sharing or obtaining cybersecurity threat information because they fear becoming targets for threat actors. Threat can lead to damage to their reputation, regulatory investigations, and legal conflicts. In~\cite{wagner2018novel}, the authors introduced a trust system designed to establish a secure environment for sharing threat intelligence. They thoroughly examined popular CTI platforms and providers and validated the effectiveness of their trust framework through practical case studies. Chadwick et al.~\cite{chadwick2020cloud} proposed a trust model comprising five distinct levels within a cloud-edge data-sharing infrastructure to secure the organization's CTI data. At Level 1, organizations exhibit complete trust in all involved parties and share their CTI data as-is in plain text. Level 2 extends trust primarily to the cloud infrastructure provider but harbors partial trust in collaborators, permitting the provider to apply anonymization or pseudonymization to CTI data. Level 3 maintains full trust in the provider while assuming untrusted collaborators, necessitating the encryption of CTI data by the provider, specifically homomorphic encryption for analysis. Level 4 signifies partial trust in all parties, allowing data sharing with the condition that sensitive fields are anonymized or pseudonymized beforehand. Lastly, at Level 5, organizations express no trust in collaborators or the cloud provider conducting the analysis, mandating the homomorphic encryption of CTI data before sharing. 
\par
Several researchers~\cite{jiang2023bfls,homan2019new,purohit2022cyber,riesco2020cybersecurity,huff2021distributed} have leveraged Blockchain technologies to achieve privacy in the sharing of CTI. Homan et al.~\cite{homan2019new} presented a CTI sharing model using Hyperledger Fabric Blockchain, addressing trust issues by trusted communities through Blockchain channels and enforcing Traffic Light Protocol (TLP) sharing rules via smart contracts within the network. DefenseChain~\cite{purohit2022cyber} employed Blockchain to enhance trust in CTI sharing among various organizations. The implementation of DefenseChain took place on an Open Cloud testbed, using Hyperledger Composer, as well as in a simulated environment. Riesco et al.~\cite{riesco2020cybersecurity} employed the Ethereum Blockchain, along with the standard CTI token, to promote the exchange and dynamic sharing of threat intelligence among various stakeholders. They integrated semantic Web standards to streamline the sharing of knowledge related to behavioral threat intelligence patterns. Huff et al.~\cite{huff2021distributed} has devised a solution that allows for secure information sharing and enables entities to participate in threat-sharing communities anonymously using a distributed Blockchain ledger. The incorporation of zk-SNARKs and Sparse Merkle Trees further strengthens anonymity and the efficient utilization of tokens. BFLS~\cite{jiang2023bfls} has integrated Federated Learning~(FL)~\cite{li2020federated} with Blockchain technology to create a decentralized approach for training a threat detection model. They enhance the Blockchain's consensus protocol to select high-quality CTI models for participation in FL. The chosen models are automatically aggregated and updated using smart contracts, improving model quality and more efficient sharing.
\par

\section{Adversarial Attack}
\label{sec:attack}
CTI is essential for timely threat warnings, but if adversaries compromise it, it loses its effectiveness. CTI is typically derived from publicly available sources such as security blogs, Twitter hacker forums, the Dark Web, and security bulletins. These platforms often provide advance notice of vulnerabilities, which can be compared to standard sources like NVD or CVE. Unfortunately, adversaries can introduce counterfeit threat data to disrupt the creation of actionable CTI.  
%However, the openness of social media can sometimes be a disadvantage, as it makes the information susceptible to 'poisoning attacks,' where adversaries spread misinformation. 
This misinformation can cause the AI cyber defense system to generate contradictory information, leading to incorrect policy updates. This, in turn, can harm network and endpoint defensive rules, leading to the selection of diversionary actions to evade attacks, ultimately putting the organization at risk. Adversarial ML techniques have become a preferred tool for malicious insiders. These techniques feed deceptive inputs to ML models, causing them to make incorrect decisions or predictions. Just as adversarial attacks have gained prominence in image recognition systems, much research focuses on malicious attacks on text data. 
\subsection{Adversarial Attacks in NLP}
In~\cite{papernot2016crafting}, Papernot et al. conducted the initial study on adversarial attacks targeting text. To perform the attack, they embraced the strategy of crafting adversarial images and produced adversarial text samples by manipulating the derivatives linked to text embeddings. Earlier, researchers adapted image-based adversarial techniques to maintain semantic and syntactic integrity in a text through methods like the Greedy Search algorithm~\cite{ebrahimi2018adversarial} and Reinforcement Learning~\cite{pattanaik2017robust}. Later, they developed specialized adversarial attacks for text due to their discrete nature. Creating adversarial samples involved changing individual characters and words, effectively preserving semantic consistency and syntactic correctness. However, this approach often resulted in a lack of diversity in the generated adversarial texts. Later, researchers shifted their focus to modifying entire texts to achieve greater diversity while maintaining semantic and syntactic integrity. These methods, in general, require careful design to ensure both effective attacks and high-quality adversarial texts.

\subsection{Adversarial Examples Generation}

\begin{figure} [htbp]
    \centering
    \includegraphics[scale=0.22]{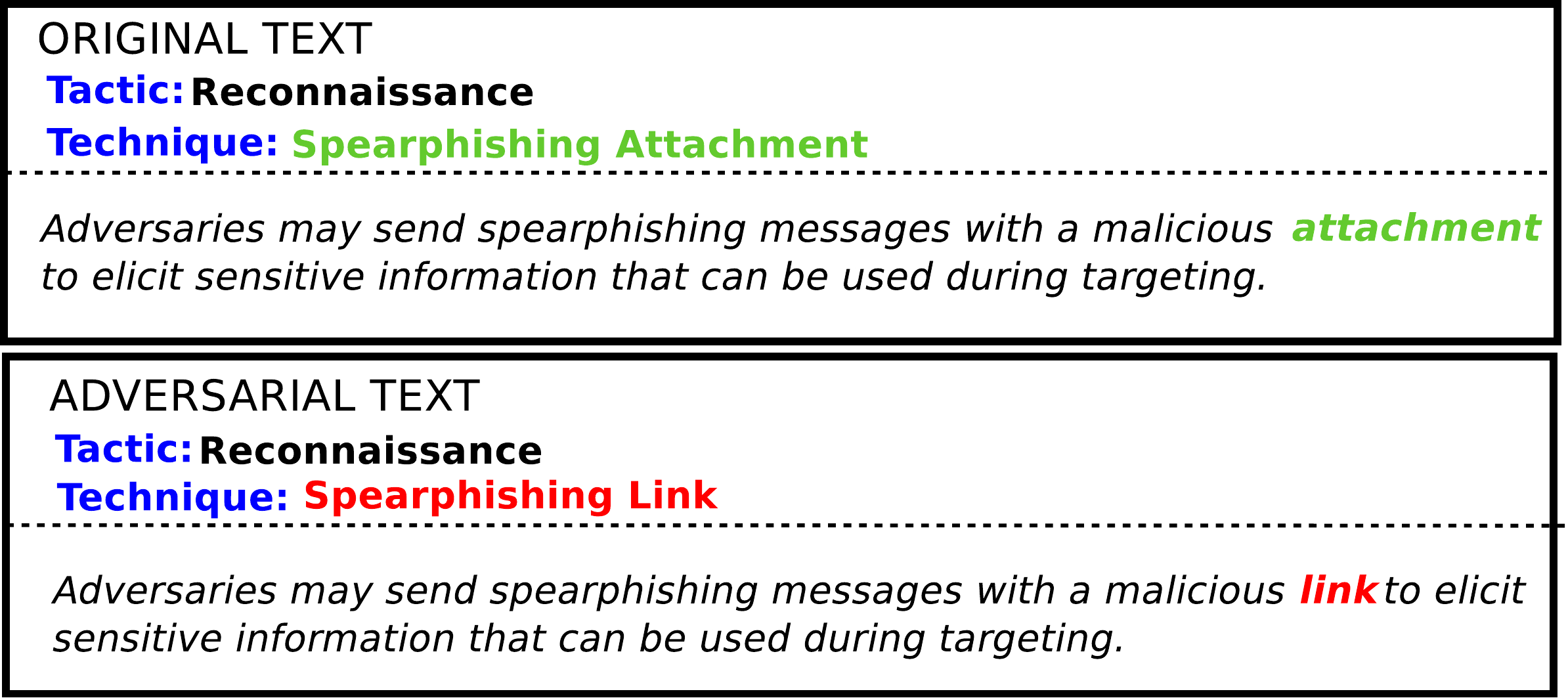}
    \caption{Adversarial example of a CTI text}
    \label{fig:adversarial_example}
\end{figure}

An adversarial example is a deceptive text resembling a benign input designed to deceive AI models. Compared to the generation of adversarial images, crafting adversarial text is more challenging. Various strategies are used in constructing adversarial instances for NLP tasks, encompassing alterations at the character, word, and sentence levels~\cite{qiu2022adversarial,alshemali2020improving}. Common strategies entail character replacement, the addition of random characters or punctuation marks, the removal of characters or punctuation, character exchange, and the random reshuffling of character order within tokens. In image perturbation, even subtle pixel changes can render the image imperceptible to the human eye.
In contrast, text-based adversarial examples can often appear visually or semantically similar to genuine content. When we apply character-level perturbation, these perturbations become more noticeable and easier to detect~\cite{zhang2020adversarial}. For instance, spell-checker tools can easily identify typographical errors, and if words or characters are altered, it can result in grammatical errors or change the meaning that can be detected. Aside from character-level changes, several strategies focus on modifying the text's semantic content. This involves substituting synonyms or antonyms for verbs, adverbs, and adjectives, as well as altering verb negations or tenses. Remarkably, even the elimination of frequently used stop-words can impact the performance of models. Another strategy entails injecting various sentence types into the text, such as grammatically correct human-approved or ungrammatical sentences. Adversarial attacks may also include rephrasing statements or intentionally removing essential sentence components. These strategies aim to provide small but effective changes to text inputs to mislead DNN models in NLP tasks. The figure~\ref{fig:adversarial_example} illustrates an adversarial example generated from the CTI text describing an attack technique, specifically \textit{spearphishing attachment}. A slight modification has been made to create this adversarial example, primarily involving the substitution of ``malicious attachment" with ``malicious link". This substitution maintains the overall meaning of the sentence, but it transforms the adversarial example to describe a different attack technique, namely, \textit{spearphishing link}. This change in wording can influence how individuals perceive and respond to the threat, emphasizing the importance of being cautious of both email attachments and suspicious links. Such adversarial attacks can impact tasks like attack tactics and technique classification.

\subsection{Poisoning Attacks and Defense Techniques}
Adversarial techniques encompass various categories, including poisoning attacks, inversion, functional extraction, and evasion. Among these, data poisoning attacks are notably prevalent and disruptive in the context of CTI. One specific example of such an attack is the VirusTotal poisoning attack, exemplified by the McAfee Advanced Threat Research team~\cite{vt_report}. Since many intrusion detection and security research efforts rely on data from VirusTotal, such attacks can significantly impact the CTI process. In a data poisoning attack, the objective is to contaminate the training data of AI systems through various methods. This malicious contamination can involve the inclusion of fake data into the AI system's training corpus. By doing so, attackers can manipulate the system's learning process, causing it to perform poorly on genuine data. For instance, in a recent study~\cite{ranade2021generating}, researchers utilized the GPT-2 model to generate fake CTI data. They collected data from security blogs, vulnerability databases, and APT reports. This collected data was used to fine-tune the GPT-2 model, which was then employed to produce fabricated CTI data. The generation process was initiated with a provided prompt, which served as input to the fine-tuned GPT-2. This model utilized tokenized prompts, normalization layers, attention layers, feed-forward neural networks, and softmax layers to produce fake CTI data. The generated fake CTI data was subsequently utilized to demonstrate a data poisoning attack on a knowledge extraction system. 

\par 
To tackle the adversarial attack, the work in paper~\cite{khurana2019preventing} proposed method employs Support Vector Classifier~(SVC) and an embedding model to assess the reputation of threat intelligence before integrating it into the AI system. Specifically, this work aims to determine the credibility of Reddit posts and `Redditor features'. The authors collected a diverse corpus of cybersecurity data and converted each Reddit post into a vector using a 'bag of words' representation. The embeddings for each word were obtained from an existing system known as `Cyber-All-Intel'. Consequently, each post is represented as the sum of cybersecurity term vectors. From a large collection of posts, a ground truth dataset was developed through manual labeling. The authors created two clusters: one consisting of `credible' posts and the other referred to as the `non-credible' set. They developed a linear SVC model, which was trained on 1206 posts using both post and Redditor features. Finally, the ensembles of the SVC model and the word embedding model were combined to compute the reputation score and validity of the posts. The proposed system achieved an accuracy rate of 71.73\% in classifying posts as `credible' or `non-credible'.
\section{Lesson Learned}
\label{sec:lesson}
In this section, we delve into the insights gained from our survey. The lessons we have extracted from our survey are as follows:
\begin{itemize}
    \item The evolving and dynamic nature of cyber threats and attacks necessitates the use of CTI for effective defense. Security experts and practitioners gather data from various sources such as social media, blogs, the Dark Web, and hacker forums to derive threat intelligence. The survey revealed that Twitter, Reddit, and GitHub are commonly used social media sources for collecting data.
    \item Researchers develop multiple data crawlers to collect information from diverse blogs, and they explore hacker forums and the Dark Web to understand the behavior and activities of key hackers.
    \item As a significant portion of threat intelligence data is unstructured, researchers employ NLP  techniques such as stop-word removal, lemmatization, stemming, and word vectorization using techniques like TF-IDF, Word2Vec, Doc2Vec, GloVe, and BERT.
    \item Data collection from open sources may lead to the inclusion of irrelevant information. Text classification plays a crucial role in generating relevant CTI by filtering out such irrelevant data.
   \item Dark Web platforms, a primary source for attack tools and malware, often use non-English languages for communication. Some studies have explored cross-lingual threat intelligence to overcome language barriers.
   \item Cyber entities are primarily extracted using a combination of models, such as BiLSTM, CRF, and BERT embedding. Researchers employ pipeline and joint entity recognition models to define relationships between these entities. The joint entity relationship techniques require further refinement.
   \item Visual representation of cyber security entities and their relationships is achieved through knowledge graphs. Researchers have employed tools such as Neo4j, Grakn, and Knowledge Graph Query Languages (Graql, SPARQL, Cypher) for this purpose.
  \item While XAI offers a valuable means to interpret model decisions, its widespread adoption remains limited. Incorporating XAI can significantly enhance the transparency and understanding of threat intelligence models.
  \item Extracted TTPs and threat intelligence are often shared in standardized formats, like STIX, to enhance collaboration among cybersecurity professionals.
  \item The establishment of standardized quality metrics and the enhancement of data-sharing privacy are imperative for building trust and promoting collaboration within the cybersecurity community. Additionally, the integration of blockchain and federated learning approaches can significantly enhance privacy.
  \item As the primary source of CTI data is social media and security blogs, it is inherently susceptible to adversarial attacks. Developing efficient adversarial defense techniques is a critical need in this field.
\end{itemize}
\section{Challenges and Open Research Directions}
\label{sec:challenges}

In this section, we delve into the primary challenges that researchers and cybersecurity professionals encounter in CTI, recognizing the complexities and intricacies involved in safeguarding our digital world. These challenges range from the handling of vast and diverse data sources to preserving privacy and from ensuring data quality to addressing the constant evolution of threat tactics. 
\subsection{Handling Unstructured Data}
CTI data primarily originates from sources like social media, blogs, and threat reports. This data is often unstructured, making it essential to efficiently analyze, aggregate, and connect various data streams. To gather this data, Web crawlers are employed, and the researchers~\cite{zhao2020timiner,zhao2021automatically} utilized multiple crawlers for this purpose. However, implementing Web crawlers for data collection poses challenges due to the technical complexities and the ever-changing nature of online content. Moreover, many threat reports contain extraneous and irrelevant information, making it challenging to extract the essential details related to attack behaviors. To address this issue, automatic solutions are needed. The researchers~\cite{kristiansen2020cti,hossen2021generating} in their study employed supervised classifiers to extract pertinent data, a process entailing the manual annotation of CTI data. Manual data annotation is both time-consuming and resource-intensive, and this becomes particularly challenging due to the dynamic nature of cyber threats, which demands continuous annotation efforts. Additionally, certain studies~\cite{kuehn2023threatcrawl} incorporated transformer-based models for classification, including cybersecurity-specific models like SecBERT, CySecBERT, and CyBERT. Nevertheless, there exists a space for research to develop models capable of effectively handling emerging cybersecurity attack terminology.
\subsection{Adversarial Data Injection}
As data primarily originates from open sources, it is imperative to acknowledge the potential threat posed by malicious actors who might deliberately inject poisoned or noisy data to disrupt the integrity of the CTI system. The impact of such fake data can be profound, affecting not only the quality of CTI but also the efficacy of threat mitigation strategies. However, only a limited number of studies have undertaken experiments to assess the impact of adversarial attacks~\cite{ranade2021generating} on CTI applications and the development of corresponding defense mechanisms~\cite{khurana2019preventing}. This highlights a notable deficiency within the CTI field, emphasizing the urgent need for robust defense mechanisms capable of shielding against poisoned data introduced by adversaries during adversarial attacks. 
\subsection{Standardized Quality Assessment} In CTI, several quality challenges have emerged, highlighting critical research directions for the future. One key challenge is the overwhelming volume of data in the digital realm, making it difficult to distinguish genuine threats from noise. Additionally, the reliance on open-source information can introduce issues of verification and accuracy, potentially leading to false alarms or missed threats. The timely nature of threat intelligence is paramount, as threat actors constantly evolve, rendering outdated information obsolete and potentially leaving systems vulnerable. Addressing bias in threat intelligence sources is another critical research direction, requiring methods for source credibility assessment. To address these challenges, ongoing monitoring, data verification, and the development of standardized quality assessment methods are necessary, ultimately enhancing the overall value of CTI for informed decision-making and proactive cybersecurity defense.
\subsection{Privacy and Ethical Considerations}
When dealing with CTI, several important considerations come into play, particularly regarding privacy and legal issues related to data sharing.  To safeguard cyber infrastructure against intentional threats, a collaborative effort involving cybersecurity professionals, researchers, and various organizations is essential. However, organizations may be hesitant to share information due to concerns about potential reputational damage resulting from disclosing details of cyberattacks. To address these concerns, initiatives like TISP provide preliminary trust-building mechanisms, but these are currently limited to group-based access control and ranking systems\cite{abu2018cyber}. Researchers are actively exploring techniques to enable organizations to share threat intelligence while preserving the confidentiality of sensitive information, employing approaches such as federated learning and Blockchain~\cite{li2020federated,homan2019new,riesco2020cybersecurity}. Federated learning, which facilitates collaborative ML model development without sharing raw data, holds promise for enhancing collective threat detection while upholding data privacy. Another noteworthy strategy, differential privacy, ensures that the inclusion or exclusion of specific data points has minimal impact on overall outcomes, making it a viable option for safeguarding individual privacy while extracting valuable insights from CTI. Furthermore, researchers are developing sophisticated access control mechanisms that restrict access to specific CTI data to authorized entities, employing fine-grained access controls and consent management. Nevertheless, further research is essential to foster trust in CTI sharing, exploring methods to establish confidence among sharing entities and ensuring the responsible and defensive use of shared data while mitigating the risk of misuse.
\subsection{Multilingual NLP for Threat Intelligence}
Enhancing the capabilities of NLP models to effectively analyze and understand threat intelligence across language barriers remains a pressing challenge. While English is commonly used, it is essential to recognize that the Dark Web, a hotspot for cyber threats, hosts content in a multitude of languages. Notably, there is a limited amount of research that delves into the linguistic analysis of non-English content on the Dark Web~\cite{ranade2018using,ebrahimi2018detecting}. The challenges are exacerbated when considering text written in languages like Russian but using English characters. This linguistic diversity poses unique challenges in deciphering and extracting meaningful threat intelligence from non-English text. The multilingual nature of the Dark Web underscores the importance of cross-lingual threat analysis, which remains an active area for open research.
\section{Summary and Conclusion}
\label{sec:conclusion}
\begin{figure}[ht]
	\centerline{
        \includegraphics[scale=0.4]{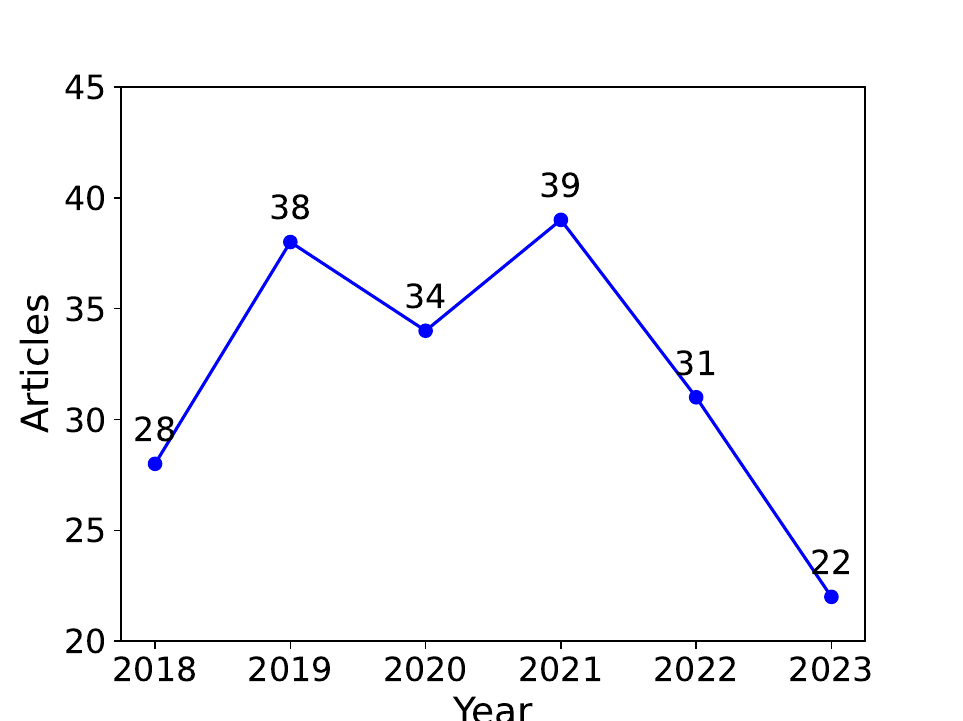}
    }
    \caption{Literature timeline} \label{fig:summaryPerYear}
\end{figure}

Cyber Threats Intelligence is made up of a powerful set of tools offering valuable insights into possible cyber threats and attacks, empowering the adoption of proactive defense measures. The continuous practice of CTI demands vigilant monitoring and constant adjustments to keep pace with the ever-changing threat environment. For this reason, recently, advanced ML methods have been applied to this field. In particular, NLP techniques have come in aid to generate robust and actionable intelligence when it comes to data collection from heterogeneous and textual data sources, data pre-processing, extraction of features, classification, clustering, and entity recognition. 
To give a contribution in this setting, in this survey, we provided a detailed review of the most significant works focusing on NLP-based techniques for CTI. In accordance with our proposed selection criteria,
we thoroughly review and discuss the most relevant and recent works, including the ones dealing with the definitions and principles of CTI, the main data crawling techniques to extract data from main CTI Web sources, and NLP's models, techniques, and applications within the CTI domain. Furthermore, the survey also addresses the topics of CTI standardization protocols and results sharing. 
Finally, we discussed the main challenges and promising future research directions.
In summary, we analyzed 192 articles published in renowned international conferences, journals, symposiums, and workshops with a focus on NLP-based techniques for CTI or related areas. Table \ref{tab:summaryPerTopic} depicts a quantitative overview of the reviewed literature divided into topics, whereas Figure \ref{fig:summaryPerYear} visualizes the analyzed number of articles published per year.

\begin{table}
\caption{Amount of papers analysed per topic}
\scriptsize
\centering
\begin{tabular}{|l|l|}
\hline
    \textbf{Topic} & \textbf{Amount of papers} \\
    \hline \hline
    Preprocessing and Tokenization & 14\\
    Text Representation Techniques & 14\\
    NLP Library and Tools & 19\\
    Crawling from Clear Web & 15\\
    Crawling from Social Media & 24\\
    Crawling from Dark/Deep Web & 26\\
    Text Classification & 21\\
    Text Similarity and Clustering & 11\\
    Text Mining and Open Information Extraction & 4\\
    Cross-Lingual Threat Intelligence & 5\\
    Topic Detection and Trend Analysis & 10\\
    Text Summarization & 4\\
    Name Entity Recognition & 30\\
    Event Identification & 10\\
    Relation Extraction & 23\\
    Knowledge Representation using Graphs & 26\\
    Explainability in CTI & 4\\
    CTI Information-sharing platforms & 9\\
    Standardization efforts & 13\\
    Quality and Privacy in CTI Sharing &13\\
    Adversarial Attack & 2\\
    \hline\hline

\end{tabular}
\label{tab:summaryPerTopic}
\end{table}

The study of the research works explored in this paper can be regarded as a foundation, as we plan to continue our investigation by diving into particular aspects only mentioned in this survey. For instance, an interesting direction can be the review of the paper exploiting existing CTI industrial platforms to give the reader a larger spectrum of diverse problems pertaining to them. Moreover, an extensive and exhaustive description of all the Authoritative Cybersecurity Databases available online is also a demanding task. 

We sincerely hope that this piece of work can aid both researchers and practitioners in capturing the pivotal elements of this domain, clarifying the most significant progress, and highlighting future research advancements.

\section*{Acknowledgment}
This work was supported in part by the project SERICS (PE00000014) under the NRRP MUR program funded by the EU-NGEU, and by the Italian Ministry of University and Research through the PRIN Project ``HOMEY: a Human-centric IOE-based framework for supporting the transition towards industry 5.0'' (code 2022NX7WKE), and by the HORIZON Europe Framework Programme through the project ``OPTIMA - Organization sPecific Threat Intelligence Mining and sharing'' (101063107).

\bibliographystyle{unsrt} 
\bibliography{biblio}

\end{document}